\documentclass[doublespacing]{lsuthesis}

\usepackage{graphicx}
\usepackage{amsmath}
\usepackage{amsthm}
\usepackage{dcolumn}
\usepackage{bm}
\usepackage[colorlinks=true, linkcolor=black, citecolor=red, urlcolor=black]{hyperref}

\usepackage{ragged2e}
\justifying
\usepackage{ulem}
\usepackage{amsfonts}
\usepackage{siunitx}
\usepackage{diagbox}
\usepackage[numbers,sort&compress]{natbib}
\usepackage{etoolbox}
\apptocmd{\thebibliography}{\setlength{\itemsep}{0pt}}{}{}

\newcommand{\bra}[1]{\langle#1|}
\newcommand{\ket}[1]{|#1\rangle}
\newcommand{\braket}[1]{\langle#1\rangle}
\newcommand{\pare}[1]{\left( #1 \right)}
\newcommand{\abs}[1]{\left\vert #1 \right\vert}
\newcommand{\cor}[1]{\left[ #1 \right]}

\DeclareMathOperator{\sinc}{sinc}
\DeclareMathOperator{\Tr}{Tr}

\usepackage{lsutitle}
\title{Multiparticle Quantum Plasmonics: Fundamentals and Applications}
\thesistype{Dissertation}
\department{The Department of Physics and Astronomy}
\soughtdegree{Doctor of Physics}
\author{Mingyuan Hong}
\degrees{B.S., University of Science and Technology of China, 2019}
\graduationdate{May 2025}

\begin{document}

\frontmatter

\maketitle

\begin{centeredpage}
\copyright\ 2025\\
Mingyuan Hong
\end{centeredpage}



\chapter{Acknowledgments}

First and foremost, I would like to express my deepest gratitude to my advisor, Prof. Omar S. Maga\~{n}a-Loaiza, for his invaluable guidance, unwavering support, and insightful mentorship throughout my PhD journey. His vision, encouragement, and patience have shaped not only this thesis but also my growth as a researcher.

I am especially thankful to Prof. Chenglong You, whose collaboration, technical expertise, and constant willingness to help have been instrumental to my work. I truly appreciate the many discussions, troubleshooting sessions, and shared ideas that made challenging tasks much more manageable.

I would also like to thank my colleagues and friends: Dr. Narayan Bhusal, Dr. Fatemeh Mostafavi, Riley B. Dawkins, Jannatul Ferdous, Dr. Ashe Miller, Justin Woodring, Ian Baum and Dr. Michelle L. J. Lollie, for their helpful conversations and support that made the lab a great place to work and grow.

I am grateful to my collaborators — Prof. Roberto de J. Le\'on-Montiel and Prof. Mario A. Quiroz-Juarez — for their contributions and shared insights throughout the projects we worked on together. Their expertise and cooperation broadened the scope of my research and strengthened the outcomes of our work.

I would like to sincerely thank Prof. Hwang Lee, Prof. Martin Feldman and Prof. Rongying Jin for their thoughtful feedback, questions, and encouragement throughout my research. Their participation in my general and final defenses was deeply appreciated.

I would also like to express my sincere appreciation to the late Prof. Jonathan P. Dowling, who gave me the opportunity to join the Ph.D. program at LSU. His support and encouragement at the beginning of this journey meant a great deal to me. I gratefully acknowledge his role in shaping the start of my academic path.

Finally, I am deeply grateful to my parents for their unconditional love and support throughout this journey. Their belief in me has always been a source of strength and inspiration.

\tableofcontents

\chapter{List of Publications}

[1] \textbf{Mingyuan Hong}, Riley B. Dawkins, Benjamin Bertoni, Chenglong You, and Omar S. Maga\~{n}a-Loaiza. Nonclassical near-field dynamics of surface plasmons. \textit{Nat. Phys.} \textbf{20}, 830–835 (2024).

\begin{itemize}
    \item Chosen as featured article and highlighted in \textit{Nature Physics News \& Views}.
    \item Selected by the Optical Society of America (now OPTICA) and highlighted in \textit{Optics \& Photonic News} December issue, as one of the most exciting works of 2024 in optics.
\end{itemize}

[2] \textbf{Mingyuan Hong}, Ashe Miller, Roberto de J. León-Montiel, Chenglong You, and Omar S. Maga\~{n}a-Loaiza. Engineering Super-Poissonian Photon Statistics of Spatial Light Modes. \textit{Laser \& Photon. Rev.} \textbf{17}, 2300117 (2023).

[3] Chenglong You, \textbf{Mingyuan Hong}, Narayan Bhusal, Jinnan Chen, Mario A. Quiroz-Juárez, Joshua Fabre, Fatemeh Mostafavi, Junpeng Guo, Israel De Leon, Roberto de J. León-Montiel, and Omar S. Magaña-Loaiza. Observation of the modification of quantum statistics of plasmonic systems. \textit{Nat. Commun.} \textbf{12}, 5161 (2021).

\begin{itemize}
    \item Chosen as featured article and highlighted in \textit{Nature Physics News \& Views}.
\end{itemize}

[4] Narayan Bhusal, \textbf{Mingyuan Hong}, Ashe Miller, Mario A. Quiroz-Juárez, Roberto de J. León-Montiel, Chenglong You, and Omar S. Magaña-Loaiza. Smart quantum statistical imaging beyond the Abbe-Rayleigh criterion. \textit{npj Quantum Information} \textbf{8}, 83 (2022). 

[5] Chenglong You, \textbf{Mingyuan Hong}, Fatemeh Mostafavi, Jannatul Ferdous, Roberto de J. León-Montiel, Riley B. Dawkins, and Omar S. Magaña-Loaiza. Isolating the classical and quantum coherence of a multiphoton system. \textit{PhotoniX} \textbf{5}, 39 (2024). 

[6] Chenglong You, \textbf{Mingyuan Hong}, Peter Bierhorst, Adriana E. Lita, Scott Glancy, Steve Kolthammer, Emanuel Knill, Sae Woo Nam, Richard P. Mirin, Omar S. Magaña-Loaiza, and Thomas Gerrits. Scalable multiphoton quantum metrology with neither pre- nor post-selected measurements. \textit{Appl. Phys. Rev.} \textbf{8}, 041406 (2021).

[7] Fatemeh Mostafavi, \textbf{Mingyuan Hong}, Riley B. Dawkins, Jannatul Ferdous, Rui-Bo Jin, Ian Baum, Roberto de J. León-Montiel, Chenglong You, and Omar S. Magaña-Loaiza. Multiphoton Quantum Imaging using Natural Light. \textit{Appl. Phys. Rev.} \textbf{12}, 011424 (2025).

[8] Jannatul Ferdous, \textbf{Mingyuan Hong}, Riley B. Dawkins, Fatemeh Mostafavi, Alina Oktyabrskaya, Chenglong You, Roberto de J. León-Montiel, and Omar S. Magaña-Loaiza. Emergence of Multiphoton Quantum Coherence by Light Propagation. \textit{ACS Photon.} \textbf{11}, 3197–3204 (2024).

[9] Riley B. Dawkins, \textbf{Mingyuan Hong}, Chenglong You, and Omar S. Magaña-Loaiza. The quantum Gaussian-Schell model: a link between classical and quantum optics. \textit{Opt. Lett.} \textbf{49}, 4242-4245 (2024).

[10] Michelle L. J. Lollie, Fatemeh Mostafavi, Narayan Bhusal, \textbf{Mingyuan Hong}, Chenglong You, Roberto de J. León-Montiel, Omar S. Magaña-Loaiza, and Mario A. Quiroz-Juárez. High-dimensional encryption in optical fibers using spatial modes of light and machine learning. \textit{Mach. Learn.: Sci. \& Technol.} \textbf{3}, 035006 (2022).

[11] Narayan Bhusal, Sanjaya Lohani, Chenglong You, \textbf{Mingyuan Hong}, Joshua Fabre, Pengcheng Zhao, Erin M. Knutson, Ryan T. Glasser, and Omar S. Magaña-Loaiza. Spatial Mode Correction of Single Photons Using Machine Learning. \textit{Adv. Quantum Technol.} \textbf{4}, 2000103 (2021).


\listoffigures



\chapter{Abstract}

Quantum plasmonics explores the interaction between light and collective charge oscillations at metal-dielectric interfaces, enabling strong light confinement and enhanced quantum effects at the nanoscale. While traditional quantum optics has primarily focused on single-photon systems, an intermediate regime exists between classical and single-photon optics—multiparticle (or multiphoton) quantum optics. In this regime, classical light sources, when analyzed through techniques such as photon-number-resolving (PNR) detection and projective measurement, can reveal nontrivial quantum correlations. This thesis investigates how multiparticle quantum plasmonics harnesses these correlations to control quantum statistical properties, enhance coherence, and enable novel applications in quantum technologies.

In this thesis, we begin by establishing the theoretical foundation of multiparticle quantum plasmonics, introducing key concepts such as photon-plasmon interactions, coherence theory, and statistical fluctuations. The first study demonstrates that multiparticle scattering can modify quantum statistics in plasmonic systems, providing a new degree of control over fluctuations traditionally assumed to be preserved. The second study explores the nonclassical near-field dynamics of surface plasmons, revealing how quantum coherence arises from bosonic and fermionic contributions within isolated subsystems. The third study focuses on quantum plasmonic sensing, where a conditional detection scheme enhances the signal-to-noise ratio of weak plasmonic signals, enabling improved phase estimation for metrological applications. In the final study, we extend the multiparticle approach to quantum imaging using natural light. By isolating multiphoton correlations from thermal light fields with PNR detection and a single-pixel imaging protocol, we demonstrate enhanced image contrast even under noisy conditions. This result shows that nonclassical features can be accessed in classical light through careful subsystem projection, expanding the scope of multiparticle techniques beyond plasmonic systems.

Together, these studies highlight the fundamental role of multiparticle interactions in controlling and applying quantum statistical properties. By bridging fundamental investigations and practical implementations, this thesis contributes to the advancement of quantum plasmonics and demonstrates how multiphoton methods can drive progress in quantum imaging, sensing, and information processing.

\mainmatter

\chapter{Introduction}

Historically, most research in optics has focused on classical light, where electromagnetic waves are well described by Maxwell’s equations. However, with the advent of the first and second quantum revolutions, the focus shifted toward the quantum nature of light and its interactions with matter \cite{glauber1963,sudarshan1963,mandel_wolf_1995}. The first quantum revolution established the foundations of quantum mechanics, providing explanations for fundamental phenomena such as wave-particle duality and the quantization of energy \cite{thouless2014quantum}. Building on this groundwork, the second quantum revolution harnessed uniquely quantum features—such as entanglement and superposition—for transformative technological applications, including quantum computing, communication, and sensing \cite{obrien2009,Lodahl2017}. A key development in this era was the emergence of quantum optics, where single-photon-level experiments became feasible through sources like quantum dots and spontaneous parametric down-conversion (SPDC) \cite{Mandel:79,BurenkovPRA2017}. These sources generate photon pairs that exhibit purely quantum behaviors such as superposition, entanglement, and antibunching \cite{arrechi1967,Henny296,SchellekensScience}.

While quantum optics has traditionally focused on isolated single-photon systems, there exists an intermediate regime between single photons and classical light fields—this is the domain of multiparticle (or multiphoton) quantum optics \cite{You2021,you2020multiparticle}. Here, classical light sources can be structured into subsystems using techniques such as photon-number-resolving detection and projective measurements, allowing the study of quantum correlations within seemingly classical fields \cite{BurenkovPRA2017,SilberbergHBT,OmarOptica17,YouPhotoniX}. This approach has revealed nontrivial quantum properties embedded in multiphoton interactions, offering a new perspective on quantum coherence and entanglement \cite{Chen2021,PhysRevAmultimodethermal,You2021APR,YouPhotoniX,RimaACS2024}. Studying the correlations within these subsystems provides valuable insights into quantum imaging, quantum simulation, and quantum plasmonics \cite{AspuruGuzik2012,you2017multiparameter,magana2016exotic,Bhusal2022}. Among these, quantum plasmonics, the main focus of this thesis, explores how multiparticle quantum effects emerge in plasmonic nanostructures and how they can be leveraged for quantum technologies \cite{tame2013quantum,Holtfrerich2016}.

Quantum plasmonics is a rapidly evolving field that explores the interaction between quantum light and collective charge oscillations, known as plasmons, at metal-dielectric interfaces \cite{chang2006quantum,Maier2007}. The field originated from early experiments showing that individual photons can excite surface plasmons while preserving their quantum properties \cite{altewischer2002plasmon,schouten2005,fasel2005energy}, leading to the development of quantum-enhanced plasmonic devices \cite{Lee2021,Mostafavi2022}. Over the years, advancements in nanophotonics and quantum optics have expanded our understanding of how plasmonic systems enable strong light-matter interactions at the nanoscale \cite{barnes2003surface,schuller2010plasmonics,novotny2011antennas}, facilitating applications in quantum information processing, sensing, and communication \cite{mejia2018plasmonic,dowran2018quantum}.

A major driving force behind the growth of quantum plasmonics is its ability to bridge the gap between classical and quantum regimes of light-matter interactions \cite{di2012quantum,Heeres2013}. Recent research has demonstrated that surface plasmon polaritons (SPPs) not only preserve quantum properties but also serve as a platform for manipulating nonclassical light states \cite{cai2014high,dheur2016single,fujii2014direct}. Experimental breakthroughs, such as single-plasmon generation in nanowires coupled to quantum dots \cite{akimov2007generation}, entanglement preservation in plasmon-assisted transmission \cite{moreno2004theory}, and the propagation of quadrature-squeezed plasmons in metallic waveguides \cite{huck2009demonstration}, have reinforced the role of plasmonic systems in quantum photonics \cite{DiMartino2014,buse2018symmetry}. Moreover, plasmonic platforms have been shown to exhibit distinct quantum statistical properties, allowing for the direct study of quantum coherence and loss mechanisms at subwavelength scales \cite{Yu2019,daniel2017surface}. A particularly exciting development is the ability of plasmonic systems to modulate spatial coherence, opening new opportunities for photonic engineering beyond traditional refractive and diffractive elements \cite{li2017strong,wang2018quantum,Hong2023}. Additionally, the heralded excitation of single plasmons confirms that plasmonic structures are not merely passive carriers of quantum states but active components in quantum state manipulation \cite{safari2019measurement}. Collectively, these advances underscore the transformative potential of quantum plasmonics, establishing it as a key player in quantum sensing, secure communication, and integrated photonic networks \cite{pooser2015plasmonic,johnson2017diamond}.

A notable development in this domain is the demonstration that plasmonic systems can modify quantum statistics, indicating that quantum fluctuations are not always strictly preserved under all conditions \cite{You2021}. This thesis presents experimental evidence of such modifications in multiparticle plasmonic systems, discussed in detail in Chapter 3. In particular, we investigate how photon–plasmon interactions influence the statistical properties of light beyond the single-particle regime. Traditional experiments in quantum plasmonics have largely focused on single-particle phenomena, often assuming that multiparticle effects do not fundamentally alter quantum fluctuations \cite{Heeres2013,vest2017anti}. Our findings challenge this assumption by revealing that, in certain plasmonic structures, multiparticle scattering pathways introduce interference effects that lead to measurable changes in the excitation modes \cite{lee2016quantum}. Through the application of quantum coherence theory, we show that plasmonic near fields offer additional degrees of freedom for manipulating quantum states at the nanoscale \cite{you2020multiparticle}. These results position plasmonic platforms as versatile and tunable environments for controlling quantum correlations, with promising implications for quantum metrology and information processing \cite{You2021APR,lee2017quantum}.

With this approach to modifying the quantum statistics of plasmonic systems, we further uncover their underlying physical properties. Chapter 4 focuses on the nonclassical near-field dynamics of surface plasmons, examining the quantum coherence properties that emerge in these systems \cite{You2021, HongNP24, OPN}. While surface plasmons are typically described as collective charge oscillations exhibiting classical behavior at macroscopic scales, our research provides direct experimental evidence that their near-field interactions display distinctly nonclassical characteristics \cite{tame2013quantum,chang2006quantum}. By isolating multiparticle plasmonic subsystems, we observed that their quantum dynamics can be governed by either bosonic or fermionic coherence effects, depending on the specific scattering conditions \cite{HasselbachNature,ButtikerHBTPRL}. This finding reveals a fundamental connection between classical and quantum descriptions of plasmonic systems, clarifying how quantum coherence emerges from underlying microscopic interactions \cite{LandauerPRB,ButtikerPRL}. Furthermore, our study demonstrates that vacuum fluctuations can excite surface plasmons in ways that give rise to measurable quantum signatures \cite{vest2018plasmonic}. These insights not only advance the fundamental understanding of surface plasmons but also highlight new opportunities for engineering quantum states in nanophotonic platforms \cite{Tame2021,OPN}.

As we uncover more fundamental physical properties of plasmonic systems, we can begin exploring their potential applications, with quantum sensing being one of the most prominent and rapidly developing fields \cite{Lee2021,Mostafavi2022}. Quantum plasmonic sensing has recently emerged as a promising approach for detecting delicate samples with high sensitivity \cite{mejia2018plasmonic,pooser2015plasmonic}. In our study introduced in Chapter 5, we explored a novel sensing protocol based on conditional plasmon detection, which provides an additional degree of freedom to manipulate quantum fluctuations in plasmonic systems \cite{You2021}. By employing a plasmon-subtracted detection scheme, we demonstrated that it is possible to enhance the signal-to-noise ratio (SNR) of weak plasmonic signals, mitigating the limiting effects of field fluctuations \cite{dowran2018quantum,lee2016quantum}. Our approach enables improved phase estimation, a key requirement in quantum metrology, and has significant implications for applications such as molecular sensing, chemical detection, and ultrasensitive biosensing \cite{Mostafavi2022,Lee2021,You2021APR}. These results highlight the potential of quantum plasmonic sensing to extend beyond classical detection limits, reinforcing the role of nonclassical light-matter interactions in the advancement of high-precision quantum technologies.

Beyond plasmonic systems, the principles of multiparticle quantum optics can be extended to a variety of physical platforms. The broader implications of this framework are explored through a study on multiphoton quantum imaging with natural light, presented in Chapter 6. This investigation demonstrates how quantum correlations can be extracted from classical thermal light fields using projective measurements and photon-number-resolving detection \cite{Maga_a_Loaiza_2019,Chen2021,Bhusal2022,Mostafavi2025APR}. While distinct from plasmonic systems in terms of physical implementation, this work shares a common methodological foundation—namely, the isolation of meaningful quantum subsystems from a classically noisy background \cite{PhysRevAmultimodethermal}. By employing a single-pixel imaging technique and conditional detection, we were able to enhance image contrast exponentially under high-noise conditions—an effect unattainable through classical imaging methods alone \cite{magana2016exotic}. This approach underscores the versatility of multiphoton-based strategies and illustrates how quantum features can emerge from structured detection, even in traditionally classical sources \cite{OmarOptica17,You2021APR}. Although this chapter focuses on a different application domain, it reinforces the central message of the thesis: that multiparticle quantum optics offers a rich and flexible framework for probing and utilizing quantum effects in realistic, scalable systems.

In summary, this thesis presents a comprehensive investigation into multiparticle quantum plasmonics, highlighting its fundamental principles and practical applications. I believe our studies contribute to the broader effort of integrating plasmonic systems into future quantum technologies. The following chapters will provide some related fundamentals about quantum optics, a detailed discussion of our findings, experimental methodologies, and their implications for advancing quantum plasmonic research.

\chapter{Fundamentals of multiparticle quantum plasmonics}

In this chapter, we provide a brief review of some of the most remarkable quantum mechanical states of the electromagnetic field. We start this section by reviewing Fock, coherent and thermal states of light \cite{glauber1963,obrien2009,anno2006}. In addition, we define the degree of second-order quantum coherence. These  definitions are used later in the thesis to describe the quantum mechanical properties of multiphoton systems.

\section{Quantum states of light}

The annihilation and creation operators can be used to describe quantum states of light. Remarkable examples include Fock states, coherent states, squeezed states and thermal states. The possibility of preparing light in these states has been extensively utilized to prepare multiparticle quantum systems \cite{loudon2000quantum}.

\subsection{Fock states}
Quantum states of light known as Fock states, or photon number states, are denoted as $|n\rangle$, where $n$ represents the number of photons in a single mode of the electromagnetic field \cite{loudon2000quantum}. Interestingly, Fock states have a well-defined number of particles. Thus, we can define the particle number operator as $\hat{n} = {\hat a^\dag }\hat a$, which satisfies 
\begin{equation}
\hat{n}\left| n \right\rangle  = n\left| n \right\rangle.
\end{equation}
Moreover, the action of the creation and annihilation operators on the Fock state $|n\rangle$ is given by 
\begin{align}
\hat{a}^{\dagger}|n\rangle &= \sqrt{n+1}|n+1\rangle,\\
\hat{a}|n\rangle &= \sqrt{n}|n-1 \rangle.
\end{align}

The number state $|n\rangle$ can be obtained from the vacuum state $|0\rangle$ :
\begin{equation}
\label{eqn:15}
|n\rangle = \frac{(\hat{a}^{\dagger})^n}{\sqrt{n!}}|0\rangle.
\end{equation}
From Eq. (\ref{eqn:15}), we can also derive some properties of these quantum states. Fock states are orthonormal:
\begin{equation}
\langle n|m \rangle = \delta_{nm},
\end{equation}
and the Fock state basis is complete
\begin{equation}
\sum\limits_{n = 0}^\infty  {\left| n \right\rangle } \left\langle n \right| =  \hat{I}.
\end{equation}

\subsection{Coherent state} 

Coherent states can be used to describe photons in an ideal laser beam. These are quantum mechanical states with classical noise properties \cite{Gerry2004}. The definition of a coherent state $\left| \alpha  \right\rangle$ is given by the annihilation operator
\begin{equation}
 \hat a\left| \alpha  \right\rangle  = \alpha \left| \alpha  \right\rangle.
\end{equation}

Therefore, a coherent state is an eigenstate of the annihilation operator. Since Fock states form a complete basis, one can represent coherent states in the number basis as
\begin{equation}
\left| \alpha  \right\rangle  = \exp \left( { - \frac{1}{2}{{\left| \alpha  \right|}^2}} \right)\sum\limits_{n = 0}^\infty  {\frac{{{\alpha ^n}}}{{\sqrt {n!} }}} \left| n \right\rangle.
\end{equation}

The photon number follows a Poissonian statistical distribution,
\begin{equation}
	p(n) = e^{-|\alpha|^2}\frac{|\alpha|^{2n}}{n!},
\end{equation}
with standard deviation of $\Delta n = |\alpha| = \sqrt{\langle n \rangle}$. Hence, the average photon number of the coherent state is
\begin{equation}
\left\langle {\hat n} \right\rangle  = \left\langle \alpha  \right|{\hat a^\dag }\hat a\left| \alpha  \right\rangle  = {\left| \alpha  \right|^2}.
\end{equation}

\subsection{Thermal state}

Now, we will focus our attention to the description of thermal states. This family of states describes common light sources such as sunlight. These sources of light are characterized by classical noise properties that can be mathematically described by statistical mixtures of number states as \cite{mandel_wolf_1995}
\begin{equation}
	{\hat \rho _{\text{th}}} = \frac{{\bar n}}{{1 + \bar n}}\sum\limits_{n = 0}^\infty  {{{\left( {\frac{{\bar n}}{{1 + \bar n}}} \right)}^n}} \left| n \right\rangle \left\langle n \right|.
\end{equation}
Here, $\bar n$ represents the mean photon number of the thermal field. Furthermore, the photon number fluctuations of the thermal state is given by
\begin{equation}
    \left\langle(\Delta n)^{2}\right\rangle=\left\langle\hat{n}^{2}\right\rangle-\langle\hat{n}\rangle^{2}=\bar{n}+\bar{n}^{2},
\end{equation}
which is larger than the mean photon number $\bar n$. Thus, thermal states show super-Poissonian photon statistics.

\section{Classical and quantum coherence -- \texorpdfstring{$g^{(2)}(0)$}{g(2)(0)}}

\begin{figure}[ht!]
\centering
\includegraphics[width=0.5\textwidth]{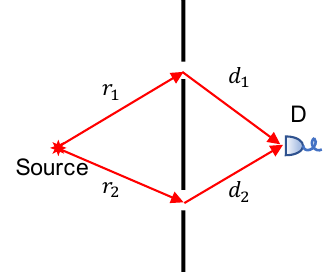}
\caption[Schematic of the Young’s double-slit setup]{Schematic of the Young’s double-slit setup. Here, $r_1$ and $r_2$ represent the distances from the source to the upper and lower slits, respectively. The distances from the upper and lower slits to the detector $D$ are represented by $d_1$ and $d_2$ respectively.}
\label{fig:Youngs}
\end{figure}

The advent of the laser gave an enormous impulse to the development of the theory of optical coherence \cite{born2013principles}. Nowadays, photonic technologies depend to an important extent on our ability to manipulate the coherence properties of the electromagnetic field. In this regard, plasmonic systems have been extensively used to engineer the spatial and temporal properties of photons \cite{tame2013quantum,chang2006quantum,Hong2023}. In this section, we provide a brief review of the concept of spatial coherence, a property that will be used in several parts of the thesis. We illustrate this concept through the famous Young's two-slit experiment. 

The double-slit experiment was introduced by Thomas Young to illustrate the wave nature of light. This beautiful experiment has also been utilized to quantify spatial coherence of light. We show a simplified version of this experiment in Fig. \ref{fig:Youngs}. We assume that a double-slit structure is illuminated with a quasi monochromatic source of light. At time $t$, the total electric field at the detector $E_d(r,t)$ is given by
\begin{align}
E_d(r,t)=a_1E(r_1,t_1)e^{ik_0d_1}+a_2E(r_2,t_2)e^{ik_0d_2},
\end{align}
which is the sum of the field amplitudes $E(r_1,t_1)$ and $E(r_2,t_2)$ produced by each of the slits. Here, $t_1=t-d_1/c$ and $t_2=t-d_2/c$ represent the times at which the photons leave the slits, and $k_0= \omega/c$ is the wavenumber in vacuum. The values for $a_1$ and $a_2$ are defined by the geometry of the slits. Then the intensity $I_{d}$ measured by the detector D is given by 
\begin{equation}
\begin{aligned}
\label{firstordercoh}
I_d(r)=&\langle E_d^*(r,t)E_d(r,t)\rangle\\
=&a_1^2I_1(r_1,t_1)+a_2^2I_2(r_2,t_2)\\
+&a_1a_2\Gamma(r_1,r_2,t_1,t_2) e^{-ik_0(d_1-d_2)}+c.c.,
\end{aligned}
\end{equation}
where $I_1=\langle E^*(r_1, t_1)E(r_1, t_1)\rangle$ and $I_2=\langle E^*(r_2, t_2)E(r_2, t_2)\rangle$. The notation $\langle\bullet\rangle$ represents an ensemble average. These quantities can be used to introduce the mutual coherence function
\begin{equation}
    \Gamma^{(1)}(r_1,r_2,t_1,t_2)=\left\langle E^{*}\left(r_{1},t_1\right) E\left(r_{2},t_2\right)\right\rangle.
\end{equation}
Indeed, it is possible to express Eq. (\ref{firstordercoh}) as
\begin{equation}
\begin{aligned}
\label{re-firstordercoh}
I_d(r)=&\langle E_d^*(r,t)E_d(r,t)\rangle\\
=&a_1^2I_1(r_1,t_1)+a_2^2I_2(r_2,t_2)\\
&+2\abs{a_1}\abs{a_2}\operatorname{Re}[\Gamma(r_1,r_2,t_1,t_2)].
\end{aligned}
\end{equation}
The first two terms in Eq. (\ref{re-firstordercoh}) are transmission contributions from the first and second slits, respectively. Furthermore, the last term in Eq. (\ref{re-firstordercoh}) describes interference. The intensities $I_1(r_1,t_1)$ and $I_2(r_2,t_2)$ provide information about self-field correlations. These can be described by the functions of first-order coherence $\Gamma^{(1)}(r_1,r_1,t_1,t_1)$ and $\Gamma^{(1)}(r_2,r_2,t_2,t_2)$. Similarly, the mutual-field correlations can be described by the function of first-order coherence $\Gamma^{(1)}(r_1,r_2,t_1,t_2)$. It is worth noting that the interference fringes are formed when the length of the spatial coherence of the illuminating beam is larger than the separation between the slits. In other words, interference fringes are produced when the properties of light are similar at the spatial locations defined by the two slits. The quality of the formed fringes can be quantified through the Rayleigh's definition of fringe visibility, 
\begin{equation}
    \mathcal{V}=\left(I_{\max }-I_{\min }\right) /\left(I_{\max }+I_{\min }\right),
\end{equation}
where $I_{\max}$ and $I_{\min}$ are the maximum and minimum intensity values in the interference pattern, respectively. The visibility $\mathcal{V}$ is equal to zero for incoherent sources. Furthermore, $\mathcal{V}=1$ describes a coherent source. Also, sources of light characterized by visibilities in the range  $0\leq\mathcal{V}\leq1$ are considered partially coherent.

A quantum formulation of coherence can be constructed using similar ideas to those described above \cite{glauber1963}. In this regard, the general first-order correlation function is defined as,
\begin{equation}
    G^{(1)}\left(r_{1}, r_{2},t_1,t_2\right)=\operatorname{Tr}\left\{\hat{\rho} \hat{E}^{(-)}\left(r_{1},t_1\right) \hat{E}^{(+)}\left(r_{2},t_2\right)\right\},
\end{equation}
where $\hat{\rho}$ is the density matrix of a quantum state, and $\hat{E}^{(+)}$ is the electric-field operator and $\hat{E}^{(-)}=[\hat{E}^{(+)}]^\dagger$ \cite{Gerry2004, scully1997quantum, loudon2000quantum}.
In addition, the normalized first-order correlation function is defined as
\begin{equation}
    g^{(1)}\left(r_{1}, r_{2},t_1,t_2\right)=\frac{G^{(1)}\left(r_{1}, r_{2},t_1,t_2\right)}{\left[G^{(1)}\left(r_{1}, r_{1},t_1,t_1\right)G^{(1)}\left(r_{2}, r_{2},t_2,t_2\right)\right]^{1 / 2}}.
\end{equation}

As discussed above, the first-order coherence function can be utilized to determine the spatial coherence of the electromagnetic field. However, additional information can be gained through the implementation of intensity correlations. In this regard, in 1956, Hanbury Brown and Twiss (HBT) performed a novel interference experiment through the use of measurements of intensity correlation \cite{brown1956correlation}. The original HBT stellar interferometer was designed to determine diameters of stars \cite{knight2005observation}. This experiment utilized two detectors located at different positions on Earth that collected light produced by independent sources on the disc of a star. 

\begin{figure}[ht!]
\centering
\includegraphics[width=0.5\textwidth]{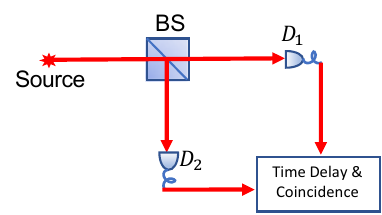}
\caption[Diagram of the Hanbury Brown and Twiss interferometer]{Diagram of the Hanbury Brown and Twiss interferometer. A beam of light is passed through a beam splitter (BS) and measured by two detectors $D_1$ and $D_2$. The time delay between the two detectors is controlled in this experiment, and the output signals produced by both detectors are correlated. This measurement is equivalent to the implementation of correlations of intensity fluctuations \cite{loudon2000quantum}.}
\label{fig:HBT}
\end{figure}

A simplified schematic of the HBT experiment is shown in Fig. \ref{fig:HBT}. Here, two detectors $D_1$ and $D_2$ are placed at the same distance from the beam splitter. The setup measures intensity correlations as a function of the time delay between the signals generated by the two detectors. Here, the coincident count rate is given by 
\begin{equation}
    C(t_1, t_2)=\langle I(t_1) I(t_2)\rangle,
\end{equation}
where $I(t_1)$ and $I(t_2)$ are the intensities measured at two detectors $D_1$ and $D_2$. The generic function for second-order coherence is defined as 
\begin{align}
\Gamma^{(2)}(t_1,t_2)=\langle E^*(t_1)E(t_1)E^*(t_2)E(t_2)\rangle
\end{align}
which describes a statistical average of the product of intensities associated to the fields $E(t_1)$ and $E(t_2)$. In general, the fields are detected at two different spatial and temporal positions. For practical purposes, we use the normalized version of the classical second-order coherence function  
\begin{align}
\gamma^{(2)}(\tau)=\frac{\langle E^*(t_1)E(t_1)E^*(t_2)E(t_2)\rangle}{\langle E^*(t_1)E(t_1)\rangle^2},
\end{align}
where $\tau=t_1-t_2$ is the time delay between the two light beams, and it is smaller than the coherence time of the source. 

Similarly to the first-order quantum coherence function, we can introduce the second-order quantum coherence function \cite{Gerry2004, scully1997quantum, loudon2000quantum},
\begin{equation}
    G^{(2)}\left(t_{1}, t_{2} ; t_{2}, t_{1}\right)=\operatorname{Tr}\left\{\hat{\rho} \hat{E}^{(-)}\left(t_{1}\right) \hat{E}^{(-)}\left(t_{2}\right) \hat{E}^{(+)}\left(t_{2}\right) \hat{E}^{(+)}\left(t_{1}\right)\right\},
\end{equation}
and the normalized second-order quantum coherence function, $g^{(2)}$ is given by,
\begin{equation}
\label{g2}
    g^{(2)}\left(t_{1}, t_{2} ; t_{2}, t_{1}\right)=\frac{G^{(2)}\left(t_{1}, t_{2} ; t_{2}, t_{1}\right)}{G^{(1)}\left(t_{1}, t_{1}\right) G^{(1)}\left(t_{2}, t_{2}\right)}.
\end{equation}
Notably, for a single-mode field, it is possible to reduce Eq. (\ref{g2}) to 
\begin{equation}
\label{reducedg2}
    g^{(2)}(\tau)=1+\frac{\left\langle(\Delta \hat{n})^{2}\right\rangle-\langle\hat{n}\rangle}{\langle\hat{n}\rangle^{2}}.
\end{equation}
We can observe that Eq. (\ref{reducedg2}) does not depend on the time difference $\tau$. The second-order quantum coherence function thus becomes a powerful tool to probe the underlying statistical properties of light. 

It is worth noting the values of $g^{(2)}(0)$ for different light sources. A laser beam described by a coherent state $|\alpha\rangle$ satisfies $g^{(2)}(0) = 1$. Furthermore, a single-mode thermal state ${\hat \rho _{\text{th}}}$ is characterized by $g^{(2)}(0)=2$. Indeed, any classical electric field satisfies $g^{(2)}(0)\geq1$. For photon number states (Fock states) represented by $|n\rangle$, for a situation in which $n\geq1$, it can be shown that $g^{(2)}(0)=1-1/n$. Particularly, for a single photon state $|1\rangle$, one expects $g^{(2)}(0)=0$. By comparing the aforementioned examples, we conclude that the measurement of $g^{(2)}$ can be used to characterize nonclassical properties of light \cite{walls2007quantum}. 

\section{Glauber–Sudarshan theory of coherence}

The Glauber--Sudarshan theory of coherence provides a quantum framework for understanding the statistical properties of light and distinguishing between classical and nonclassical states \cite{glauber1963, sudarshan1963}. It extends the classical theory of optical coherence by describing light in terms of quantum states and phase-space distributions, particularly through the Glauber--Sudarshan P representation. In this formalism, the density matrix of a quantum state is expressed as a weighted sum over coherent states, allowing a direct comparison with classical optics. A key aspect of this theory is its classification of light based on the P function: if $P(\alpha)$ behaves like a classical probability distribution (smooth and non-negative), the light is classical, whereas singular or negative values indicate nonclassical behavior, such as squeezing or antibunching. Glauber’s pioneering work laid the foundation for quantum optics, introducing concepts like coherent states and providing a rigorous mathematical framework to analyze the coherence properties of lasers and quantum light sources, which has since become essential in quantum information science and photonics.

Mathematically, the P representation expresses the density matrix $\hat{\rho}$ in terms of coherent states $|\alpha\rangle$ as:
\begin{equation}
    \hat{\rho} = \int P(\alpha) |\alpha\rangle \langle \alpha| \, d^2\alpha.
\end{equation}
Here, $P(\alpha)$ serves as a quasi-probability distribution that generalizes classical probability to quantum optics. For a general state, the expectation value of an operator  $\hat{O}$ can be computed as:
\begin{equation}
    \langle \hat{O} \rangle = \int P(\alpha) \langle \alpha | \hat{O} | \alpha \rangle \, d^2\alpha.
\end{equation}
If $P(\alpha)$ is well-behaved, the corresponding state can be considered classical. However, for highly nonclassical states, such as squeezed and Fock states, $P(\alpha)$ may take negative values or become highly singular, signaling purely quantum properties. The P representation is closely related to other phase-space representations, including the Wigner function and the Husimi Q function. The Wigner function provides a quasi-probability distribution in phase space and is defined as: 
\begin{equation} 
W(\alpha) = \frac{2}{\pi} \int \langle \alpha - \beta | \hat{\rho} | \alpha + \beta \rangle e^{i(\alpha \beta^*-\alpha^*\beta)} d^2\beta. 
\end{equation} 
Unlike the P and Q functions, the Wigner function can take negative values, making it particularly useful for identifying nonclassical features in quantum states. The Q function, by contrast, is always non-negative and is given by: 
\begin{equation} 
Q(\alpha) = \frac{1}{\pi} \langle \alpha | \hat{\rho} | \alpha \rangle. 
\end{equation}

The Glauber--Sudarshan P representation remains a cornerstone of quantum optics, providing a powerful tool to analyze coherence properties and distinguish between classical and quantum states of light. By offering a phase-space description of quantum states, it plays a crucial role in modern quantum technologies, including quantum communication, quantum sensing, and the study of nonclassical light-matter interactions.

\section{Quantum Gaussian-Schell Model}

In the previous section, the classical and quantum coherence properties of single quantum states have been introduced. To characterize the coherence of multiphoton wavepackets, we present the quantum Gaussian-Schell model (GSM) \cite{Foley1978, Rodenburg2014, omar2019Multiphoton, dawkins2024quantum}. The quantum theory of the electromagnetic field uncovered that classical forms of light were indeed produced by distinct superpositions of nonclassical multiphoton wave packets \cite{glauber1963,walls2007quantum}. This situation prevails for partially coherent light, the most common kind of classical light. We demonstrate the extraction of the constituent multiphoton quantum systems of a partially coherent light field. We shift from the realm of classical optics to the domain of quantum optics via a quantum representation of partially coherent light using its complex-Gaussian statistical properties. Our formulation of the quantum Gaussian-Schell model unveils the possibility of performing PNR detection to isolate the constituent multiphoton wave packets of a classical light field.

We compute the correlation properties between multiphoton wavepackets at the spatial locations $s_1$ and $s_2$. For the state with density matrix in the Fock state basis as $\hat{\rho}_{QGS}^D = \sum_{n,m,k,l=0}^\infty p_{\text{nmkl}}|n,m\rangle\langle k,l|$, we can define a multiphoton wavepacket correlation function $\tilde{g}^{(2)}(N,M)$ as follows
\begin{equation}\label{eq8}
    \tilde{g}^{(2)}(N,M) = \frac{p_{\text{NMNM}}}{\left(\sum_{m=0}^\infty p_{\text{NmNm}}\right)\left(\sum_{n=0}^\infty p_{\text{nMnM}}\right)}.
\end{equation}
Here, $p_{\text{NMKL}} = \text{Tr}\left[\hat{\rho}_{QGS}^D\hat{n}_{\text{NMKL}}\right]$ is the probability associated with the Fock-projection operator $|N, M\rangle\langle K, L|$. Given that the probability of observing a specific multiphoton wavepacket is proportional to its number of occurrences, $\tilde{g}^{(2)}(N,M)$ effectively becomes the standard coherence function. Specifically, $\tilde{g}^{(2)}(N,M)$ characterizes the coherence of $N$-photon wavepackets at the $s_1$-detector with $M$-photon wavepackets at the $s_2$-detector. Thus, the coherence function $\tilde{g}^{(2)}(N,M)$ is crucial for demonstrating the underlying nonclassical multiphoton coherence in partially coherent light sources, which can be critical for various applications in quantum information sciences \cite{dawkins2024quantum}.

\section{Summary}

This chapter provided a foundational overview of the quantum states of light and the theoretical framework required to describe coherence and statistical fluctuations. We reviewed Fock, coherent, and thermal states, and discussed first- and second-order coherence functions, the Glauber–Sudarshan P representation, and the quantum Gaussian–Schell model for partially coherent fields \cite{glauber1963, sudarshan1963, Gerry2004, walls2007quantum, scully1997quantum, dawkins2024quantum}. These concepts form the theoretical basis for understanding multiparticle interactions in quantum plasmonics and will serve as essential tools throughout this thesis.

\chapter{Observation of the modification of quantum statistics of plasmonic systems}

In this chapter, we report the first observation of the modification of quantum statistical properties in plasmonic systems. For almost two decades, it has been widely assumed that the quantum statistics of bosons are preserved in these platforms. This idea has been supported by experimental results showing that nonclassical correlations can survive light–matter interactions mediated by photon-plasmon scattering. Furthermore, it has been believed that similar dynamics ensure the conservation of the quantum fluctuations that characterize the nature of light sources. In contrast, our work demonstrates that quantum statistics are not always preserved in plasmonic systems. We show that multiparticle scattering effects, induced by confined optical near fields, can alter the excitation modes of plasmonic systems. These results are validated through the quantum theory of optical coherence for both single-mode and multimode systems. Altogether, the findings introduced in this chapter challenge established assumptions and open new directions for the control of multiparticle quantum systems using plasmonic platforms.

\section{Introduction}

The observation of the plasmon-assisted transmission of entangled photons gave birth to the field of quantum plasmonics almost twenty years ago \cite{altewischer2002plasmon}. Then years later, the coupling of single photons to collective charge oscillations at the interfaces between metals and dielectrics led to the generation of single surface plasmons \cite{akimov2007generation}. These findings unveiled the possibility of exciting surface plasmons with quantum mechanical properties \cite{tame2013quantum, you2020multiparticle}. In addition, such experiments demonstrated the possibility of preserving the quantum properties of photons as they scatter into surface plasmons and vice versa \cite{di2012quantum, fasel2005energy, huck2009demonstration, daniel2017surface, lawrie2013, chang2006quantum, safari2019measurement}. This research stimulated the investigation of other exotic quantum plasmonic states \cite{ di2012quantum, fasel2005energy,vest2018plasmonic, lawrie2013}. Ever since, the preservation of the quantum statistical properties of plasmonic systems has constituted a well-accepted tenant of quantum plasmonics \cite{tame2013quantum, you2020multiparticle}.

In the realm of quantum optics, the underlying statistical fluctuations of photons establish the nature of light sources \cite{glauber1963,Mandel:79}. These quantum fluctuations are associated to distinct excitation modes of the electromagnetic field that define quantum states of photons and plasmons \cite{anno2006, omar2019Multiphoton, tame2013quantum}. In this regard, prior plasmonic experiments have demonstrated the preservation of quantum fluctuations while performing control of quantum interference and transduction of  correlations in metallic nanostructures \cite{vest2017anti, dheur2016single, li2017strong, daniel2017surface, chang2006quantum, safari2019measurement, lee2016quantum, dowran2018quantum, holtfrerich2016toward, fasel2005energy}. Despite the dissipative nature of plasmonic fields, the additional interference paths provided by optical near fields have enabled the harnessing of quantum correlations and the manipulation of spatial coherence \cite{vest2017anti, buse2018symmetry, li2017strong, schouten2005, magana2016exotic}. So far, this exquisite degree of control has been assumed independent of the excitation mode of the interacting particles in a plasmonic system \cite{tame2013quantum}. Moreover, the quantum fluctuations of plasmonic systems have been considered independent of other properties such as polarization, temporal, and spatial coherence \cite{di2012quantum, li2017strong, schouten2005, buse2018symmetry, magana2016exotic}. This assumption has served as a foundation for the use of plasmonic systems in quantum control, sensing, and information processing technologies \cite{dowran2018quantum, anno2006, holtfrerich2016toward, vest2017anti, tame2013quantum, you2020multiparticle, obrien2009}. 

\begin{figure}[htbp]
  \centering
  \includegraphics[width=0.99\textwidth]{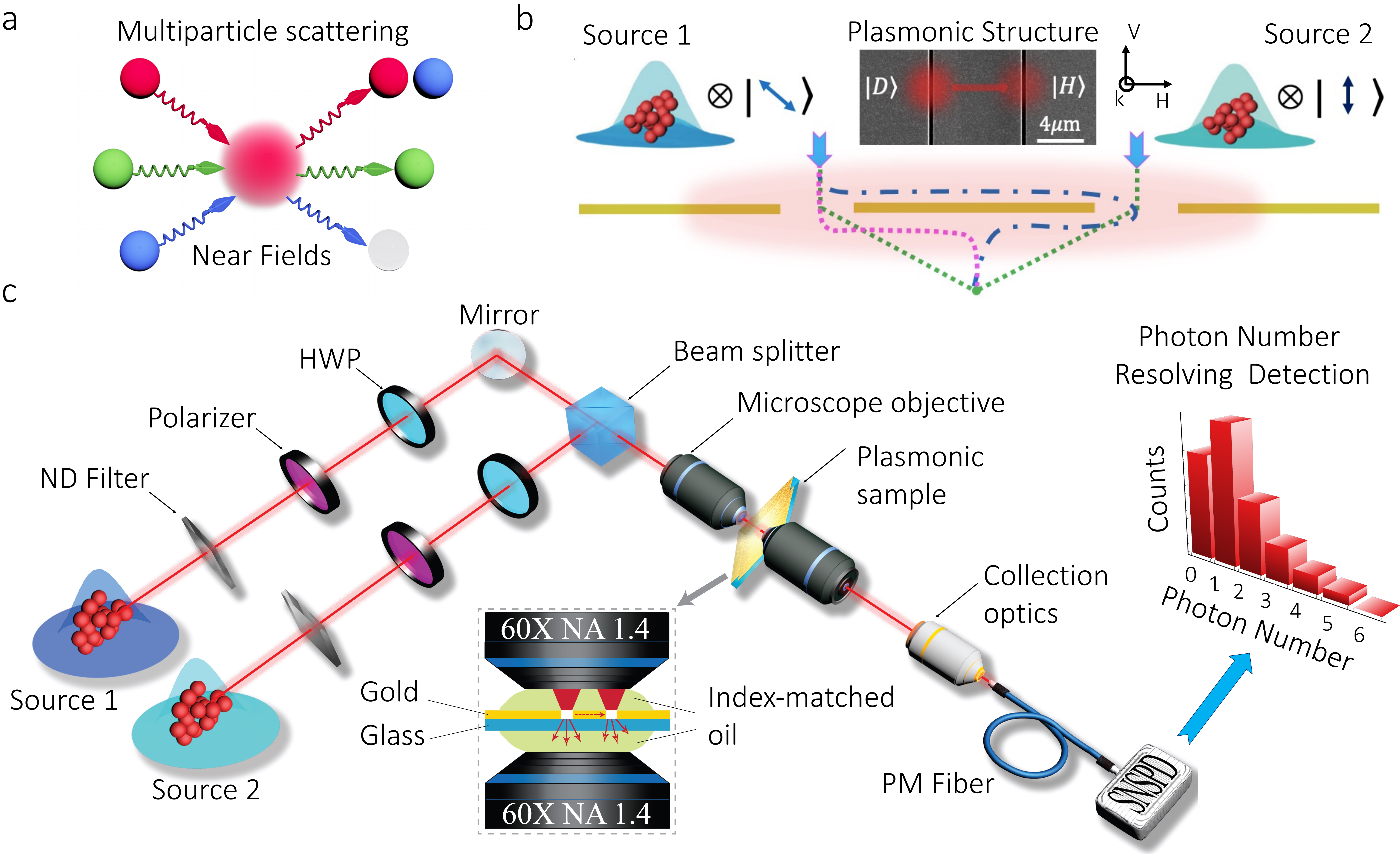}
  \caption[Multiparticle scattering in plasmonic systems]{\textbf{Multiparticle scattering in plasmonic systems.} The diagram in \textbf{a} illustrates the concept of multiparticle scattering mediated by optical near fields. The additional interference paths induced by confined near fields leads to the modification of the quantum statistics of plasmonic systems. This idea is implemented through the plasmonic structure shown in \textbf{b}. The dotted lines represent additional scattering paths induced by confined optical fields in the plasmonic structure \cite{magana2016exotic}. Our metallic structure consists of a 110-nm-thick gold film with slit patterns. The width and length of each slit are 200 nm and 40 $\mu$m, respectively. The slits are separated by 9.05 $\mu$m. The fabricated sample is illuminated by either one or two thermal sources of light with specific polarizations. The strength of the plasmonic near-fields is controlled through the polarization of the illuminating photons. The plasmonic near-fields are only excited with photons polarized along the horizontal direction. The experimental setup for the observation of the modification of quantum statistics in plasmonic systems is shown in \textbf{c}. We prepare either one or two independent thermal wavepackets with specific polarizations. The polarization state of each of the wavepackets is individually controlled by a polarizer (Pol) and half-wave plate (HWP). The two photonic wavepackets are injected into a beam splitter (BS) and then focused onto the gold sample through an oil-immersion objective. The refractive index of the immersion oil matches that of the glass substrate creating a symmetric index environment around the gold film. The transmitted photons are collected with another oil-immersion objective. We measure the photon statistics in the far field using a superconducting nanowire single-photon detector (SNSPD) that is used to perform photon-number-resolving detection \cite{OmarOptica17, You2020Indentification}. }
\label{Fig1}
\end{figure}

The work presented in this chapter re-examines this longstanding assumption. Experimental evidence is provided to show that quantum statistical properties are not universally conserved in plasmonic systems. Specifically, it is shown that scattering among photons and plasmons induces multiparticle interference effects that can lead to modifications in the excitation modes of the system and, consequently, in the statistical properties of the transmitted field \cite{You2021}. The multiparticle dynamics that take place in plasmonic structures can be controlled through the strength of the confined optical near fields in their vicinity. In addition, it is demonstrated that external degrees of freedom—such as spatial mode structure—can influence the underlying quantum statistics of the field. Given the growing interest in multimode plasmonic platforms for quantum information processing, the results discussed here are extended to a two-mode system comprising independent wavepackets \cite{lawrie2013, vest2017anti,safari2019measurement, daniel2017surface}. The experimental findings are interpreted using the quantum theory of optical coherence, providing a theoretical foundation for the observations \cite{glauber1963}. These results demonstrate that plasmonic platforms enable direct access to a new level of quantum state control, where the manipulation of quantum fluctuations becomes a functional tool in nanophotonic quantum systems \cite{tame2013quantum, you2020multiparticle}.

\begin{figure}[ht!]
  \centering
  \includegraphics[width=0.99\textwidth]{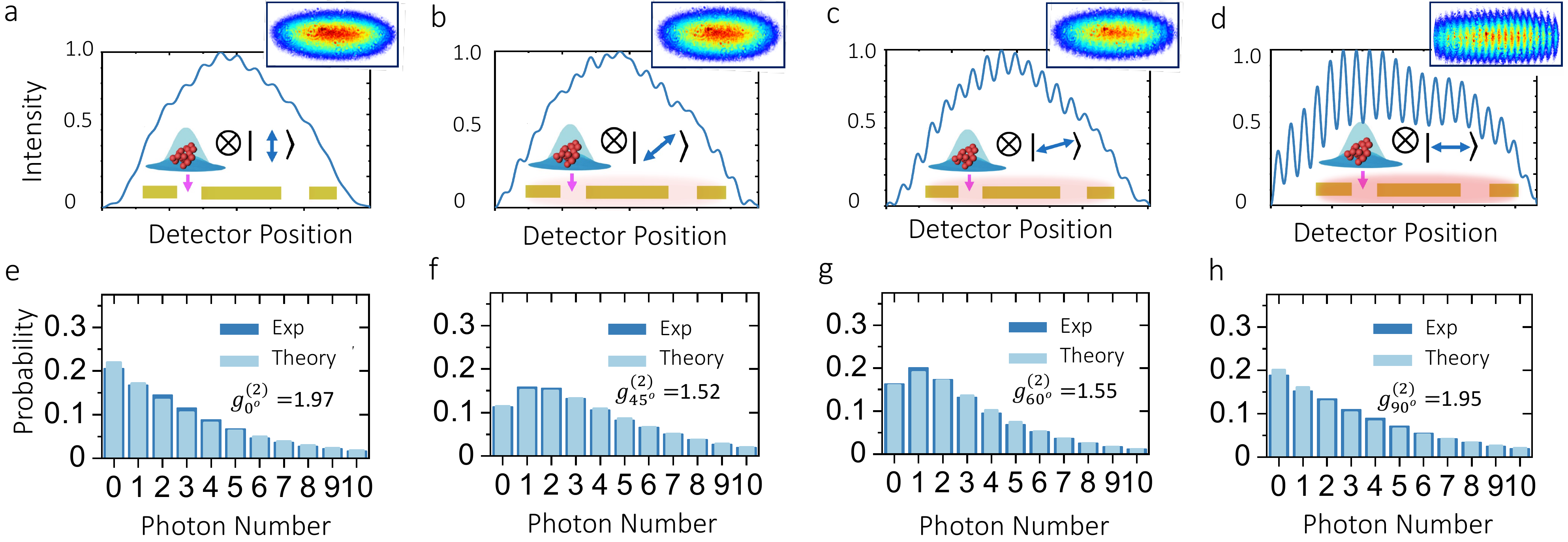}
  \caption[Experimental observation of plasmon-induced interference and the modification of quantum statistics]{\textbf{Experimental observation of plasmon-induced interference and the modification of quantum statistics.} The formation of interference fringes in a plasmonic structure with two slits is shown in panels \textbf{a} to \textbf{d}. Panel \textbf{a} shows the spatial distribution of a thermal wavepacket of photons transmitted by a single slit. In this case, the contribution from optical near fields is negligible and no photons are transmitted through the second slit. As shown in panel \textbf{b}, a rotation of the photon's polarization increases the presence of optical near fields in the plasmonic structure. In this case, photon-plasmon scattering processes induce small changes to the spatial distribution of the transmitted photons. Part \textbf{c} shows that the increasing excitation of plasmons is manifested through the increasing visibility of the interference structure. Panel \textbf{d}, shows  a clear modification of spatial coherence induced by optical near fields. Remarkably, the modification of spatial coherence induced by the presence of plasmons is also accompanied by the modification of the quantum statistical fluctuations of the field as indicated in panels \textbf{e} to \textbf{h}. Each of these photon-number distributions corresponds to the spatial profiles above from \textbf{a} to \textbf{d}.
  The photon-number distribution in \textbf{e} demonstrates that the  photons transmitted by the single slit preserve their thermal statistics. Remarkably, multiparticle scattering induced by the presence of near fields modifies the photon-number distribution of the hybrid system as shown in \textbf{f} and \textbf{g}. These probability distributions resemble those of coherent light sources. Interestingly, as demonstrated in \textbf{h}, the photon-number distribution  becomes thermal again when photons and plasmons are polarized along the same direction.  }
\label{Fig2}
\end{figure}

We now introduce a theoretical model to describe the global dynamics experienced by a multiphoton system as it scatters into surface plasmons and vice versa. Interestingly, these photon-plasmon interactions can modify the quantum fluctuations that define the nature of a physical system \cite{glauber1963, anno2006, omar2019Multiphoton}. As illustrated in Fig. \ref{Fig1}\textbf{a}, the multiparticle scattering processes that take place in plasmonic structures can be controlled through the strength of the confined near fields in their vicinity. The near-field strength defines the probability of inducing individual phase jumps through scattering \cite{safari2019measurement}. These individual phases establish different conditions for the resulting multiparticle dynamics of the photonic-plasmonic system.

\section{Results and discussion}

We investigate the possibility of modifying quantum statistics in the plasmonic structure shown in Fig. \ref{Fig1}\textbf{b}. The scattering processes in the vicinity of the multi-slit structure leads to additional interference paths that affect the quantum statistics of multiparticle systems \cite{vest2017anti,magana2016exotic}. The gold structure in Fig. \ref{Fig1}\textbf{b} consists of two slits aligned along the \emph{y}-direction (see Appendix A). The structure is designed to excite plasmons when it is illuminated with thermal photons polarized along the \emph{x}-direction \cite{magana2016exotic, schouten2005}. For simplicity, we will refer to light polarized in the \emph{x}- and \emph{y}-direction as horizontally $\ket{H}$ and vertically $\ket{V}$ polarized light, respectively. Our polarization-sensitive plasmonic structure directs a fraction of the horizontally polarized photons to the second slit when the first slit is illuminated with diagonally $\ket{D}$ polarized photons. As depicted in Fig. \ref{Fig1}\textbf{b}, this effect is used to manipulate the quantum statistics of a mixture of photons from independent wave packets \cite{arrechi1967,smith2018}.

The modification of quantum statistics induced by the scattering paths in Fig. \ref{Fig1}\textbf{b} can be understood through the Glauber-Sudarshan  theory of coherence \cite{glauber1963, sudarshan1963}. For this purpose, we first define the P-function associated to the field produced by the indistinguishable scattering between the two horizontally-polarized modes. These represent either photons or plasmons emerging through each of the slits. 
\begin{equation}\label{Eq:Ps}
    \text{P}_{\text{Pl}}(\alpha) = \int\text{P}_{{1}}(\alpha - \alpha')\text{P}_{{2}}(\alpha')d^{2}\alpha'.
\end{equation}
The P-function for a thermal light field is given by $ \text{P}_{{\text{i}}}(\alpha) = (\pi\bar{n}_{\text{i}})^{-1}\exp{(-\abs{\alpha}^{2}/\bar{n}_{\text{i}})}$. Here, $\alpha$ describes
the complex amplitude as defined for coherent states $\ket{\alpha}$. The mean particle number of the two modes is represented by $\bar{n}_1=\eta\bar{n}_{\text{s}}$ and $\bar{n}_2=\bar{n}_{\text{pl}}$. Moreover,
the mean photon number of the initial illuminating photons is represented by
 $\bar{n}_{\text{s}}$, whereas $\bar{n}_{\text{pl}}$ describes the mean photon number of scattered plasmon fields. 
The parameter $\eta$ is defined as $\cos^{2}{\theta}$. The polarization angle, $\theta$, of the illuminating photons is defined with respect to the vertical axis. Note that the photonic modes in Eq. (\ref{Eq:Ps}) can be produced by independent multiphoton wavepackets. Furthermore, we make use of the coherent state basis to represent the state of the combined field as $\rho_{\text{Pl}} = \int\text{P}_{\text{Pl}}\pare{\alpha}\ket{\alpha}\bra{\alpha}d^{2}\alpha$ \cite{sudarshan1963}. This expression enable us to obtain the number distribution $\text{p}_{\text{Pl}} (n)$ for the scattered photons and plasmons with horizontal polarization. We can then write the combined number distribution for the multiparticle system at the detector as $\text{p}_{\text{det}}\pare{n} = \sum_{m=0}^{n} \text{p}_{\text{Pl}}\pare{n-m}\text{p}_{\text{Ph}}\pare{m}$. The distribution $\text{p}_{\text{Ph}}\pare{m}$ accounts for the vertical polarization component of the illuminating multiphoton wavepackets.
Thus, we can describe the final photon-number distribution after the plasmonic structure as
\begin{equation}\label{Eq:final_p}
\text{p}_{\text{det}}\pare{n} = \sum_{m=0}^{n}\frac{(\bar{n}_{\text{pl}}+\eta\bar{n}_{s})^{n-m}\cor{(1-\eta)\bar{n}_{\text{s}}}^m}{(\bar{n}_{\text{pl}}+\eta  \bar{n}_{s}+1)^{n-m+1}\cor{(1-\eta)\bar{n}_{\text{s}}+1}^{m+1}}.
\end{equation}
Note that the quantum statistical properties of the photons scattered from the sample are defined by the strength of the plasmonic near fields $\bar{n}_{\text{pl}}$. As illustrated in Fig. \ref{Fig1}\textbf{a}, the probability function in Eq. (\ref{Eq:final_p}) demonstrates the possibility of modifying the quantum statistics of photonic-plasmonic systems.
 
 \begin{figure}[ht!]
  \centering
  \includegraphics[width=0.99\textwidth]{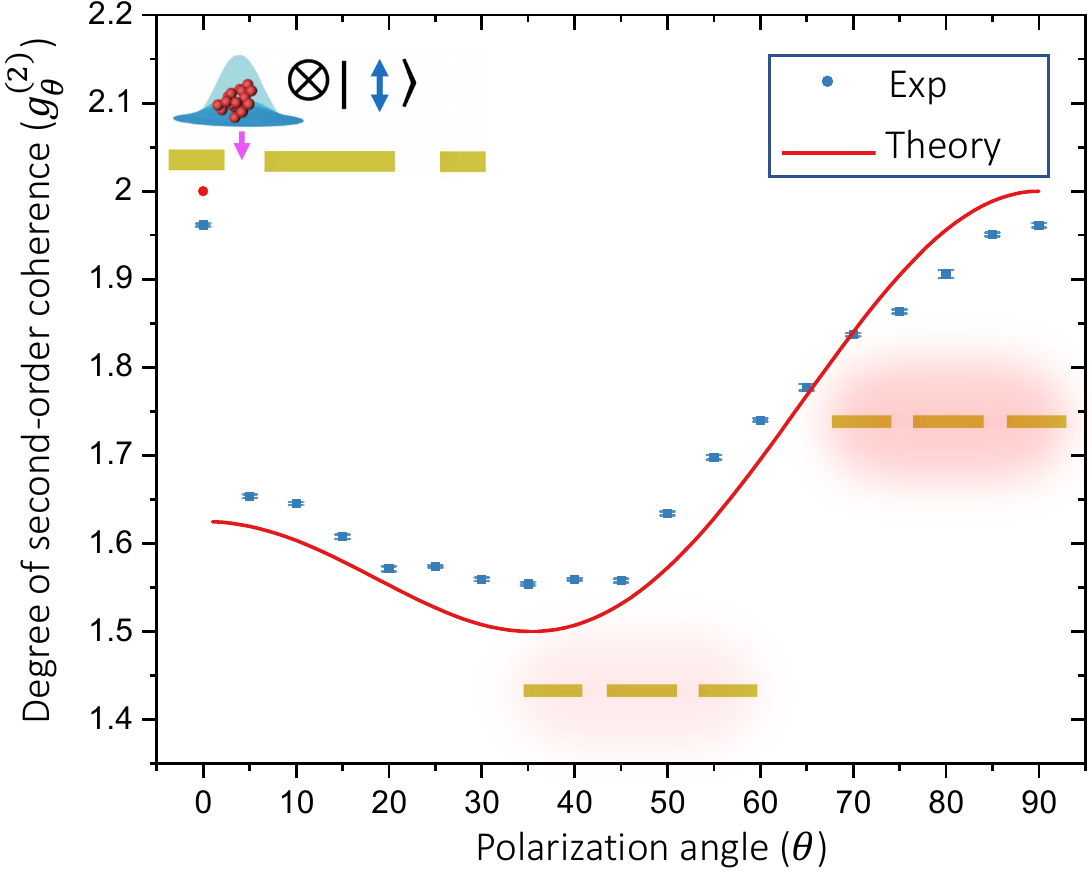}
  \caption[Modification of the quantum statistics of a single multiparticle system in a plasmonic structure]{\textbf{Modification of the quantum statistics of a single multiparticle wavepacket in a plasmonic structure.} The experimental data is plotted together with the theoretical prediction for the degree of the second-order correlation function. The theoretical model is based on the photon-number distribution described by Eq. (\ref{Eq:final_p}) for a situation in which $\bar{n}_\text{s}=3\bar{n}_{\text{pl}}$.}
  \label{Fig3}
\end{figure}
 
We first explore the modification of quantum statistics in a plasmonic double-slit structure illuminated by a thermal multiphoton wavepacket \cite{arrechi1967, smith2018}. The experimental setup is depicted in Fig. \ref{Fig1}\textbf{c}. We focus thermal photons onto a single slit and measure the far-field spatial profile and the quantum statistics of the transmitted photons. As shown in Fig. \ref{Fig2}, we perform multiple measurements corresponding to different polarization angles of the illuminating photons. As expected, the spatial profile of the transmitted photons does not show interference fringes when the photons are transmitted by the single slit (see Fig. \ref{Fig2}\textbf{a}). However, the excitation of plasmonic fields increases the spatial coherence of the scattered photons. As indicated by Figs. \ref{Fig2}\textbf{b} to \textbf{d}, the increased spatial coherence leads to the formation of interference structures. This effect has been observed multiple times \cite{schouten2005, magana2016exotic, li2017strong,daniel2017surface}. Nonetheless, it had been assumed independent of the quantum statistics of the hybrid photonic-plasmonic system \cite{tame2013quantum,  you2020multiparticle}. However, as demonstrated by the probability distributions from Fig.  \ref{Fig2}\textbf{e} to \textbf{h}, the modification of spatial coherence is indeed accompanied by a modification of the quantum fluctuations of the plasmonic system. In our experiment, the mean photon number of the photonic source $\bar{n}_\text{s}$ is three times the mean particle number of the plasmonic field $\bar{n}_{\text{pl}}$ (see Appendix A). The theoretical prediction for the photon-number distributions in Fig. \ref{Fig2} were obtained using Eq. (\ref{Eq:final_p}) for a situation in which $\bar{n}_\text{s}=3\bar{n}_{\text{pl}}$.

The sub-thermal photon-number distribution shown in Fig. \ref{Fig2}\textbf{f} demonstrates that the strong confinement of plasmonic fields can induce anti-thermalization effects \cite{kondakci2015}. Here, the scattering among photons and plasmons attenuates the chaotic fluctuations of the injected wavepacket, characterized by a thermal photon-number distribution, as indicated by Fig. \ref{Fig2}\textbf{e}. Conversely, the transition in the photon-number distribution shown from Fig.  \ref{Fig2}\textbf{g} to \textbf{h} is mediated by a thermalization effect. In this case, the individual phase jumps induced by photon-plasmon scattering increases the chaotic fluctuations of the multiparticle system \cite{safari2019measurement}, leading to the thermal state in Fig. \ref{Fig2}\textbf{h}. As shown in  Fig. \ref{Fig3}, the quantum statistics of the photonic-plasmonic system show an important dependence on the strength of the optical near fields surrounding the plasmonic structure.  The photon-number-distribution dependence on the polarization angle of the illuminating photons is quantified through the degree of second-order coherence $g^{(2)}$ in Fig. \ref{Fig3}. The remarkable agreement between theory and experiment validates our observation of the modification of quantum statistics of plasmonic systems.

\begin{figure}[ht!]
  \centering
  \includegraphics[width=0.99\textwidth]{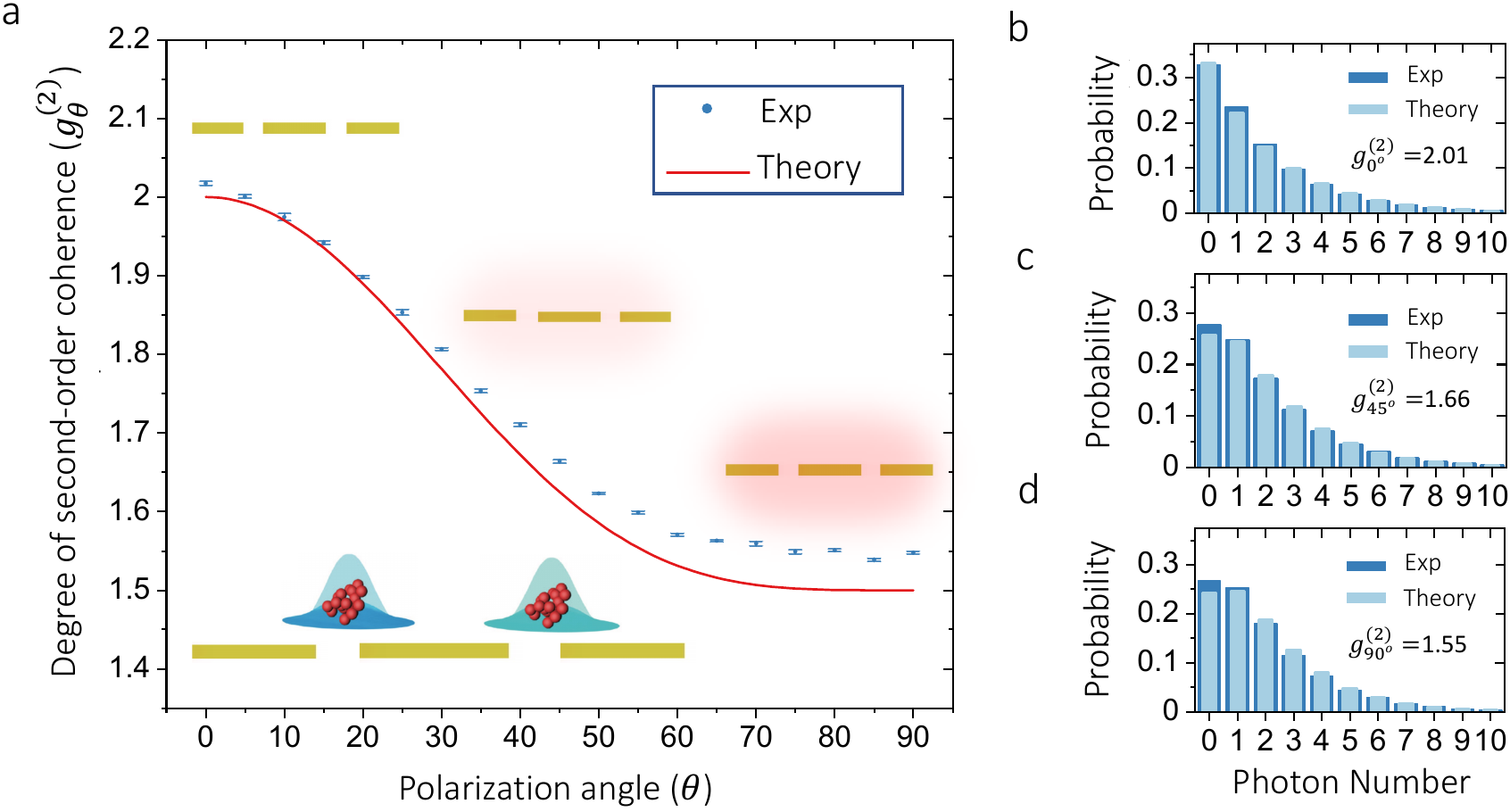}
  \caption[Modification of the quantum statistics of a multiphoton system comprising two sources]{\textbf{Modification of the quantum statistics of a multiphoton system comprising two sources.} The modification of the quantum statistics of a multimode plasmonic system composed of two independent multiphoton wavepackets is shown in panel \textbf{a}. Here, we plot experimental data together with our theoretical prediction for the degree of second-order coherence. The theoretical model is based on the photon-number distribution predicted by Eq. (\ref{Eq:final_p}) for two independent wavepackets with thermal statistics and same mean photon numbers satisfying $\bar{n}_\text{s}=\bar{n}_{\text{pl}}$. As demonstrated in \textbf{b}, the photon-number distribution is thermal for the scattered wavepackets in the absence of near-fields ($\theta=0^{\circ}$). However, an anti-thermalization effect takes place as the strength of the plasmonic near fields is increased ($\theta=45^{\circ}$), this is indicated by the probability distribution in \textbf{c}. Remarkably, as reported in \textbf{d}, the degree of second-order coherence of the hybrid photonic-plasmonic system is 1.5 when plasmonic near fields are strongly confined ($\theta=90^{\circ}$). These results unveil the possibility of using plasmonic near fields to manipulate coherence and the quantum statistics of multiparticle systems. 
  }
\label{Fig4}
\end{figure}

The multiparticle near-field dynamics observed for a thermal wavepacket can also induce interactions among independent wavepackets. We demonstrate this possibility by illuminating the plasmonic structure with two independent thermal wavepackets. This is illustrated in the central inset of Fig. \ref{Fig1}\textbf{c}. In this case, both multiphoton sources are prepared to have same mean photon numbers. In Fig. \ref{Fig4}, we report the modification of the quantum statistics of a multiphoton system comprising two modes that correspond to two independent wavepackets. As shown in Fig. \ref{Fig4}\textbf{a}, the confinement of electromagnetic near fields in our plasmonic structure modifies the value of the second-order correlation function. In this case, the shape of the second-order correlation function is defined by the symmetric contributions from the two thermal wavepackets. As expected, the quantum statistics of the initial thermal system with two sources remains thermal (see Fig.  \ref{Fig4}\textbf{b}). However, as shown in Figs.  \ref{Fig4}\textbf{b} to \textbf{d},  this becomes sub-thermal as the strength of the plasmonic near fields increase. The additional scattering paths induced by the presence of plasmonic near fields modify the photon-number distribution as demonstrated in Fig. \ref{Fig4}\textbf{c} \cite{magana2016exotic}. The strongest confinement of plasmonic near fields is achieved when the polarization of one of the sources is horizontal ($\theta=90^{\circ}$). As predicted by Eq. (\ref{Eq:final_p}) and reported in Fig.  \ref{Fig4}\textbf{d}, the second-order coherence function for this particular case is $g^{(2)}=1.5$. 

\section{Summary}

The results presented in this chapter have significant implications for the study and development of multimode plasmonic systems \cite{tame2013quantum, you2020multiparticle}. Recent efforts have explored the transduction of quantum statistical fluctuations in multimode fields for applications in quantum imaging and plasmonic networks \cite{lawrie2013, Holtfrerich2016, Bhusal2022}, yet the active modification of quantum statistics remained largely unexplored \cite{tame2013quantum}. Our work demonstrates that plasmonic near fields offer deterministic mechanisms for tailoring photon statistics through external control parameters such as polarization \cite{You2021}. This framework enables new methods for exploring thermalization and anti-thermalization in arbitrary light fields, with possible extensions to other disordered systems \cite{kondakci2015}.

For nearly two decades, physicists and engineers have assumed that the quantum statistical properties of plasmonic systems are always preserved \cite{altewischer2002plasmon, akimov2007generation, tame2013quantum, you2020multiparticle, di2012quantum, fasel2005energy, huck2009demonstration, daniel2017surface, lawrie2013, chang2006quantum, safari2019measurement}. Our work reported in this chapter challenges that view by demonstrating that quantum statistics can indeed be modified through multiparticle scattering in plasmonic structures. These results were validated through the quantum theory of coherence, revealing new ways to control the excitation modes of plasmonic systems. Our findings offer promising opportunities for manipulating multiphoton dynamics and contribute to the advancement of quantum photonics, quantum many-body systems, and quantum information science \cite{tame2013quantum, anno2006, obrien2009}.

\chapter{Nonclassical near-field dynamics of surface plasmons}

Following the results in Chapter 3, which demonstrated the modification of quantum statistical properties in plasmonic systems, this chapter focuses on the underlying near-field quantum dynamics that were uncovered utilizing such a technique. The coupling of photons to collective charge oscillations at the surface of a metal to form surface-plasmon polaritons enables strong confinement of electromagnetic near fields in the vicinity of photonic nanostructures. Even though surface plasmons are formed from bosons and fermions, this kind of near-field wave exhibits bosonic properties in the limit of many electrons. In this chapter, it is shown that the classical near-field dynamics of surface plasmons are defined by nonclassical processes of scattering among their constituent multiparticle subsystems. By isolating multiparticle plasmonic subsystems, it is demonstrated that their quantum dynamics are governed by either bosonic or fermionic processes of coherence. The quantum-coherence properties of plasmonic fields excited by the vacuum fluctuations of the electromagnetic field are also discussed. Our findings reported in this chapter uncover multiparticle properties of electromagnetic near fields with important implications for quantum technology.

\section{Introduction}

There has been an enormous interest in understanding surface plasmon polaritons at their most fundamental level \cite{tame2013quantum, you2020multiparticle}. For more than two decades, extensive research has been conducted with the goal of uncovering the quantum properties of this kind of quasiparticles resulting from the coupling of bosons and fermions \cite{altewischer2002plasmon, akimov2007generation,tame2013quantum,you2020multiparticle,di2012quantum, fasel2005energy,huck2009demonstration, daniel2017surface, lawrie2013, chang2006quantum, safari2019measurement, Pres2023, Dai2022}. These studies have cast interest in the physics of evanescent plasmonic fields and its potential to unlock novel forms of quantum coherence in photonic systems \cite{ di2012quantum, fasel2005energy,vest2018plasmonic, lawrie2013, buse2018symmetry, vest2017anti,DiMartino2014, Heeres2013, You2021, Maly2023}. Moreover, plasmonic near fields have facilitated the exploration of the quantum vacuum fluctuations of the electromagnetic field \cite{Kardar1999, Volokitin2007}. For example, this research has unveiled manifestations of vacuum fluctuations in nano-optical systems \cite{Lambrecht2002, Ford1993, Laliotis2014, Intravaia2005, Dodonov2020, Rodriguez2011}. Despite the relevance of these effects for the foundations of quantum physics and the development of quantum technologies, most of this kind of research has been theoretical \cite{Ford1993, Intravaia2005,Dodonov2020, Rodriguez2011}. This is mainly due to the experimental challenges involved in the measurement of isolated quantum dynamics of plasmonic systems \cite{anno2006, You2021APR, You2021, You2020Indentification, you2020multiparticle, tame2013quantum, YouPhotoniX}. Indeed, previous experimental work on quantum plasmonics has been devoted to the investigation of the collective dynamics of open plasmonic systems \cite{tame2013quantum, you2020multiparticle}.

The advent of quantum light sources enabled the preparation of smaller plasmonic systems with nonclassical properties \cite{tame2013quantum, you2020multiparticle}. One can trace back the birth of quantum plasmonics to the observation of the plasmon-assisted transmission of entangled photons almost twenty years ago \cite{altewischer2002plasmon}. This work motivated interest in the generation of single surface plasmons and the investigation of their wave-particle duality \cite{akimov2007generation, kolesov2009wave, dheur2016single}. Furthermore, plasmonic fields have been exploited to harness entanglement and to control two-photon interference \cite{fasel2005energy, huck2009demonstration, di2012quantum, daniel2017surface, lawrie2013, chang2006quantum, safari2019measurement, dheur2016single, li2017strong, chang2006quantum, safari2019measurement, lee2016quantum, holtfrerich2016toward}. Recently, it was shown that multiparticle scattering among plasmons and photons can lead to the modification of the quantum statistical properties of plasmonic systems \cite{You2021, Tame2021}. Despite the fundamental importance of previous investigations, these effects are produced by the collective dynamics of the constituent particles of  plasmonic systems \cite{dowran2018quantum, anno2006, holtfrerich2016toward, vest2017anti, tame2013quantum, you2020multiparticle, obrien2009}. As such, the investigation of the isolated quantum dynamics responsible for the emergence of known macroscopic behaviors of surface plasmons remains unexplored.

\begin{figure}[htbp]
  \centering
 \includegraphics[width=0.85\textwidth]{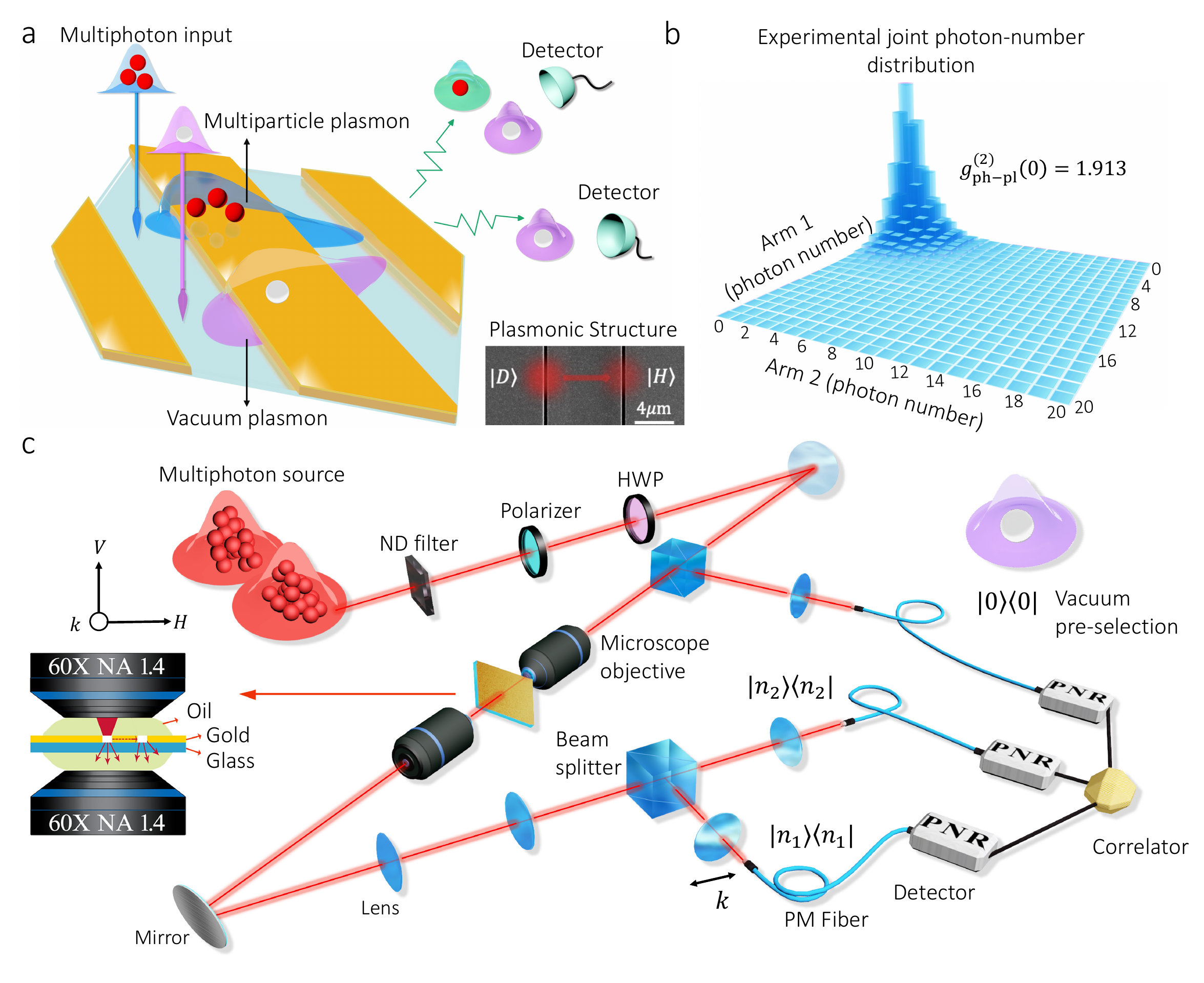}
\caption[Quantum near-field dynamics of plasmonic fields]{\textbf{Quantum near-field dynamics of plasmonic fields.} The diagram in \textbf{a} illustrates a plasmonic nanostructure hosting multiparticle plasmonic scattering. The yellow region represents the gold layer, whereas the transparent surface indicates the glass substrate. In this case, each constituent state of the illuminating thermal system excites plasmonic fields with different number of particles. The embedded photograph shows a view of the sample from above. Our metallic structure consists of a 110-nm-thick gold film with slit patterns. The width and length of the slits are 200 nm and 40 $\mu$m, respectively. The inter-slit separation is 9.05 $\mu$m. The plot in \textbf{b} reports the experimental joint photon-number distribution of the classical hybrid system after the plasmonic sample. As depicted in \textbf{c}, we control the polarization state of the thermal multiphoton system by means of a polarizer and half-wave plate (HWP). We use a beam splitter (BS) to pre-select on the vacuum component of the multiphoton system \cite{Nunn2022, Nunn2023}. Then, we focus the photons onto one of the slits through an oil-immersion objective. The inset illustrates this perspective from the side. The refractive index of the immersion oil matches that of the glass substrate creating a symmetric index environment around the gold film. The transmitted photons are collected with another oil-immersion objective. We enlarge and split the far field of the scattered photons by means of a 4-f system and BS, respectively. In addition, we use another lens to access the far field ($k$-plane) of the sample. These photons are collected by two polarization-maintaining (PM) single-mode fibers.  Here, we keep the fiber of one of the arms fixed while we scan the fiber of the second arm. We formalize the experiment by performing multiphoton correlations among the three PNR detectors.}
\label{Fig6}
\end{figure}

In our work introduced in this chapter, the near-field dynamics of a plasmonic system are investigated by projecting them into their constituent multiparticle subsystems \cite{HongNP24, OPN}. This approach enables isolation of the nonclassical dynamics within these subsystems, allowing direct study of their quantum coherence behavior. Interestingly, the scattering effects hosted by these isolated plasmonic subsystems are defined by either bosonic or fermionic properties of coherence. This experimental capability makes it possible to access dynamics associated with plasmons excited by vacuum fluctuations of a multiphoton system \cite{Ford1993,Laliotis2014, Intravaia2005, Rodriguez2011, Dodonov2020}. In addition, we discuss how the collective contribution of these multiparticle quantum near-field effects lead to the classical bosonic properties of plasmons. Notably, these quantum effects are not limited to a specific platform but can be observed across plasmonic systems where multiparticle scattering is not constrained \cite{You2021, Tame2021, tame2013quantum, you2020multiparticle}. These observations are validated through the quantum theory of coherence for multiparticle systems. Altogether, the results discussed in this chapter unveil universal properties of plasmonic near fields with important implications for the preparation of quantum many-body systems \cite{anno2006,omar2019Multiphoton,thouless2014quantum, AspuruGuzik2012, obrien2009}.

\begin{figure}[htbp]
  \centering
  \includegraphics[width=0.82\textwidth]{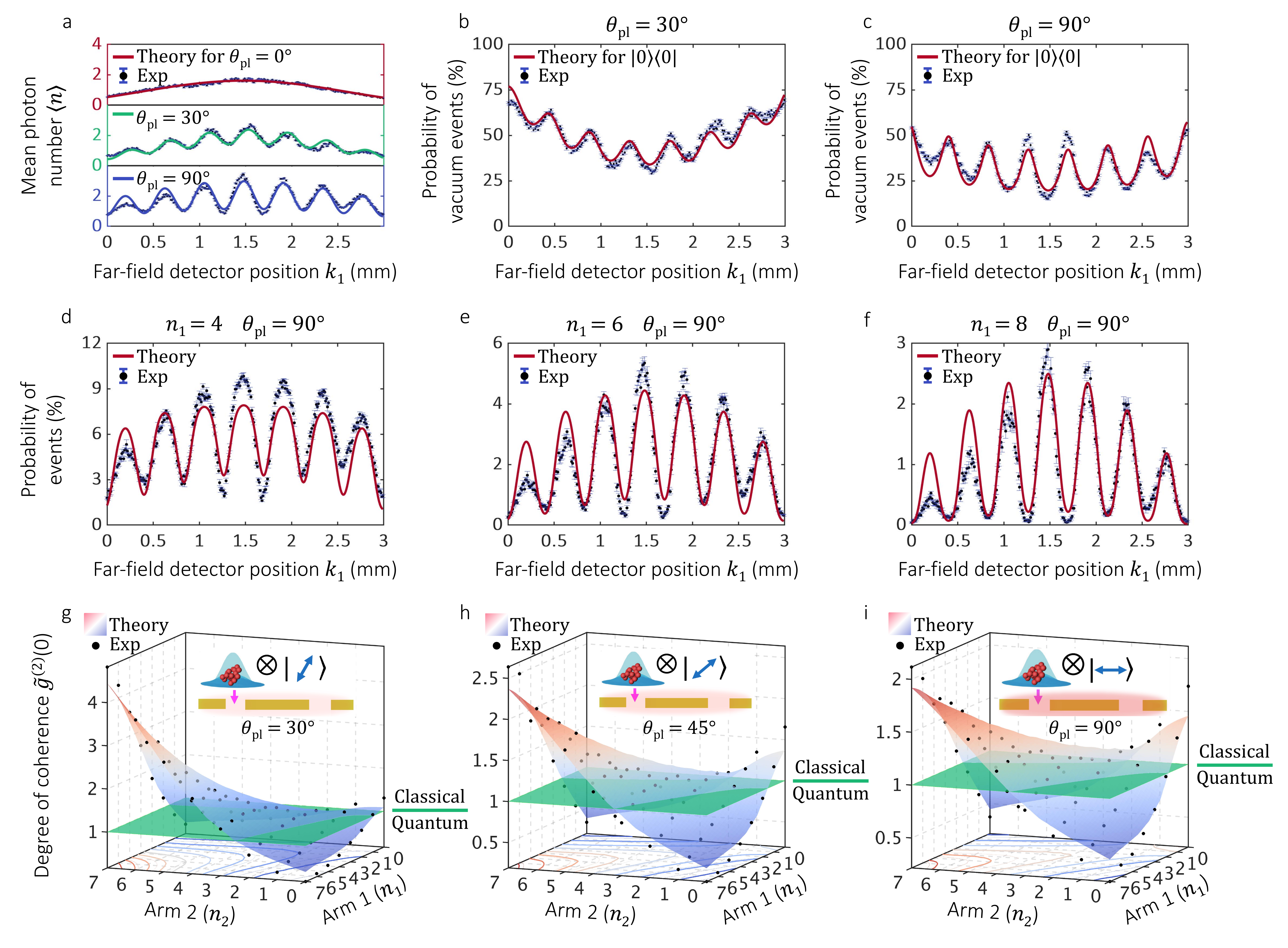}
  \caption[Collective and multiparticle-subsystem dynamics of surface plasmons]{\textbf{Collective and multiparticle-subsystem dynamics of surface plasmons.} The first panel in \textbf{a} shows the far-field intensity pattern formed in the absence of surface plasmons. The second and third panels therein show fringes produced by the collective interference between photons and plasmons. The visibility of these classical interference structures depends on the strength of the plasmonic waves, which is controlled through the polarization angle $\theta_{\text{pl}}$. Notably, the projection of this plasmonic system into its quantum constituents multiparticle subsystems unveils different scattering processes. These quantum dynamics are reported from \textbf{b} to \textbf{f} for subsystems with different number of particles. The theoretical predictions were obtained using Eq. (\ref{eq:first-order}). The projection of the hybrid plasmonic-photonic system into its vacuum component leads to the quantum interference pattern in \textbf{b}, which is modulated by a dip. As shown in \textbf{c}, one can increase the visibility of the interference pattern by increasing the presence of plasmonic fields in the nanostructure through a larger $\theta_{\text{pl}}$. The formation of these interference structures is produced by the scattering of vacuum plasmons with the vacuum-fluctuation of the illuminating field. Further, we show the multiparticle dynamics for a four-particle system in panel \textbf{d}. Remarkably, larger multiparticle systems, such as those shown in \textbf{e} and \textbf{f}, lead to the formation of quantum interference fringes with almost perfect visibility. These dynamics suggest the presence of a form of quantum coherence that we quantify through $\tilde{g}^{(2)}(0)$. In \textbf{g} we demonstrate that while most of the multiparticle constituents of the plasmonic system show classical properties of coherence ($\tilde{g}^{(2)}(0)>1$), there are subsystems showing quantum dynamics. This is indicated by values of $\tilde{g}^{(2)}(0)$ below one. Interestingly, as reported from  \textbf{g} and \textbf{i}, these properties depend on the strength of the plasmonic field described by $\theta_{\text{pl}}$. The theoretical surface is estimated based on Eq. (\ref{eq:rho}). The error bars represent the standard deviation of ten datasets, each consisting of approximately 100,000 photon-number-resolving measurements.}
\label{Fig7}
\end{figure}

\section{Results and discussion}

The gold nanostructure used for the exploration of nonclassical near-field dynamics of surface plasmons is depicted in Fig. \ref{Fig6}\textbf{a}. In order to excite classical plasmonic waves, we focus a thermal multiphoton field into one of the two slits of the plasmonic sample. It is worth noticing that thermal light fields represent the most classical kind of optical fields \cite{arrechi1967}. Indeed, thermal fields are formed by the statistical mixture of vacuum and multiphoton wavepackets \cite{smith2018}. Our nanostructure enables us to control the strength of the excited plasmonic field through the polarization state of the thermal multiphoton field \cite{You2021, Tame2021}. Indeed, the plasmonic near fields are only excited in our nanostructure by photons polarized along the horizontal direction. As shown in Fig. \ref{Fig6}\textbf{a}, the excited plasmonic fields propagate through the plasmonic sample inducing processes of scattering that we then measured using a pair of PNR detectors \cite{OmarOptica17, You2020Indentification, Bhusal2022}. The collective photonic-plasmonic multiparticle system in the near field of the metallic nanostructure is described by the following density matrix
\begin{equation}
    \begin{aligned}
    \label{Eq:density}
        \hat{\rho}_{\text{ph-pl}} =& \int d^2\tau P(\tau,\bar{n}_\text{H},\bar{n}_\text{V}) |\tau s_{\theta_{\text{pl}}}\rangle\langle\tau s_{\theta_{\text{pl}}}|^{\text{ph}}_\text{V}\\&\otimes|\tau c_{\psi}c_{\theta_{\text{pl}}},\tau s_{\psi}c_{\theta_{\text{pl}}}\rangle\langle\tau c_{\psi}c_{\theta_{\text{pl}}},\tau s_{\psi}c_{\theta_{\text{pl}}}|^{\text{ph-pl}}_\text{H}.
    \end{aligned}
\end{equation}
\noindent
Here, we are using the Glauber P-function representation of the state with $\tau$ as the coherent amplitude, where we used the shorthands $c_x \equiv \cos(x)$ and $s_x\equiv \sin(x)$ \cite{glauber1963}. In this expression, $\bar{n}_{\text{H/V}}$ represents the mean photon number for the mode with polarization H or V. The angle $\psi$ determines the plasmonic splitting ratio for the horizontal mode, and $\theta_{\text{pl}}$ is the polarization angle of the input state. A detailed description of this state as well as an explicit form for $P(\tau,\bar{n}_\text{H},\bar{n}_\text{V})$ can be found in Appendix B.1. The lower indices ``H'' and ``V'' represent the horizontally- and vertically-polarized components of the multiphoton system, respectively. Furthermore, the photonic modes are described by the index ``ph'', whereas the plasmonic modes are described by the index ``pl''.

The hybrid photonic-plasmonic system, described by Eq. (\ref{Eq:density}), produces the joint photon-number distribution reported in Fig. \ref{Fig6}\textbf{b}. Given the thermal nature of the illuminating field, the joint-photon number distribution of the hybrid system is classical. This is characterized by a degree of second-order coherence $g^{(2)}_{\textrm{ph-pl}}(0)$, equal to 1.91 \cite{Gerry2004}. In our experiment, we are capable of detecting multiphoton wavepackets with up to forty particles. The setup we used to perform these measurements is shown in Fig. \ref{Fig6}\textbf{c}. This experimental arrangement enables us to focus our thermal multiphoton system into different parts of the plasmonic sample. As discussed below, we keep track of the vacuum-fluctuation component of our multiphoton system by pre-selecting on vacuum events. This technique has been studied in the context of zero-photon subtraction \cite{Nunn2022, Nunn2023}. The bosonic and fermionic correlations of the plasmonic system are investigated by performing projective measurements in the far field of the sample using PNR detection.

We now discuss the contrasting dynamics between classical plasmonic waves and their constituent multiparticle subsystems. As indicated in Fig. \ref{Fig6}\textbf{c}, we illuminate one plasmonic slit with a thermal light field to generate surface plasmons. The excited plasmonic waves travel to the second slit to produce an interference pattern in the far field of the sample. As reported in previous experiments, the coherence between the interfering photons and plasmons defines the visibility of the interference fringes \cite{you2020multiparticle, You2021}. We confirm this behavior in  
Fig. \ref{Fig7}\textbf{a}. In this case, the strength of the plasmonic field is controlled through the polarization angle $\theta_{\text{pl}}$. These measurements were collected in the far field of the sample ($k-$plane) using the scanning arm in Fig. \ref{Fig6}\textbf{c}. Interestingly, this classical interference effect is produced by many quantum-mechanical interactions among the constituent multiparticle subsystems of the plasmonic waves. This can be observed by analyzing the intensity of the system described by $\langle\hat{n}(k)\rangle$
\begin{equation}
    \begin{aligned}
    \label{eq:first-order}
    \langle\hat{n}(k)\rangle&\propto  \operatorname{sinc}^{2}\left(\frac{k}{\alpha}\right) \Big(\bar{n}_{\text{V}} + \bar{n}_{\text{H}}\big[1 + \gamma\sin(2\psi)\cos(2\beta k)\big]\Big),
\end{aligned}
\end{equation}
where the parameter $\gamma$ is a scaling factor. Furthermore, $\beta=\pi d/\lambda D$ and $\alpha=\lambda D/\pi b$, where $d$ is the distance between the slits, $D$ is the propagation distance, and $\lambda$ is the wavelength of our source. A more detailed derivation and discussion of Eq. (\ref{eq:first-order}) can be found in Appendix B.1 and B.2. Importantly, since Eq. (\ref{Eq:density}) can be written as a sum over multiparticle Fock states, we see that $\langle\hat{n}(k)\rangle$ is the sum over contributions from each of its multiparticle subsystems.

We isolate the multiparticle subsystems that contribute to the classical behavior of $\langle\hat{n}(k)\rangle$ by performing projective measurements. These measurements enable us to explore the quantum dynamics of surface plasmons \cite{You2021}. We can theoretically access these subsystems by defining the density matrix of the system in the far field of the sample ($k$-plane) as
\begin{equation}
\label{eq:rho}
\hat{\rho}_{\text{ph-pl}}\left(k_1, k_2\right)=\hat{\rho}_a \otimes \hat{\rho}_b+\left(1 - \zeta \sin^2(\beta(k_1-k_2))\right)\left(\hat{\rho}_J-\hat{\rho}_a \otimes \hat{\rho}_b\right).
\end{equation}

\noindent Here, $\zeta$ depends on both $\bar{n}_\text{V}$ and the plasmon-photon couplings. The joint photon distribution between the two detectors is $\hat{\rho}_J= \int d^2\tau P(\tau)\left|\tau c_\theta, \tau s_\theta\right\rangle\left\langle \tau c_\theta, \tau s_\theta\right|$ where $P(\tau)$ is the P-function of a thermal state and $\theta$ measures the difference in intensity between the detectors. Furthermore, $\hat{\rho}_a=\Tr_b\left[\hat{\rho}_J\right]$, and $\hat{\rho}_b=\Tr_a\left[\hat{\rho}_J\right]$. The P-function $P(\tau)$ depends on the intensity $\langle\hat{n}(k)\rangle$  collected by the two detectors. The projection of the system into a subsystem with a number of particles $n_1$ leads to the probability distribution $p_a(n_1)=\Tr{[\hat{\rho}_{\text{ph-pl}}(k_1,k_2)\hat{n}_{an_1}]}$ where $\hat{n}_{an_1} = |n_1\rangle_a\langle n_1|_a$. These subsystems uncover new multiparticle dynamics that exhibits a surprising opposite behavior with respect to the classical plasmonic waves shown in Fig. \ref{Fig7}\textbf{a}. For example, the projection of the hybrid plasmonic-photonic system into its vacuum component leads to the quantum
interference pattern in Fig. \ref{Fig7}\textbf{b}, which is modulated by a dip. Remarkably, the stronger confinement of plasmonic near fields in the nanostructure increases vacuum coherence. 
The formation of these interference
structures is produced by the interaction of vacuum plasmons with the vacuum-fluctuation of the illuminating field \cite{Ford1993}. This effect leads to the formation of an interference pattern with higher visibility such as the one reported in Fig. \ref{Fig7}\textbf{c}. As discussed in Appendix B.3, the conditional projective measurements of vacuum events guarantee that approximately $83\%$ of the measured events in Fig. \ref{Fig7}\textbf{b} and \textbf{c} are indeed induced by the vacuum fluctuations of the illuminating field. 
Thus, only a small portion of these events were excited by photons that were lost \cite{Nunn2022,Nunn2023}.

\begin{figure}[htbp]
  \centering
  \includegraphics[width=0.99\textwidth]{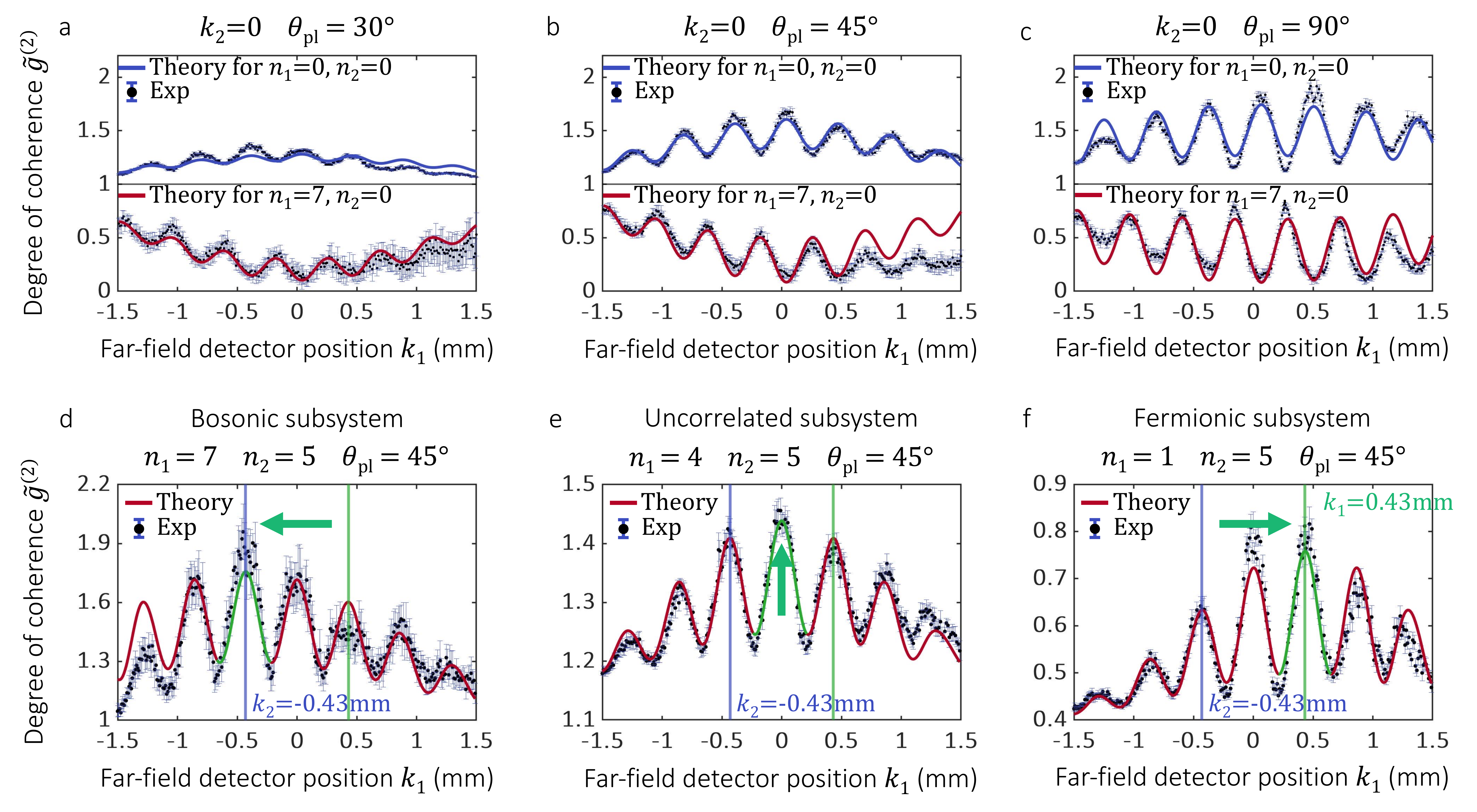}
  \caption[Observation of bosonic- and fermionic-like processes of coherence in a plasmonic system]{\textbf{Observation of bosonic- and fermionic-like processes of coherence in a plasmonic system.} We access to the correlated and anti-correlated processes of scattering in the plasmonic system by projecting it into the multiparticle systems $n_1$ and $n_2$. This operation is implemented by placing two PNR detectors in the far-field plane of the sample at positions $k_1$ and $k_2$.  We discuss the vacuum dynamics of the plasmonic system from \textbf{a} to \textbf{c}. Interestingly, the correlation of the vacuum components of the system show classical bosonic features characterized by $\tilde{g}^{(2)}>1$.  However, the correlation of vacuum with a seven-particle subsystem shows fermionic features characterized by $\tilde{g}^{(2)}<1$. The visibility of the interference fringes reported from \textbf{a} to \textbf{c} is controlled through the strength of the plasmonic near fields in the sample. Remarkably,
  the displacement of the fixed detector to the position $k_2=-0.43$ mm unveils more interesting scattering dynamics mediated by either bosonic or fermionic processes of coherence. The projection of the plasmonic system into the twelve-particle subsystem in \textbf{d} leads to the formation of a correlated interference structure centered at $k_1=-0.43$ mm. This pattern is mediated by bosonic coherence. It is also possible to isolate uncorrelated dynamics of multiparticle systems. This is reported in \textbf{e} for a nine-particle plasmonic subsystem. For the first time, we were able to isolate anti-correlated scattering events among the multiparticle constituents of a plasmonic system. Surprisingly, the anti-correlated pattern produced by the six-particle subsystem in \textbf{f} is produced by the fermionic coherence hosted by our plasmonic system. The theoretical predictions were estimated based on Eq. (\ref{eq:last}). The errors indicate the standard deviation calculated from ten datasets, where each of them consists of approximately 100,000 photon-number-resolving measurements.}
\label{Fig8}
\end{figure}

The multiparticle subsystems carried by the plasmonic waves in the nanostructure show interesting dynamics that differ from those in Fig. \ref{Fig7}\textbf{a}. Specifically, the four-particle interference pattern in Fig. \ref{Fig7}\textbf{d} shows higher visibility than the patterns in Fig. \ref{Fig7}\textbf{a}. Indeed, larger multiparticle subsystems show higher visibility, this is reported from Fig. \ref{Fig7}\textbf{d} to \textbf{f}. Interestingly, the projection of the surface plasmon into the eight-particle subsystem in Fig. \ref{Fig7}\textbf{f} enables us to extract an interference pattern with nearly perfect visibility. These interference structures suggest the presence of quantum coherence mediating the dynamics of the multiparticle subsystems \cite{mandel_wolf_1995}. Indeed, the observation of quantum coherence is reported from Fig. \ref{Fig7}\textbf{g} to \textbf{i}. 
We explore the role of quantum coherence through the conditional degree of second-order coherence defined as $\tilde{g}^{(2)}\left(k_{1}, k_{2}\right)=\operatorname{Tr}\left[\hat{\rho}_{\text{ph-pl}}\left(k_1, k_2\right)\hat{n}_{an_1} \hat{n}_{bn_2}\right]/\left(p_a(n_1)p_b(n_2)\right)$. This function takes values larger than one for bunching processes that characterize the dynamics of classical bosonic systems. However, $\tilde{g}^{(2)} (0)$, for the same detector position $k_{1}=k_{2}=0$, can be smaller than one for quantum antibunching processes \cite{Jeltes2007}. Interestingly, it has been identified that this kind of nonclassical processes is mediated by fermionic coherence \cite{Jeltes2007, Crespi2013}. Additional details regarding the properties of $\tilde{g}^{(2)}$ can be found in Appendix B.4. Furthermore, it is worth noticing that bunching and anti-bunching effects observed in photonic systems are commonly ascribed to bosonic and fermionic statistics, respectively \cite{vest2017anti, Crespi2013, Li2021, Li2021}. This idea has been applied to describe interference phenomena produced by photonic Bell wavefunctions with different symmetries \cite{Crespi2013}. Similarly, fermionic, and bosonic statistics have recently been used to describe bosonic and fermionic photon-photon interactions \cite{Li2021}. In our experiment, we detect photons rather than plasmons. However, as described by Eq. (\ref{Eq:density}) and Eq. (\ref{eq:rho}), the measured interference patterns are dictated by the coherence properties of the plasmonic field surrounding our metallic structure. Thus, we attribute observed bunching effects to the presence of bosonic-like coherence, while anti-bunching phenomena are associated with fermionic-like coherence of the plasmonic subsystem \cite{vest2017anti, Jeltes2007, Crespi2013, Li2021}. Remarkably, the conditional degree of second-order coherence $\tilde{g}^{(2)}(0)$ reported from Fig. \ref{Fig7}\textbf{g} to \textbf{i} demonstrates that the multiparticle dynamics of plasmonic systems is indeed mediated by bosonic- and fermionic-like processes of coherence \cite{Jeltes2007, Crespi2013, vest2017anti}. The properties of coherence are defined by the number of particles in the measured wavepacket. As such, the multiparticle coherence is controlled by the strength of the plasmonic field surrounding the metallic structure. While the multiparticle subsystems in Fig. \ref{Fig7}\textbf{g} to \textbf{i} may exhibit classical or quantum coherence, most wavepackets exhibit classical bosonic properties. As described by Eq. (\ref{eq:first-order}), the collective contributions from all the constituent multiparticle subsystems produce the classical bosonic behavior observed in macroscopic surface plasmons \cite{tame2013quantum}.

Now we demonstrate that multiparticle subsystems with bosonic or fermionic coherence in Fig. \ref{Fig7}\textbf{g} to \textbf{i} induce either correlated or anti-correlated scattering processes, respectively \cite{Jeltes2007, Crespi2013, vest2017anti}. This novel correlated dynamics of plasmonic subsystems, comprising $n_1+n_2$ number of particles, can be described as follows
\begin{equation}
    \label{eq:last}
    \begin{aligned}
    \tilde{g}^{(2)}\left(k_1, k_2 \right)&=\sinc^2\left(\frac{k_{1}-k_{2} + k'}{\sigma}\right)\left( 1 + \left(1 - \zeta \sin^2(\beta(k_1-k_2))\right)\Big[\tilde{g}_{\text{th}}^{(2)}(n_1,n_2) - 1\Big]\right).
    \end{aligned}
\end{equation}
Here, the $\tilde{g}_{\text{th}}^{(2)}(n_1,n_2) - 1$ term describes the coherence properties of the detected multiparticle subsystem. These properties of coherence dictate the visibility of the bosonic- or fermionic-like processes of scattering reported in Fig. \ref{Fig8}. Specifically, the interference of $n_1$ with $n_2$ particles is described by $1 - \zeta \sin^2(\beta(k_1-k_2))$. This scattering process is modulated by $\sinc^2\left((k_{1}-k_{2} + k')/\sigma\right)$. Further discussion and derivation of this formula can be found in Appendix B.1. We now discuss the observation of scattering effects induced by bosonic- and fermionic-like processes of coherence in Fig. \ref{Fig8}. In this case, we scan the position of detector one, $k_1$, while the position of detector two, $k_2$, is fixed. We first explore the correlated bosonic dynamics of the vacuum component of the plasmonic system. As shown from Fig. \ref{Fig8}\textbf{a} to \textbf{c}, the correlation of vacuum events unveils the formation of bosonic interference \cite{Jeltes2007}. The visibility of the interference pattern increases with the strength of the near fields surrounding the plasmonic structure. Interestingly, the correlation of a multiparticle subsystem with the vacuum component of the plasmonic system produces nonclassical interference fringes with fermionic features \cite{Crespi2013}. The visibility of these patterns is also controlled through the confinement of plasmonic near fields. Remarkably, the displacement of the fixed detector makes the role of the bosonic or fermionic coherence of the plasmonic system more evident. In this case, we set the position of the fixed detector $k_2$ to $-0.43$ mm. This configuration unveils the bosonic correlations of the twelve-particle subsystem in Fig. \ref{Fig8}\textbf{d}. In this case, the highest peak of the correlated pattern is also centered at $k_1=-0.43$ mm. However, the strength of the correlation can be attenuated by projecting the plasmonic system into an uncorrelated subsystem. This is reported in Fig. \ref{Fig8}\textbf{e} for a nine-particle subsystem. Surprisingly, the fermionic coherence that characterizes the six-particle subsystem in Fig. \ref{Fig8}\textbf{f} induces the formation of an anti-correlated interference pattern centered at $k_1=+0.43$ mm. The small deviation is due to experimental imperfections. These observations unveil novel nonclassical dynamics of surface plasmons. 

\section{Summary}

In this chapter, we demonstrated that the classical behavior of surface plasmons is indeed produced by nonclassical scattering processes among their constituent multiparticle subsystems \cite{HongNP24, OPN}. These novel dynamics of plasmonic systems were experimentally demonstrated by isolating multiparticle subsystems through the implementation of projective measurements. This capability enabled the exploration of the vacuum dynamics of surface plasmons. In addition, it was shown that the scattering dynamics of the multiparticle subsystems that form a plasmonic wave are mediated by bosonic and fermionic processes of coherence. Altogether, these findings reveal fundamental quantum properties of plasmonic quasiparticles with broad implications across disciplines, including condensed matter physics, nuclear physics, and quantum information science \cite{anno2006,omar2019Multiphoton,thouless2014quantum, AspuruGuzik2012, obrien2009}.

\chapter{Conditional multiparticle quantum plasmonic sensing}

Building upon the previous chapter, which examined the nonclassical nonclassical dynamics of plasmonic near fields, this chapter explores how those properties can be leveraged for important applications. In particular, we focus on multiparticle quantum plasmonic sensing as a strategy to improve precision measurements at the nanoscale.

The possibility of using weak optical signals to perform sensing of delicate samples constitutes one of the main goals of quantum photonic sensing. Furthermore, the nanoscale confinement of electromagnetic near fields in photonic platforms through surface plasmon polaritons has motivated the development of highly sensitive quantum plasmonic sensors. Despite the enormous potential of plasmonic platforms for sensing, this class of sensors is ultimately limited by the quantum statistical fluctuations of surface plasmons. Indeed, the fluctuations of the electromagnetic field severely limit the performance of quantum plasmonic sensing platforms in which delicate samples are characterized using weak near-field signals. Furthermore, the inherent losses associated with plasmonic fields levy additional constraints that challenge the realization of sensitivities beyond the shot-noise limit. In this chapter, we introduce a protocol for quantum plasmonic sensing based on the conditional detection of plasmons. It is demonstrated that the conditional detection of plasmonic fields, via plasmon subtraction, provides a new degree of freedom to control quantum fluctuations of plasmonic fields. This mechanism enables improvement of the signal-to-noise ratio of photonic sensors relying on plasmonic signals that are comparable to their associated field fluctuations. Consequently, the possibility of using weak plasmonic signals to sense delicate samples, while preserving the sample properties, has important implications for molecule sensing and chemical detection.

\section{Introduction}

The possibility of controlling the confinement of plasmonic near-fields at the subwavelength scale has motivated the development of a variety of extremely sensitive nanosensors \cite{NanoToday, Hwang, Maier, Lee2021}. Remarkably, this class of sensors offers unique resolution and sensitivity properties that cannot be achieved through conventional photonic platforms in free space \cite{You2021APR,Slussarenko2017,Lee2021, Polino}. In recent decades, the fabrication of metallic nanostructures has enabled the engineering of surface plasmon resonances to implement ultrasensitive optical transducers for detection of various substances ranging from gases to biochemical species \cite{NanoToday,Hwang,Lee2021}. Additionally, the identification of the quantum mechanical properties of plasmonic near-fields has prompted research devoted to exploring mechanisms that boost the sensitivity of plasmonic sensors \cite {altewischer2002plasmon, akimov2007generation, tame2013quantum, you2020multiparticle, You2021}. 

The scattering paths provided by plasmonic near-fields have enabled robust control of quantum dynamics \cite{vest2017anti,schouten2005,magana2016exotic, You2021}. Indeed, the additional degree of freedom provided by plasmonic fields has been used to harness the quantum correlations and quantum coherence of photonic systems \cite{buse2018symmetry,You2021,li2017strong,schouten2005}. Similarly, this exquisite degree of control made possible the preparation of plasmonic systems in entangled and squeezed states \cite{Vest,huck2009demonstration,fasel2005energy,di2012quantum}. Among the large variety of quantum states that can be engineered in plasmonic platforms \cite{tame2013quantum,you2020multiparticle}, entangled systems in the form of N00N states or in diverse forms of squeezed states have been  used to develop quantum sensors \cite{Heeres2013,Chen:18,Lee2021,dowran2018quantum,lee2016quantum}. In principle, the sensitivity of these sensors is not constrained by the quantum fluctuations of the electromagnetic field that establish the shot-noise limit \cite{lee,Polino}. However, due to inherent losses of plasmonic platforms, it is challenging to achieve sensitivities beyond the shot-noise limit under realistic conditions \cite{You2021APR}. Despite existing obstacles, recent work demonstrates the potential of exploiting nonclassical properties of plasmons to develop quantum plasmonic sensors for detection of antibody complexes, single molecules, and to perform spectroscopy of biochemical substances \cite{mejia2018plasmonic,Kongsuwan,alashnikov,Mauranyapin2017}. 

In this chapter, a new scheme for quantum sensing is presented, based on plasmon-subtracted thermal states \cite{Dakna, omar2019Multiphoton, OmarOptica17}. This approach provides an alternative to existing protocols that rely on entangled or squeezed plasmonic states \cite{Vest,huck2009demonstration,fasel2005energy,di2012quantum,Heeres2013,Chen:18,Lee2021,dowran2018quantum,lee2016quantum}. A sensing architecture based on a nanoslit plasmonic interferometer is used \cite{biosensordeleon}, which provides a direct relationship between the light exiting the interferometer and the phase shift induced in one of its arms by the substance to be sensed (analyte). The protocol involves a conditional quantum measurement applied to the interfering plasmonic fields via plasmon subtraction. This process enables the reduction of quantum fluctuations in the sensing field and increases the mean occupation number of the plasmonic sensing platform \cite{omar2019Multiphoton, OmarOptica17}. Furthermore, plasmon subtraction provides a method for manipulating the signal-to-noise ratio (SNR) associated with the measurement of phase shifts. It is shown that the reduced fluctuations of plasmonic fields lead to an enhancement in phase estimation, quantified through the uncertainty associated with phase measurements. Reduced uncertainties in phase lead directly to improved sensitivity in the sensing protocol. The analysis is carried out using a quantum mechanical model that incorporates realistic loss effects in a plasmonic nanoslit sensor. The probabilities of successfully implementing the protocol are reported as a function of the plasmonic field occupation number and the losses of the nanostructure. The results presented in this chapter suggest that the proposed protocol offers practical advantages for lossy plasmonic sensors operating with weak near-field signals \cite{Maga_a_Loaiza_2016}. Consequently, our platform can have important implications for plasmonic sensing of delicate samples such as molecules, chemical substances or, in general, photosensitve materials \cite{mejia2018plasmonic,Kongsuwan,alashnikov,Mauranyapin2017}.

\section{Results and discussion}

\begin{figure}[ht!]
	\centering
	\includegraphics[width=1\textwidth]{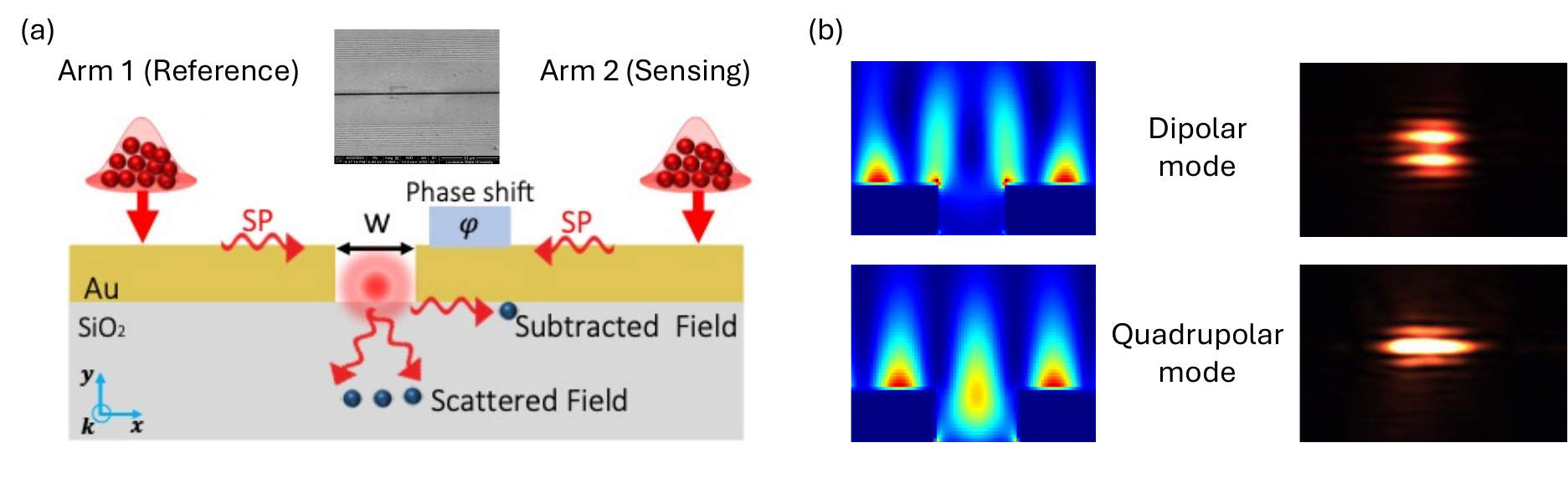}
	\caption[The plasmonic nanoslit for conditional quantum sensing and its performance]{The diagram in (a) shows the design of our simulated plasmonic sensor, comprising a slit of width w in a $200$ nm gold thin film. Here, the plasmonic structure is illuminated by two thermal multiphoton sources that excite two plasmonic fields with super-Poissonian statistics (the input grating couplers are not shown in the figure). The two counter-propagating surface plasmon (SP) modes  interfere at the interface between the gold layer and the $\text{SiO}_2$ substrate. The interference conditions are defined by the phase shift $\varphi$ induced in one of the plasmonic modes by the substance that we aim to sense. The inset is our well fabricated nanoslit structure. The performance of our sample is examined and shown in (b). We implement both dipolar and quadrupolar plasmonic excitations. In the left panel we show our simulation while in the right panel we present the corresponding experimental images captured in the near field with a CCD camera. By comparing the images, we can clearly see that the experimentally observed patterns inside the central nanoslit closely match the simulated results, confirming the accuracy of our setup and theoretical predictions.}
	\label{Fig9}
\end{figure}

We begin by discussing the theoretical model used to describe conditional quantum measurements in a thermal plasmonic system. The system is based on a plasmonic nanoslit that supports interactions between one photonic mode and two plasmonic modes. Specifically, a photonic input mode, denoted by the operator $\hat{b}$, is coupled with two surface plasmon modes, represented by $\hat{a}$ and $\hat{c}$. These interactions occur within the nanoslit structure, which acts as a plasmonic tritter—splitting and coupling the incoming modes into three distinct output channels. After the interaction, the output photonic mode is described by the operator $\hat{e}$, while the output plasmonic modes are labeled as $\hat{f}$ and $\hat{d}$. As indicated in Fig. \ref{Fig9}\textbf{a}, and throughout this chapter, we study the conditional detection of the output modes $\hat{d}$ and $\hat{e}$ for a situation in which only the input plasmonic modes of $\hat{a}$ and $\hat{c}$ are excited in the nanostructure. Thus, the photonic mode $\hat{b}$ is assumed to be in a vacuum state. In this case, the plasmonic tritter can be simplified to a two-port device described by the following $2\times 2$ matrix
\begin{equation}
\left(
\begin{array}{c}
\hat{d}\\
\hat{e}\\

\end{array}
\right) =  
\left(
\begin{array}{cc}
\kappa& r\\
\tau&\tau
\end{array}
\right )
\\
\left(
\begin{array}{c}
\hat{a}\\
\hat{c}
\end{array}
\right ).
\label{eq1}
\end{equation}

The photonic mode $\hat{e}$ is transmitted through the slit and its transmission probability is described by $2\lvert {\tau}\lvert ^2=T_\text{ph}$. Here, $T_\text{ph}$ represents the normalized intensity of the transmitted photons. Moreover, the plasmon-to-plasmon coupling at the output of the nanostructure is given by $\lvert{\kappa}\lvert^2+\lvert{r}\lvert^2=T_\text{pl}$. Here, the renormalized transmission (after interference and considering loss) for the plasmonic fields is described by $T_\text{pl}$. From Fig. \ref{Fig9}\textbf{a}, we note that the interference supported by the plasmonic nanoslit shares similarities with those induced by a conventional Mach-Zehnder interferometer (MZI) . More specifically, the two plasmonic modes, $\hat{a}$ and $\hat{c}$, interfere at the location of the nanoslit, which in turn scatters the field to generate the output\cite{biosensordeleon}. The interference conditions are defined by the phase shift induced by the analyte. Plasmonic sensors with nanoslits have been extensively investigated in the classical domain, showing the possibility of ultrasensitive detection using minute amounts of analyte \cite{biosensordeleon,NanoToday, Hwang, Maier, Lee2021,Chen:18}. The performance of the plasmonic sample is validated through both simulation and experiment, as shown in Fig. \ref{Fig9}\textbf{b}. We excite the structure using both dipolar and quadrupolar plasmonic modes. The left panel displays the simulated field distributions, while the right panel presents near-field images recorded with a CCD camera. The strong agreement between the simulated and measured patterns—particularly within the central nanoslit—demonstrates the reliability of the experimental setup and the accuracy of the theoretical model.

We now consider a situation in which a single-mode thermal light source  is coupled to the nanostructure in Fig. \ref{Fig9}\textbf{a} exciting two counter-propagating surface plasmon modes. This can be achieved by
using a pair of grating couplers (not shown in the figure) \cite{biosensordeleon}. The statistical properties of this thermal field can be described by the Bose-Einstein statistics as 
$\rho_{th}= \sum_{n=0}^{\infty} \text{p}_{pl}(n) {\ket n} {\bra n}$, where $\text{p}_\text{pl}(n)=\bar n^{n}/(1+\bar n)^{1+n}$, and $\bar{n}$ represents the mean occupation number of the field. Interestingly, the super-Poissonian statistics of 
thermal light can be modified through conditional measurements \cite{omar2019Multiphoton,Dakna,OmarOptica17}.
As discussed below, it is also possible to modify the quantum statistics of plasmonic fields. The control of plasmonic statistics can be implemented by subtracting/adding bosons from/to thermal plasmonic systems \cite{Mizrahi,Parigi}. In this work, we subtract plasmons from the transmitted field formed by the superposition of the surface plasmon modes propagating through the reference
and sensing arms of the interferometer. This is the transmitted mode $\hat{e}$ conditioned on the output of the field $\hat{d}$. The successful subtraction of plasmons boosts the signal of the sensing platforms. This feature is particularly important for sensing schemes relying on dissipative plasmonic platforms.

The conditional subtraction of $L$ plasmons from the mode $\hat{d}$ leads to the modification of the quantum statistics of the plasmonic system, this can be described by
\begin{equation}
\text{p}_{\text{pl}}(n) =  \frac{(n+L)!\bar n_{\text{pl}}^{n}}{ n! L!(1+\bar n_{\text{pl}})^{L+1+n}}, 
\label{eq3prime}
\end{equation} 
where $\bar{n}_\text{pl}$ represents the mean occupation number of the scattered field in mode $\hat{e}$. We quantify the modification of the quantum statistics through the degree of second-order correlation function $g^{(2)}(0)$ for the mode $\hat{e}$ as 
\begin{equation}
\begin{split} 
g^{(2)}_L (0) = {\frac{L+2}{L+1}}.
\end{split} 
\label{eqg2}
\end{equation}
We note that the conditional subtraction of plasmons induces anti-thermalization effects that attenuate the fluctuations of the plasmonic thermal system used for sensing. Indeed, the $g^{(2)}_L (0)$ approaches one with the increased number of subtracted plasmons, namely large values of $L$. This effect produces bosonic distributions resembling those of coherent states \cite{omar2019Multiphoton}. Recently, similar anti-thermalization effects have been explored in photonic lattices \cite{kondakci2015}. 

\begin{figure}[ht!]
	\centering \includegraphics[width=1\textwidth]{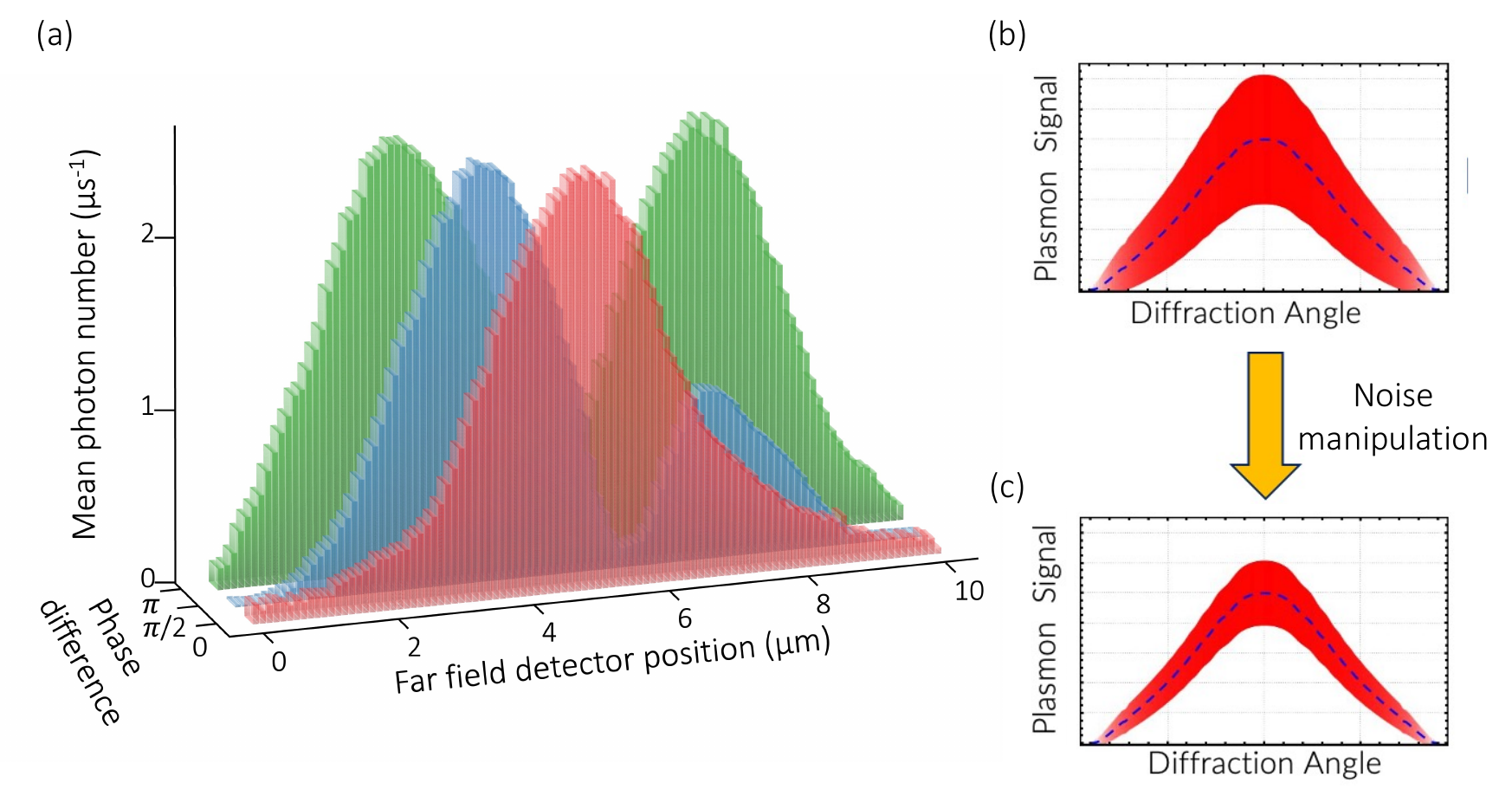}
	\caption[Far-field intensity distribution scattered by the plasmonic nanoslit]{Far-field intensity distribution scattered by the plasmonic nanoslit. Panel (a) presents the experimentally measured far-field intensity distributions corresponding to three different phase differences: $\varphi = 0$ (red bars), $\varphi = \pi/2$ (green bars), and $\varphi = \pi$ (blue bars). The data illustrates the evolution of the interference pattern as the phase between the input modes is varied. Panel (b) shows the simulated interference pattern produced by the field transmitted through the 320-nm-wide slit, corresponding to mode $\hat{e}$. Panel (c) displays the corresponding theoretical prediction after conditional detection of plasmons with 3 photons subtracted, highlighting the enhancement in signal-to-noise ratio. The dashed line in both (b) and (c) represents the intensity profile associated with quadrupolar near-field symmetries for $\varphi = 0$, respectively. The red shaded region in (b) and (c) indicates the standard deviation for $\bar{n} = 3.75$.} 
	\label{Fig10}
\end{figure}

The aforementioned plasmon subtraction can be implemented in the plasmonic nanoslit interferometer  shown in Fig. \ref{Fig9}\textbf{a}. It consists of a 200 nm thick gold film deposited on a glass substrate \cite{biosensordeleon}. This thickness is large enough to enable decoupled plasmonic modes on the top and bottom surfaces of the film, as required.  The gold film features a 320 nm slit, defining the reference arm of the interferometer to its left and the sensing arm (holding the analyte) to its right. The analyte then induces a phase difference $\varphi$  relative to the reference arm, thereby creating the output ($\hat{d}, \hat{e}$ and $\hat{f}$) that depends on this parameter. To verify the feasibility of our conditional measurement approach, we perform a finite-difference time-domain (FDTD) simulation of the plasmonic nanoslit using a wavelength of $\lambda=810$ nm for the two counter-propagating surface plasmon modes ($\hat{a}$ and $\hat{c}$). The nanoslit is designed to support two localized surface plasmon (LSP) modes, one with dipolar symmetry and other with quadrupolar symmetry. Depending on the phase difference $\varphi$, these two LSP modes can be excited with different strengths by the fields interfering at the nanoslit. being the dipolar (quadrupolar) mode optimally excited with $\varphi = 0$ ($\varphi= \pi$). This is due to the fact that the near-field symmetries of the interfering field are well-matched to the dipolar and quadrupolar fields for those values of $\varphi$\cite{biosensordeleon}. Remarkably, we experimentally measured the far-field intensity, which is depicted in Fig. \ref{Fig10}\textbf{a}. This dataset reveals how the interference pattern changes as a function of the phase difference between the input modes. By comparing the intensity distributions for $\varphi = 0$, $\pi/2$, and $\pi$, we can clearly observe the phase-dependent modulation of the far-field signal. The dashed lines in Fig. \ref{Fig10}\textbf{b} and \textbf{c} indicate the far-field angular distributions of the transmitted intensity associated with the quadrupolar LSP mode. Only a small angular range of the far-field distribution is used as the sensing signal \cite{biosensordeleon}.
The transmission parameters of our sensor are estimated from FDTD simulations. Specifically, the transmission values for the photonic and plasmonic modes are $T_{\text{ph}}\approx 0.076$ and $T_{\text{pl}}\approx0.0176$ for $\varphi= \pi$. However, our subtraction scheme is general and valid for any phase angle  $\varphi$  in the range of $0 \leq \varphi \leq   2 \pi$. Moreover, the total amount of power coupled to modes $\hat{e}$ and $\hat{d}$ normalized to the input power of the plasmonic structure is defined as $\gamma=T_{\text{ph}}+T_{\text{pl}}\approx0.0941$.
For the results shown in Fig. \ref{Fig10}\textbf{b} and \textbf{c}, we assume a mean occupation number of $\bar n=3.75$ for the input beam. As shown in Fig. \ref{Fig10}\textbf{b}, the output signals, calculated from Eq. \eqref{eq3prime} and represented by the red shaded region, exhibit strong quantum fluctuations. Surprisingly, after performing plasmon subtraction, the quantum fluctuations decrease, as indicated in Fig. \ref{Fig10}\textbf{c}. Evidently, this confirms that our conditional measurement protocol can indeed boost the output signal and consequently improve the sensing performance of a plasmonic device. However, due to the probabilistic nature of our protocol and the presence of losses, it is important to estimate the probability rates of successfully performing plasmon subtraction. In Table \ref{tab:table1} we list the degree of second-order correlation $ g^{(2)}_L (0)$, and the probability of successfully subtracting one, two, and three plasmons for different occupation numbers of the plasmonic fields used for sensing. It is worth mentioning that conditional measurements in photonic systems have been experimentally demonstrated with similar efficiencies \cite{OmarOptica17,omar2019Multiphoton}.

\begin{table} [!htbp]
		\centering
	\caption{The estimated probability of plasmon subtraction and the corresponding degree of second-order coherence $ g^{(2)}_L (0)$. The losses of the plasmonic nanostructure reduce the probability of subtracting multiple plasmons $L$ from the scattered field with an occupation number of $\bar{n}$. In this case, we assume $\varphi=\pi$.}
	\begin{tabular}{p{1.5cm}p{2cm}p{2cm}p{2cm}}
		$\bar{n}$& $ L=1$ &$ L=2 $ &$ L=3$\\
		\hline
		2&$1.0 \times10^{-2}$ &$1.0 \times10^{-4}$ &$1.1 \times10^{-6}$\\
		1&$5.2 \times10^{-3}$ &$2.7\times10^{-5}$ &$1.4 \times10^{-7}$\\
		0.5&$2.6 \times10^{-3}$&$7.0 \times10^{-6}$&$1.8 \times10^{-8}$\\
		0.3&$1.5\times10^{-3}$&$2.5\times10^{-6} $&$4.0 \times10^{-9} $\\
		\hline
		$ g^{(2)}_L (0)$&1.5&1.33&1.25\\
		\hline
	\end{tabular}
	\label{tab:table1}
\end{table}

The quantities reported in Table \ref{tab:table1} were estimated for a phase shift given by $\varphi=\pi$. This table considers realistic parameters for the losses associated to the propagation of the plasmonic sensing field, and the limited efficiency $\eta_{\text{ph}}$ and $\eta_{\text{pl}}$ of the single-photon detectors used to collect photonic and plasmonic mode respectively. In this case, we assume $\eta_{\text{ph}}=0.3$ and $\eta_{\text{pl}}=0.3$. The latter value is obtained from our simulation, whereas the former corresponds to the efficiency of commercial single-photon detectors \cite{OmarOptica17,Marsili2013}.  In general, the value for $\varphi$ determines how strongly the dipolar and quadrupolar LSP modes are excited, and consequently their far-field angular distributions. However, the process is applicable for other phases $\varphi$. Our predictions suggest that plasmonic subtraction can be achieved at reasonable rates using a properly designed nanostructure.
 
 \begin{figure}[ht!]
 	\centering \includegraphics[width=0.5\columnwidth]{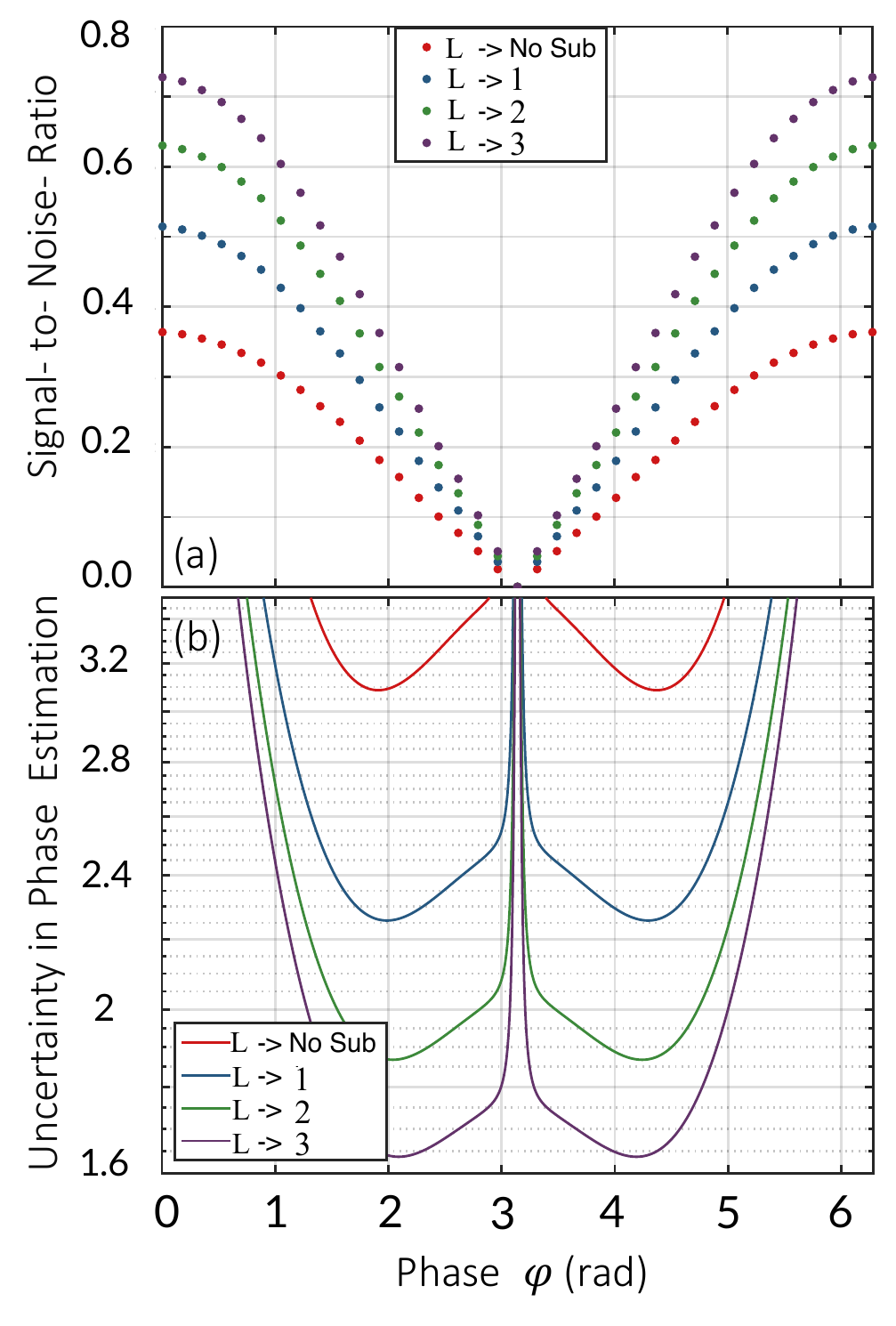}
 	\caption[SNR for the conditional detection of the plasmonic modes and uncertainties in the estimation of phase shifts induced by analytes]{ The panel in (a) reports the signal-to-noise ratio (SNR) as a function of $\varphi$ for the conditional detection of the plasmonic modes transmitted by a 320-nm nanoslit. The red dots represent the unconditional SNR. Furthermore, the blue, green, and purple dots indicate the SNR for the subtraction of one, two, and three plasmons, respectively. This plot shows the possibility of improving the SNR of our plasmonic sensor through the subtraction of plasmons. The panel in (b) indicates that an increasing SNR leads to lower uncertainties in the estimation of phase shifts induced by analytes. The lower uncertainties described by $\Delta\varphi$ imply higher sensitivities of our plasmonic sensor. }
 	\label{Fig11}
 \end{figure}
 
We now quantify the performance of our conditional scheme for plasmonic sensing through the SNR associated to the estimation of a phase shift. The SNR is estimated as the ratio of the mean occupation number to its standard deviation. This is defined as
\begin{equation}
\begin{split} 
\text{SNR} =  \sqrt[]{\frac{(1+L) \bar n \gamma\eta_{\text{ph}} \xi \cos^{2}\frac{\varphi}{2}}{1+ \bar n  \gamma(\xi\eta_{\text{ph}}+(1-\xi)\eta_{\text{pl}}) \cos^{2}\frac{\varphi}{2} }}.
\end{split} 
\label{eq4}
\end{equation}
\noindent 
Here, the parameter $\xi=T_{\text{ph}}/(T_{\text{ph}}+T_{\text{pl}})=0.80 $ 
represents the normalized transmission of the photonic mode. In Fig. \ref{Fig11}\textbf{a}, we report the increasing SNR of our plasmonic sensor through the process of plasmon subtraction by plotting the SNR for the subtraction of one, two, and three plasmons for different phase shifts $\varphi$. In addition, for sake of completeness, we evaluate the improvement in sensitivity using error propagation \cite{Hwanglee}. More specifically, we calculate the uncertainty of a phase measurement $\Delta\varphi$. This parameter is estimated as 
\begin{equation}
\Delta \varphi=\sqrt{\left\langle \hat{n}^{2}\right\rangle-\langle \hat{n}\rangle^{2}}/\left|\frac{d\langle \hat{n}\rangle}{d \varphi}\right|
\label{eqg5}
\end{equation}

Here, the observable $\hat{n}$ corresponds to the conditional intensity measurement within an angular range of the far-field distribution. In the field of quantum metrology, the reduced uncertainty of a phase measurement $\Delta\varphi$ is associated to an improvement in the sensitivity of a quantum sensor \cite{Hwanglee,phaseEstimation}. In this regard, the conditional detection of plasmons increases the sensitivity of our plasmonic sensor. This enhancement is reported in Fig. \ref{Fig11}\textbf{b}. Here, we demonstrate that the attenuation of the fluctuations of a weak plasmonic field, through the subtraction of up to three plasmons, leads to lower uncertainties in the sensing of photosensitive analytes. 

\section{Summary}

In this chapter, a new method for quantum plasmonic sensing based on the conditional subtraction of plasmons was investigated. The performance of this scheme was quantified under realistic loss conditions by considering the design of a real plasmonic nanoslit sensor. It was shown that conditional measurements offer an important path for controlling the statistical fluctuations of plasmonic fields for sensing. The analysis focused on the regime in which the sensing field contains a mean plasmon number below two. In this regime, it was shown that attenuating the quantum fluctuations of plasmonic fields increases the mean occupation number of the sensing field. Interestingly, this effect leads to larger signal-to-noise ratios for the sensing protocol. Furthermore, this feature enables sensitive plasmonic sensing using weak signals \cite{NanoToday, Hwang, Maier, Lee2021}. Overall, the results presented in this chapter offer an alternative strategy for boosting signals in quantum plasmonic platforms operating under loss and in the few-particle regime \cite{tame2013quantum, you2020multiparticle}.

\chapter{Multiphoton quantum imaging using natural light}

As a continuation of the previous chapters on quantum plasmonics, this chapter explores the broader applicability of multiparticle quantum optics by implementing a quantum imaging protocol based on natural light. It is thought that schemes for quantum imaging are fragile against realistic environments in which the background noise is often stronger than the nonclassical signal of the imaging photons. Unfortunately, it is unfeasible to produce brighter quantum light sources to alleviate this problem. In this chapter, we address this limitation by developing a quantum imaging scheme that relies on the use of natural sources of light. This is achieved by performing conditional detection on the photon number of the thermal light field scattered by a remote object. Specifically, the conditional measurements in our scheme enable us to extract quantum features of the detected thermal photons to produce quantum images with improved signal-to-noise ratios. This technique shows an exponential enhancement in the contrast of quantum images. Moreover, the measurement strategy enables imaging that originates from the vacuum fluctuations of the light field. This is experimentally demonstrated through the implementation of a single-pixel camera with photon-number-resolving capabilities. These results, presented in this chapter, introduce a new approach to quantum imaging and highlight the potential of combining natural light sources with nonclassical detection schemes for the development of robust quantum technologies.

\section{Introduction}

The use of nonclassical correlations of photons to produce optical images in a nonlocal fashion gave birth to the field of quantum imaging almost three decades ago \cite{PhysRevApticalimaging, PhysRevLettTwoPhotonCoincidence, Maga_a_Loaiza_2019,Bhusal2022}. Interestingly, it was then discovered that exploiting the quantum properties of the light field enables improving the resolution of optical instruments beyond the diffraction limit \cite{TwoPhotonLithography, PhysRevLettScalableSpatial, PhysRevLettDowling, PhysRevXSuperresolution, Bhusal2022}. It was also shown that schemes for quantum imaging allow for the formation of images with sub-shot-noise levels of precision \cite{Brida2010NaturePhotonics, Quantumsecuredimaging, Lemosundetectedphotons}. These features have been exploited to demonstrate the formation of few-photon images with high contrast \cite{APlMagana-Loaiza2013, Morris2015NatureComm, BarretoLemosmetrology}. Furthermore, the compatibility of quantum imaging techniques with protocols for quantum cryptography have cast interest in the development of schemes for quantum-secured imaging \cite{Quantumsecuredimaging, Maga_a_Loaiza_2019}. Despite the enormous potential of quantum imaging for microscopy, remote sensing, and astronomy, schemes for quantum imaging remain fragile against realistic conditions of loss and noise \cite{ref1Microscopy, PhysRevAGhostimaging, PhysRevLettPlenopticImaging, Maga_a_Loaiza_2019, Genovese}. Unfortunately, these limitations render the realistic application of quantum imaging unfeasible \cite{ref1Sciencespatialcorrelations, Maga_a_Loaiza_2019}.

Sharing similarities with other quantum technologies, existing techniques for quantum imaging rely on the use of nonclassical states of light \cite{obrien2009, Lawrie2019SqueezedLight, Zhai:13,Defienne2021}. However, the brightness of available quantum light sources is generally low \cite{omar2019Multiphoton, Zhai:13, Hong2023}. For example, existing sources of nonclassical light allow for the preparation of few-photon states that exhibit fragile quantum correlations \cite{PhysRevLettThermalLight,smith2018, ref1Sciencespatialcorrelations}. This situation leads to common scenarios where environmental noise is typically larger than the signal of photons produced by processes of spontaneous parametric down-conversion or four-wave mixing \cite{anno2006, Zhai:13}. Unfortunately, it is not feasible to produce brighter quantum light sources to overcome these limitations. Moreover, losses and noise cannot be avoided in realistic scenarios \cite{Maga_a_Loaiza_2019}. Thus, it would be beneficial to use ubiquitous natural sources of light, such as thermal light for quantum imaging. 

In this chapter, quantum images are extracted from classically noisy images generated by thermal light sources \cite{You2023CittertZernike, Dakna,RimaACS2024,Mostafavi2025APR}. The approach relies on isolating multiphoton subsystems within thermal light fields to enhance the signal-to-noise ratio of an imaging system. This protocol is implemented through a single-pixel quantum camera with photon-number-resolving capabilities \cite{You2020Indentification}. This quantum camera enables the extraction of information from the vacuum-fluctuation components of thermal light sources to produce quantum images with improved contrast. This technique shows an exponential improvement in the contrast of quantum images. Additionally, it is demonstrated that correlated multiphoton subsystems can be used to form high-contrast images even when background noise is comparable to the thermal signal. These results can only be explained using quantum physics \cite{glauber1963, sudarshan1963}. The findings presented in this chapter highlight the potential of combining natural light with nonclassical detection techniques for the development of robust quantum technologies.

\section{Results and discussion}

\begin{figure}[htbp]
\centering	
\includegraphics[width=1\textwidth]{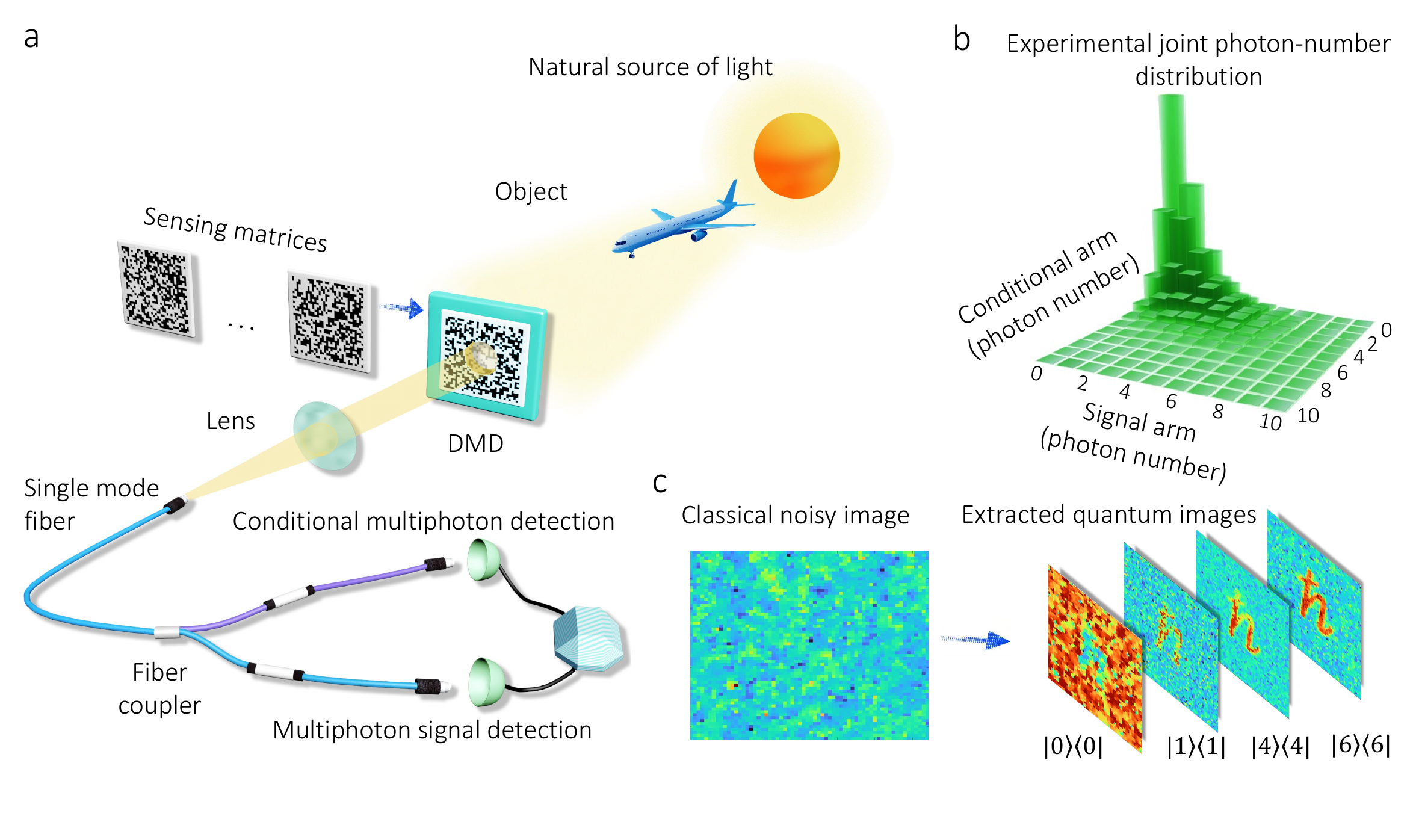}
	\caption[Multiphoton quantum imaging using natural sources of light]{\textbf{Multiphoton quantum imaging using natural sources of light.} 
The schematic in \textbf{a} depicts the implementation of a quantum camera with PNR capabilities. Here the thermal light field reflected off a target object is projected into a series of random binary matrices and then coupled into a single-mode fiber (SMF). The binary sensing matrices are displaced onto a digital micromirror device (DMD). Further, the thermal light field coupled into the SMF is split by a 50:50 fiber coupler and measured by two PNR detectors. We report the experimental joint photon-number distribution of our thermal light source in \textbf{b}. In this case, the degree of second-order coherence, $g^{(2)}(0)$, of the thermal light source is equal to 2. The series of PNR measurements for different binary sensing matrices enables us to use compressive sensing (CS) to demonstrate a single-pixel camera with PNR capabilities. As shown in \textbf{c}, our ability to measure the multiphoton subsystems, represented by the elements of the joint photon-number distribution of the thermal source, enables us to demonstrate quantum imaging even in situations in which noise prevents the formation of the classical image of the object. Specifically, the environmental noise in \textbf{c} forbids the imaging of the character $\hbar$. However, the projection of the thermal light field into its vacuum  component reveals the presence of the object. The projection into larger multiphoton subsystems enables the extraction of quantum images of the object that was not visible in the classical image.} 
	\label{Fig13}
 \end{figure}

In Fig. \ref{Fig13}\textbf{a} we illustrate the experimental implementation of our scheme for multiphoton quantum imaging. We generate thermal light by passing the coherent light from a continuous-wave laser at 633 nm through a rotating ground glass \cite{Arecchi65PRL,You2020Indentification}. The thermal light is then collected into a single-mode fiber and collimated to illuminate the target object. Then, the thermal light reflected off a target object is projected onto a digital micromirror device (DMD) where a series of random binary patterns are displayed. The thermal photons from the DMD are collected by a single-mode fiber (SMF) and then probabilistically split by a fiber coupler, with the photons in each arm measured by a PNR detector \cite{You2020Indentification, Bhusal2022}. The random sensing matrices displayed on the DMD are used to implement a single-pixel camera \cite{PRLMirhossein2014, tutorial, Montaut2018, PhysRevAGhostimaging}. Further, our photon counting scheme enables us to project the coupled thermal light field into its constituent multiphoton subsystems. The joint photon-number distribution of the thermal source is reported in Fig. \ref{Fig13}\textbf{b}. The classical nature of our thermal source is certified by the degree of second-order coherence $g^{(2)}$, which is equal to 2 in our experiment \cite{Gerry2004}. It is important to note that, for historical reasons, this kind of light is also referred to as natural light \cite{shih2020introduction}.
Each element in this joint probability distribution represents a multiphoton subsystem that we can isolate through the implementation of projective measurements \cite{PRLMirhossein2014, tutorial, Montaut2018, PhysRevAGhostimaging}. This measurement approach lies at the heart of our protocol for multiphoton quantum imaging. 
\begin{figure}[ht!]
\centering	
\includegraphics[width=1\textwidth]{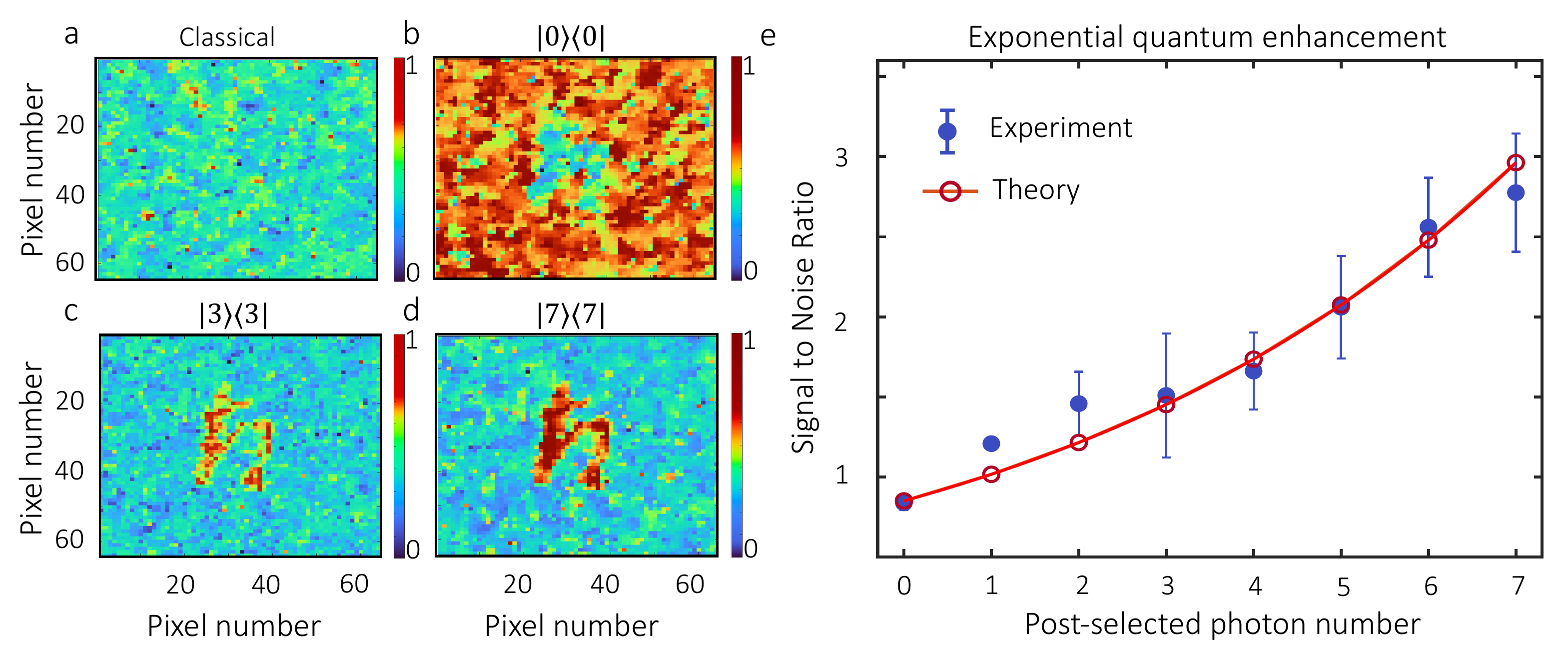}
\caption[Extraction of quantum images from a classical CS reconstruction]{\textbf{Extraction of quantum images from a classical CS reconstruction.} The reconstructed image using our single-pixel camera for classical thermal light is shown in \textbf{a}. In this case, environmental noise is higher than the signal and consequently the reconstructed image shows a low contrast that prevents the observation the object. The projection of the light field into its vacuum component boosts the contrast of the image, this is reported in \textbf{b}. Naturally, the formation of this image cannot be understood using classical optics. As demonstrated in \textbf{c}, the projection of the thermal source of light into three-photon events enables the extraction of a quantum image with an improved signal-to-noise ratio (SNR). The projection of the detected thermal field into seven-particle subsystems leads to the formation of the high-contrast quantum image in \textbf{d}. As reported in \textbf{e}, and in agreement with Eq. (\ref{Eq:post}), the improvement in the SNR is exponential with the number of projected photons.  These results were obtained using 25$\%$ of the total number of measurements that can be used in our CS algorithm, which took approximately 30 minutes. Furthermore, the mean photon number $\bar{n}_t$ of the thermal light source is 0.8.} 
\label{Fig14}
\end{figure}

As shown in Fig. \ref{Fig13}\textbf{c}, the projection of thermal light scattered by a target object into its constituent multiphoton subsystems enables the formation of high-contrast quantum images. This effect enables extracting quantum images of a target object, even when environmental noise prevents the formation of its classical image through intensity measurements. We now describe the multiphoton quantum processes that make this effect possible. For the sake of simplicity, we assume the uniform illumination of the object $\Vec{\boldsymbol{s}}_0$ by a thermal light field. As depicted in Fig. \ref{Fig13}\textbf{a}, 
the projection of the object into random sensing matrices, represented by the covector $\Vec{\boldsymbol{Q}}_t$, enables us to discretize the object into $X$ pixels. The label $t$ indexes the different sensing matrices. All such matrices can be represented by the $M\times X$ matrix  $\boldsymbol{Q} = \bigoplus_{t=1}^M \Vec{\boldsymbol{Q}}_t$, where $M$ is the number of sensing matrix configurations. Then, each filtering configuration results in a thermal state with a mean photon number given by $\bar{n}_t = \Vec{\boldsymbol{Q}}_t\cdot \Vec{\boldsymbol{s}}_0$. The multiphoton state after the fiber coupler can be written in terms of the Glauber-Sudarshan $P$ function as \cite{sudarshan1963,glauber1963}
\begin{equation}
\begin{aligned}
\label{Eq:state}
      \hat{\boldsymbol{\rho}}_{\boldsymbol{Q}} =&
      \bigoplus_{t=1}^M\int d^2\alpha \frac{1}{\pi\bar{n}_t}e^{-\frac{\left|\alpha\right|^2}{\bar{n}_t}}\\
      &\times\left|\alpha \cos(\theta),i\alpha\sin(\theta)\rangle\langle\alpha \cos(\theta),i\alpha\sin(\theta)\right|_{a,b}.
\end{aligned}
\end{equation}
The indices $a$ and $b$ denote the output modes of the fiber coupler. Furthermore, the parameter $\theta$ describes the splitting ratio between the two output ports. 

Next, we describe the signal-to-noise ratio (SNR) and how this quantity is modified by projecting the thermal field into its constituent multiphoton subsystems. To account for noise, we must consider photocounting with quantum efficiencies $\eta_{a/b}$ and noise counts $\nu_{a/b}$ \cite{omar2019Multiphoton,PhysRevATruephoton,PhysRevAmultimodethermal}. Specifically, the joint photon-number distribution reported in Fig. \ref{Fig13}\textbf{b}, can be mathematically described as
\begin{equation}
\begin{aligned}
\label{Eq:joint}
               \Vec{\boldsymbol{p}}_{\boldsymbol{Q}}(n, m)=&\bigoplus_{t=1}^M\left\langle: \frac{\left(\eta_a \hat{n}_a+\nu_a\right)^n}{n !} e^{-\left(\eta_a \hat{n}_a+\nu_a\right)} \otimes \frac{\left(\eta_b \hat{n}_b+\nu_b\right)^m}{m !} e^{-\left(\eta_b \hat{n}_b+\nu_b\right)}:\right\rangle\\
               =&\bigoplus_{t=1}^M\frac{e^{-\nu_a-\nu_b}}{\bar{n}_t n! m!}\sum_{i=0}^n\sum_{j=0}^m \binom{n}{i}\binom{m}{j}(i+j)!\\
               &\times\frac{\eta_a^i\eta_b^j\nu_a^{n-i}\nu_b^{m-j}}{\left(\frac{1}{\bar{n}_t}+\eta_a\cos^2(\theta)+\eta_b\sin^2(\theta)\right)^{1+i+j}}\cos^{2i}(\theta)\sin^{2j}(\theta),
\end{aligned}
\end{equation}  

\noindent
where $\hat{n}_{a/b}$ is the photon number operator, and $:\cdot:$ represents the normal ordering prescription. We write the $t^{\text{th}}$ component of this vector as $p_{\boldsymbol{Q},t}(n,m)$. Additionally, when there is no signal and only noise is measured, we will have the probability distribution $p_{n,i}(k) = e^{-\nu_i}\frac{\nu_i^k}{k!}$ in each arm, where $i = a,b$. The joint probability distribution in this case is then given by $p_n(k,l) = p_{n,a}(k)p_{n,b}(l)$.
 
The two-mode multiphoton system described by Eq. (\ref{Eq:joint}) enables two schemes for projective measurements that lead to different scaling factors for the SNR of quantum images. First, we project one of the arms into a particular multiphoton subsystem \cite{dawkins2024quantum}. In other words, we ignore arm $b$ and implement a photon-number-projective measurement in arm $a$. For such post-selection on a multiphoton subsystem with $N$ photons, the SNR scales with 
\begin{equation}
      \label{Eq:post}
      \overrightarrow{\textbf{SNR}}_{\text{post}} = \frac{\sum_{m=0}^\infty \Vec{\boldsymbol{p}}_{\boldsymbol{Q}}(N,m)}{p_{n,a}(N)}=\frac{\Vec{\boldsymbol{p}}_{\boldsymbol{Q}}(N)}{p_{n,a}(N)}.
\end{equation} 
This expression follows an exponentially increasing trend with respect to $N$, meaning that post-selection can significantly reduce the noise of a quantum image.

\begin{figure}[ht!]
\centering	
\includegraphics[width=1\textwidth]{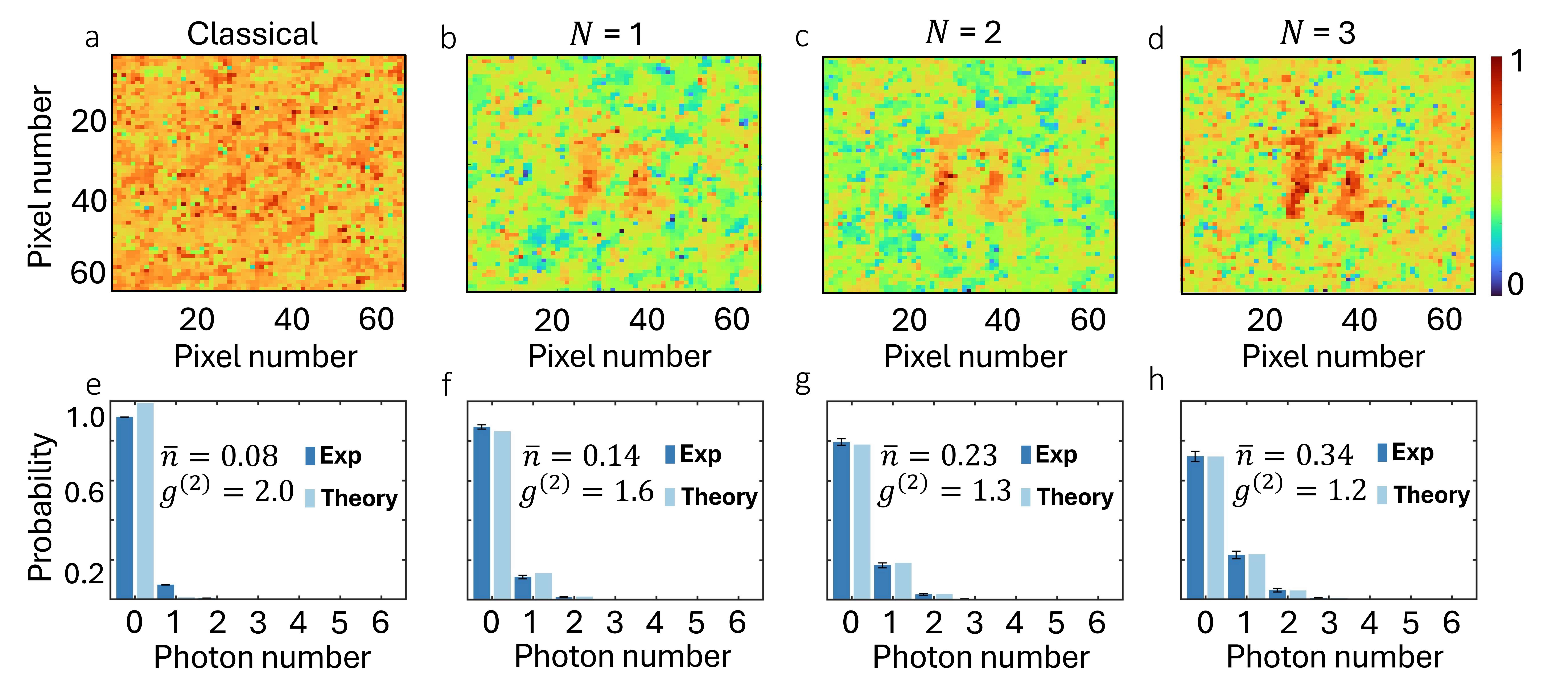}
\caption[Photon-subtracted multiphoton quantum imaging]{\textbf{Photon-subtracted multiphoton quantum imaging.} The noise accompanying a signal reflected off a target object produces the classical image reported in \textbf{a}. Here, it is not possible to identify the object of interest with a classical single-pixel camera \cite{APlMagana-Loaiza2013}. The mean photon number $\bar{n}_t$ of our thermal light source is $0.08$. Interestingly, the subtraction of one photon improves the contrast of the image leading to the CS reconstruction in \textbf{b}. Furthermore, our single-pixel camera with PNR capabilities enables multiphoton subtraction to produce the quantum images shown in  \textbf{c} and \textbf{d}. In these cases, we subtracted two and three photons, respectively. These images were produced using only 12$\%$ of the total number of measurements that can be used in our CS algorithm, which took approximately 15 minutes. The advantage provided by our protocol for photon-subtracted quantum imaging can be understood through the photon-number distributions reported from \textbf{e} to \textbf{h}. The unconditional detection of the weak thermal light signal produces the histogram in \textbf{e}. This histogram unveils the overwhelming presence of vacuum and single-photon events used to reconstruct the image in \textbf{a}. Furthermore, as shown in \textbf{f}, the process of one-photon subtraction increases the mean photon number of the thermal signal while reducing its degree of second-order coherence $g^{(2)}$. The subtraction of two-photon events leads to a stronger signal characterized by the histogram in \textbf{g}. This conditional signal produces the enhanced image of the object in \textbf{c}. Notably, the implementation of three-photon subtraction leads to the optical signal with nearly coherent statistics reported in \textbf{h}. This boosted signal enables the reconstruction of the high-contrast image in \textbf{d}. } 
\label{Fig15}
\end{figure}

The second approach relies on the subtraction of $N$ photons from the thermal multiphoton system in Eq. (\ref{Eq:joint}) \cite{Dakna, Mostafavi2022, OmarOptica17}. This procedure entails measuring photon events in arm $a$ conditioned on the detection of $N$ photons in arm $b$. Using Eq. (\ref{Eq:joint}), the intensity in arm $a$ is then given by $\langle\hat{\boldsymbol{n}}_a\rangle_N = \bigoplus_{t=0}^M\left(\sum_{k=0}^\infty k p_{\boldsymbol{Q},t}(k,N)\right)/\left(\sum_{k=0}^\infty p_{\boldsymbol{Q},t}(k,N)\right)$. Additionally, the photon-subtracted noise can be written as 
\begin{equation}
\langle\hat{n}_a\rangle_{N,0} = \bigoplus_{t=0}^M\left(\sum_{k=0}^\infty k p_n(k,N)\right)/\left(\sum_{k=0}^\infty p_n(k,N)\right).
\end{equation} 
This scheme leads to the following expression for the SNR:
\begin{equation}
 \label{Eq:sub}
    \overrightarrow{\textbf{SNR}}_{\text{sub}} = \frac{\langle\hat{\boldsymbol{n}}_a\rangle_N}{\langle\hat{n}_a\rangle_{N,0}}.
\end{equation}
The quantum enhancement for the SNR in this case is linearly increasing with respect to $N$. Therefore, photon-subtraction is also an effective means for noise-reduction.

The series of spatial projective measurements described by the vector $\Vec{\boldsymbol{Q}}_t$ enables implementing a single-pixel camera with photon-number resolving capabilities via compressive sensing (CS)\cite{PRLMirhossein2014, tutorial,Montaut2018, PhysRevAGhostimaging}. This technique permits the reconstruction of  multiphoton quantum images described by $\Vec{\boldsymbol{s}}'$ via the minimization of the following quantity with respect to $\Vec{\boldsymbol{s}}'$: 
\begin{equation}
\label{eq:cs}
     \sum_{i=0}^X \lVert\nabla s_i'\rVert_{l_1} + \frac{\mu}{2}\lVert \boldsymbol{Q}\Vec{\boldsymbol{s}}' - \langle\hat{\boldsymbol{n}}\rangle\rVert_{l_2}.
\end{equation}
As described above, $\langle\hat{\boldsymbol{n}}\rangle$ could be either $\Vec{\boldsymbol{p}}_{\boldsymbol{Q}}(N)$ or $\langle\hat{\boldsymbol{n}}_a\rangle_N$. Moreover, the $1$- and $2$-norm are denoted by $\lVert\cdot\rVert_{l_1}$ and $\lVert\cdot\rVert_{l_2}$, respectively. The discrete gradient operator is described by $\nabla$, and the penalty factor by $\mu$  \cite{tutorial, PRLMirhossein2014,Montaut2018, PhysRevAGhostimaging}. 

We now discuss the experimental process of quantum-image extraction from classical images. This was implemented using one PNR detector. In Fig. \ref{Fig14}\textbf{a}, we show the CS reconstruction of a classical image for a situation in which environmental noise is comparable to the signal. In this case, the level of noise forbids the observation of the object. The mean photon number $\bar{n}_{t}$ of the thermal light source is 0.8.  The projection of the thermal signal into its vacuum component reveals the presence of the object. As such, the quantum image in Fig. \ref{Fig14}\textbf{b} is formed by the vacuum-fluctuation component of the electromagnetic field and cannot be explained using classical physics \cite{Genovese,sudarshan1963,glauber1963, You2023CittertZernike}. This nonclassical reconstruction, obtained from the measurement of vacuum events, is determined by calculating the percentage of measurements that register zero photons. Consequently, it is expected that the vacuum signal will decrease in regions with higher mean photon numbers, and increase in regions with lower mean photon numbers. This reconstruction demonstrates that the process of projecting the thermal light signal into one of its constituent quantum subsystems, such as the vacuum, modifies the SNR as established by Eq. (\ref{Eq:post}). As suggested by the reconstruction in Fig. \ref{Fig14}\textbf{c}, the post-selection on higher multiphoton events leads to quantum images with an improved contrast. Interestingly, the projection of the thermal light signal into seven-photon subsystems leads to a dramatic improvement of the contrast of the image. This effect becomes evident in the quantum image shown in Fig. \ref{Fig14}\textbf{d}. The exponential growth of the SNR with the number of projected multiphoton subsystems is summarized in Fig. \ref{Fig14}\textbf{e}. These results demonstrate that our single-pixel camera with PNR capabilities enables the extraction of multiphoton quantum images with unprecedented degrees of contrast \cite{Brida2010NaturePhotonics,Bhusal2022, Morris2015NatureComm, Genovese, Maga_a_Loaiza_2019}. 

\begin{figure}[ht!]
\centering
\includegraphics[width=0.7\linewidth]{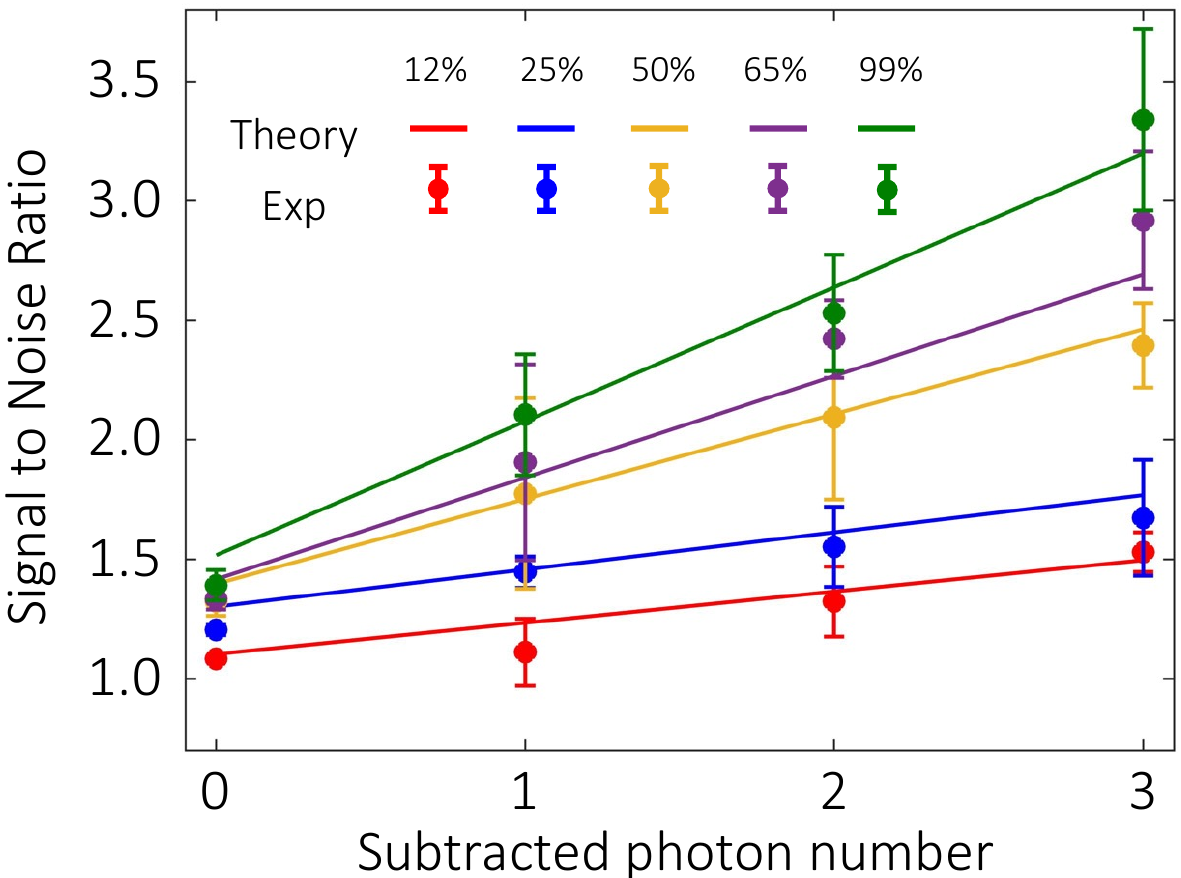}
\caption[Performance of photon-subtracted multiphoton quantum imaging]{\textbf{Performance of photon-subtracted multiphoton quantum imaging.} The SNR of the photon-subtracted quantum images shows a linear dependence on the number of subtracted photons. This behavior is in good agreement with Eq. (\ref{Eq:sub}). Interestingly, the collection of larger sets of data leads to faster improvements of the SNR for our multiphoton quantum imaging scheme. The percentages represent the number of CS measurements with respect to the total number of pixels in the image. The error bars represent the standard deviation of the SNR of image reconstructions using five different datasets, where each dataset contains millions of PNR measurements. } 
\label{Fig16}
\end{figure}

While the projection of thermal light into its constituent multiphoton subsystems enables the extraction of quantum images with high contrast, it is also possible to correlate photon events to improve the SNR of a quantum imaging protocol \cite{White2007PRL}. This feature also enables us to perform quantum imaging at low light levels. We now experimentally demonstrate this possibility by implementing a scheme for photon subtraction on our single-pixel camera with PNR capabilities. In this case, the mean photon number $\bar{n}_t$ is equal to 0.08, one order of magnitude lower than the brightness of the source used for the experiment discussed in Fig. \ref{Fig14}. As illustrated in Fig. \ref{Fig13}\textbf{a}, this quantum imaging scheme utilizes two PNR detectors \cite{omar2019Multiphoton,You2021APR}. First, we use the noisy thermal signal to reconstruct the classical image shown in Fig. \ref{Fig15}\textbf{a}. Here, the large levels of noise forbid the observation of the target object. The subtraction of one photon from the thermal noisy signal reveals the presence of the object in Fig. \ref{Fig15}\textbf{b}. As predicted by Eq. (\ref{Eq:sub}), the process of multiphoton subtraction leads to enhanced quantum images. Specifically, two-photon subtraction leads to the improved image in Fig. \ref{Fig15}\textbf{c}. Furthermore, the CS reconstruction of the three-photon subtracted quantum image reported in Fig. \ref{Fig15}\textbf{d} shows a significant  improvement of the contrast with respect to the classical image in Fig. \ref{Fig15}\textbf{a}. The physics behind our scheme for quantum imaging can be understood through the increasing mean photon number that characterizes the histograms shown from Fig. \ref{Fig15}\textbf{e} to \textbf{h}. Moreover, the thermal fluctuations of the detected field are reduced by subtracting photons \cite{Dakna, Maga_a_Loaiza_2019, OmarOptica17}. This effect is indicated by the decreasing degree of second-order coherence $g^{(2)}$ corresponding to the photon-number distributions in Fig. \ref{Fig15}. Finally, it should be noted that the photon-subtracted multiphoton quantum imaging scheme does not work with coherent light, since the subtraction of a photon does not alter its photon statistics \cite{photon-sub}.

The improvement in the SNR of the experimental photon-subtracted quantum images is quantified in Fig. \ref{Fig16}. In agreement with Eq. (\ref{Eq:sub}), the contrast of the filtered images, as a function of the number of subtracted photons, follows a linear dependence. Although, the benefits of our photon-subtracted scheme for multiphoton quantum imaging are evident for small and incomplete sets of data, the rate at which the SNR increases can be further amplified by collecting larger sets of data. It is worth noting that the exponential and linear mechanisms, reported in Fig. \ref{Fig14} and Fig. \ref{Fig16}, for improving the SNR of weak and noisy imaging signals have the potential to enable the realistic application of robust quantum cameras with PNR capabilities \cite{Bhusal2022,Cheng2023}. As such, these findings could lead to novel quantum techniques for multiphoton microscopy and remote sensing \cite{Morris2015NatureComm, Maga_a_Loaiza_2019, Genovese}.

Enhancing image contrast by combining classical light sources with nonclassical PNR detection offers distinct advantages over purely quantum approaches, which depend exclusively on quantum states of light and single-photon detection \cite{Maga_a_Loaiza_2019, Moreau2019, Defienne2024}. Our hybrid technique combines the strengths of both classical and quantum methods, achieving improved contrast while avoiding the limitations and demanding requirements of fully quantum systems \cite{OPN,WubsNews}. In recent years, quantum imaging has gained substantial interest for its potential to reduce noise, particularly through the use of correlated photon pairs generated via spontaneous parametric down-conversion \cite{TwoPhotonLithography, PhysRevLettDowling}. For instance, differential quantum imaging has increased the SNR of quantum images by over 30\% \cite{Brida2010NaturePhotonics}. Additionally, the unique features of N00N states have been used to achieve supersensitive imaging \cite{Ono2013, Israel2014}. Notably, Ono and colleagues demonstrated an SNR of 1.35$\pm$0.12 using an N=2 N00N state in entangled-enhanced microscopy \cite{Ono2013}. Subsequently, an SNR of approximately 1.5 was achieved with an N=3 state \cite{Israel2014}. By comparison, our hybrid scheme achieves even greater performance. As illustrated in Fig. \ref{Fig14}\textbf{e} and Fig. \ref{Fig16}, we achieved an SNR close to 3, highlighting the effectiveness of our approach.

The robustness of thermal light against losses enables us to outperform schemes based on N00N states, where the loss of a single photon is sufficient to eliminate all information from a quantum state \cite{ThomasPeter2011}. This vulnerability of quantum technologies that rely on N00N states was initially explored by Walmsley and colleagues and later experimentally verified in the context of quantum metrology \cite{You2021APR, ThomasPeter2011}. Furthermore, in contrast to imaging schemes that depend on N00N states, the sensitivity of our technique can be enhanced through the generation and detection of high-order multiphoton systems (see Fig. \ref{Fig14}\textbf{e} and Fig. \ref{Fig16}). This distinctive feature of our protocol makes it scalable. However, despite the robust characteristics of our imaging approach, it is important to note that its advantages become practical in scenarios where the imaging signal is weak or comparable to the background. Such conditions are often found in LIDAR applications \cite{Cohen2019}, where the performance of rangefinder systems is constrained by the ability to distinguish the detection signal from environmental noise. Remarkably, our imaging protocol demonstrates potential to address this challenge. Additionally, there is growing interest in using optical microscopy to identify light emitters \cite{Polino}. In this context, our technique can effectively form images of these optical emitters. While our approach can also be applied in scenarios with strong imaging signals, it is important to note that post-selection on large multiphoton systems could be easily implemented in these cases. However, it is worth to emphasizing that imaging under these conditions can be achieved without our specific protocol. In general, imaging a strong signal could, in principle, be accomplished using classical detection methods or classical imaging protocols.

\section{Summary}

In this chapter, a multiphoton quantum imaging scheme was developed to overcome the fragility of quantum imaging techniques in realistic environments where background noise is comparable to the nonclassical signal of imaging photons \cite{ref1Microscopy, PhysRevAGhostimaging, PhysRevLettPlenopticImaging, Maga_a_Loaiza_2019, Genovese, ref1Sciencespatialcorrelations}. This issue has traditionally limited the applicability of quantum imaging for microscopy, remote sensing, and astronomy \cite{Morris2015NatureComm, Maga_a_Loaiza_2019, Genovese}. The scheme relies on the use of natural thermal light sources and is implemented via a single-pixel camera with photon-number-resolving capabilities. This quantum detection strategy enables projection of thermal light fields into constituent multiphoton subsystems, allowing high-contrast quantum images to be extracted from classically noisy backgrounds.

Our technique demonstrates exponential enhancement in image contrast and reveals the formation of quantum images arising from the vacuum-fluctuation components of thermal light. In addition, correlated multiphoton subsystems are shown to generate high-contrast quantum images even under high-noise conditions. Altogether, the results presented in this chapter suggest a new paradigm in quantum imaging by demonstrating how structured quantum measurements applied to classical sources can yield robust performance in practical scenarios \cite{Maga_a_Loaiza_2019, Genovese, obrien2009, Lawrie2019SqueezedLight}. Furthermore, this approach highlights the potential of combining natural light sources with nonclassical detection schemes for the advancement of scalable quantum technologies \cite{Maga_a_Loaiza_2019, Genovese, obrien2009, Lawrie2019SqueezedLight, anno2006, FaccioQRC, YouPhotoniX, Zhang:24, Lollie2022, Bhusal2021a}.

\chapter{Conclusion}

This thesis has presented a comprehensive exploration of multiparticle quantum plasmonics, from foundational theory to experimental realization and practical applications. Through six chapters, we have established a framework in which plasmonic systems can be used to manipulate and probe quantum states of light at the nanoscale, offering new opportunities for both fundamental quantum science and emerging quantum technologies \cite{you2020multiparticle,tame2013quantum,chang2006quantum}.

In Chapter 2, we introduced the theoretical foundations of multiparticle quantum plasmonics, including quantum statistical states of light, coherence theory, and the Gaussian-Schell model \cite{glauber1963,sudarshan1963,Gerry2004,dawkins2024quantum}. These concepts provided the basis for interpreting the nontrivial quantum behaviors observed in later chapters.

Chapter 3 presented the first experimental observation of the modification of quantum statistics in plasmonic systems. This result challenged the long-held assumption that quantum fluctuations are always preserved in such platforms and demonstrated that multiparticle scattering in plasmonic near fields can reshape excitation modes \cite{You2021, safari2019measurement, lawrie2013}.

In Chapter 4, we investigated the nonclassical near-field dynamics of surface plasmons. By isolating multiparticle subsystems and performing projective measurements, we showed that the observed plasmonic coherence arises from underlying quantum scattering processes - some of which exhibit either bosonic or fermionic characteristics \cite{HongNP24, OPN, HasselbachNature, ButtikerHBTPRL}.

Chapter 5 extended these insights to practical applications, introducing a conditional plasmonic detection protocol that enables quantum-enhanced sensing in low-signal regimes. This sensing strategy improves signal-to-noise ratios and demonstrates how quantum statistical control can be applied to real-world plasmonic devices, even in the presence of loss \cite{You2021APR, lee2016quantum, dowran2018quantum, pooser2015plasmonic, Mostafavi2022}.

Chapter 6 broadened the scope of the thesis by applying the multiparticle framework to a different platform: quantum imaging using thermal light. We demonstrated that quantum correlations can be extracted from classical sources using photon-number-resolving detection, revealing that nonclassical imaging with effective noise cancellation is achievable even with natural light \cite{Maga_a_Loaiza_2019, Mostafavi2025APR, Bhusal2022}.

Together, these chapters show that quantum plasmonic systems are not only capable of preserving quantum properties but can also be engineered to modify, enhance, and exploit them \cite{you2020multiparticle, tame2013quantum}. The ability to access and manipulate multiphoton coherence at the nanoscale positions plasmonic platforms as promising candidates for integrated quantum technologies \cite{obrien2009, johnson2017diamond}. The results presented here in this thesis pave the way for future investigations into hybrid quantum systems, advanced quantum sensors, and scalable photonic architectures. As the field of quantum plasmonics continues to mature, the tools and techniques developed in this thesis will support efforts to bridge the gap between fundamental quantum optics and real-world quantum applications.

\appendix

\chapter{Supplementary information for Chapter 3 - Observation of the modification of quantum statistics of plasmonic systems}

In this appendix, we present the design and fabrication of the sample we used in the work reported in Chapter 3. In addition, the details of the experiment is provided.

\section{Sample Design}

Full-wave electromagnetic simulations were conducted using a Maxwell’s equation solver based on the finite difference time domain method (Lumerical FDTD). The dispersion of the materials composing the structure was taken into account by using their frequency-dependent permittivities. The permittivity of the gold film was obtained from ref. \cite{Johnson1972}, the permittivity of the glass substrate (BK7) was taken from the manufacturer’s specifications, and the permittivity of the index matching fluid was obtained by extrapolation from the manufacturer’s specification.

\begin{figure}[ht!]
  \centering
  \includegraphics[width=0.99\textwidth]{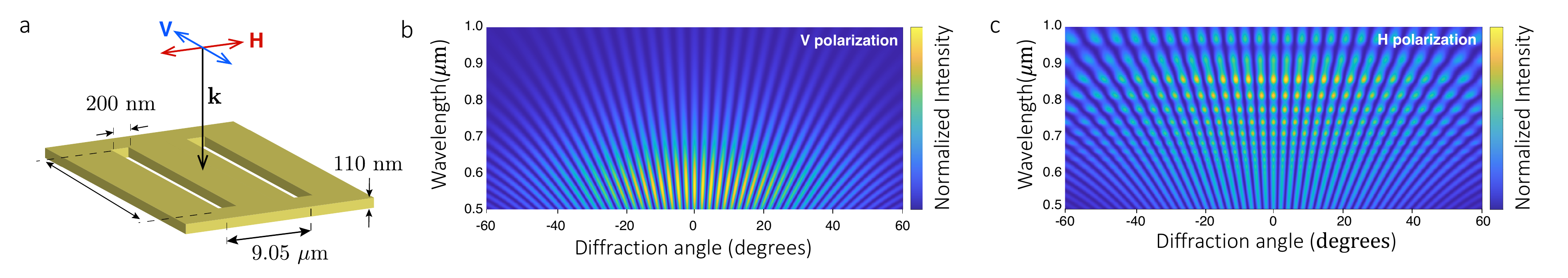}
  \caption[Design of plasmonic nanostructures]{\textbf{Design of plasmonic nanostructures.} The design of our plasmonic sample is shown in shown in \textbf{a}. The wavelength-dependent far-field interference pattern as a function of the diffraction angle is shown in \textbf{b}. In this case, the structure is illuminated with vertically-polarized photons and no plasmonic near-fields are excited. The figure in \textbf{c} shows a modified interference structure due to the presence of plasmonic near-fields. In this case, the illuminated photons are polarized along the horizontal direction.}
\label{Fig5}
\end{figure}

As shown in Fig. \ref{Fig5}, our nanostructure shows multiple plasmonic resonances at different wavelengths. This enables the observation of the multiphoton effects studied in this chapter at multiple wavelengths.

\section{Sample Fabrication}

The sample substrates are made from SCHOTT D 263 T eco Thin Glass with a thickness of $\sim$175 $\mu$m, polished on both sides to optical quality. The glass substrates were subsequently rinsed with acetone, methanol, isopropyl alcohol, and deionized water. The substrates were dried in nitrogen gas flow and heated in the clean oven for 15 minutes. Then we deposited 110-nm-thickness gold thin films directly onto the glass substrates using a Denton sputtering system with 200 W DC power, 5 mTorr argon plasma pressure, and 180 seconds pre-deposition conditioning time. The slit patterns were structured by Ga ion beam milling using a Quanta 3D FEG Dual beam system. The slit pattern consisted of 40 $\mu$m long slits with a separation of 9.05 $\mu$m. While fabricating the different slit sets, each slit is machined separately. To ensure the reproducibility, proper focusing of the FIB was checked by small test millings and if needed the FIB settings were readjusted accordingly to provide a consistent series of testing slits. 

\section{Experiment}

As shown in Fig. \ref{Fig1}\textbf{c}, we utilize a continuous-wave (CW) laser operating at a wavelength of 780 nm. We generate two independent sources with thermal statistics by dividing a beam with a 50:50 beam splitter. Then we focus the beams onto two different locations of a rotating ground-glass \cite{arrechi1967}. The two beams are then coupled into single-mode fibers to extract a single transverse mode with thermal statistics. We attenuate the two beams with neutral-density (ND) filters to tune their mean photon numbers. 
The polarization state of the two thermal beams is controlled with a pair of polarizers and half-wave plates. The two prepared beams are then combined using a 50:50 beam splitter. The combined beam is weakly focused onto the plasmonic structure that is mounted on a motorized three-axis translation stage. This enables us to displace the sample in small increments. Furthermore, we use an infinity-corrected oil-immersion microscope objectives (NA=1.4, magnification of 60X, working distance of 130 mm) to focus and collect light to and from the plasmonic structure. The light collected by the objective is then filtered using a 4f-imaging system to achieve specific particle number conditions for the photonic $n_{\text{s}}$ and plasmonic $n_{\text{pl}}$ modes. The experiment is formalized by coupling light into a polarization maintaining fiber that directs photons to a superconducting nanowire single-photon detector (SNSPD) that performs photon 
number resolving detection \cite{OmarOptica17,You2020Indentification}.

\chapter{Supplementary information for Chapter 4 - Nonclassical near-field dynamics of surface plasmons}

In this appendix, we provide (i) the derivation of equations in Chapter 4; (ii) the derivation of the diffraction envelope; (iii) the conditional detection of vacuum events; (iv) the properties of the degree of second-order coherence $\tilde{g}^{(2)}$; and (v) the spectral-spatial response of the plasmonic sample.

\section{Derivation of Equations in Chapter 4}

As described in the main body of Chapter 4, our experiment uses a thermal input beam, which illuminates one of the metallic slits \cite{dawkins2024quantum}. The horizontally polarized portion of the beam will excite surface plasmons, which will travel to the second slit where they will convert back into photons.  The initial state which illuminates the slits is given by
\begin{equation}
  \hat{\rho}_{\text{th}}=\sum_n \frac{\bar{n}^n}{\left(1+\bar{n}\right)^{n+1}}\ket{n,0}\bra{n,0}^{\text{ph-pl}} \equiv \int d^2\tau \frac{1}{\bar{n}\pi}e^{-\frac{|\tau|^2}{\bar{n}}}|\tau,0\rangle\langle \tau,0|^{\text{ph-pl}}, 
\end{equation}
where $\bar{n}$ is the mean photon number of the input light, and where in the last step we are using the Glauber P-function representation with $\tau$ as the coherent amplitude \cite{glauber1963}. Additionally, the upper index ``ph-pl'' serves as a label for the two modes, the photonic mode (light from the first slit) and the plasmonic mode (light from the second slit). It is important to note that the input light is linearly polarized, and that only light with polarization perpendicular to the slits will interact with our plasmonic sample. Since the input light can be decomposed into horizontally (H) and vertically (V) polarized components, we can describe the input beam mode using the annihilation operators $\hat{a}_{\text{H}}^{\text{ph}}$ and $\hat{a}_{\text{V}}^{\text{ph}}$. We can then describe the plasmonic double-slit interaction via a splitting of the horizontal mode. Here, we ignore any loss associated with the plasmonic interaction because it is not relevant to the subsequent derivation. Describing the photons at the second slit by the annihilation operator $\hat{a}_{\text{H}}^{\text{pl}}$, the plasmonic splitting effect can be written as $\hat{a}_{\text{H}}^{\text{ph}} \rightarrow \hat{a}_{\text{H}}^{\text{ph}} \cos (\psi)+\hat{a}_{\text{H}}^{\text{pl}} \sin (\psi)$ where $\psi$ determines the efficiency with which photons are converted to plasmons. The state of the system immediately after the excitation of surface plasmons can be approximated by 
\begin{equation}
\begin{aligned}
\hat{\rho}_{\text{ph-pl}}& = \int d^2\tau \frac{1}{(\bar{n}_\text{H}+\bar{n}_\text{V})\pi}e^{-\frac{|\tau|^2}{\bar{n}_\text{H} + \bar{n}_\text{V}}}|\tau\cos(\psi)\cos(\theta_{\text{pl}}),\tau\sin(\psi)\cos(\theta_{\text{pl}})\rangle \\
&\text{ }\text{ }\text{ }\text{ }\text{ }\text{ }\text{ }\text{ }\text{ }\text{ }\text{ }\text{ }\text{ }\text{ }\text{ }\text{ }\text{ }\text{ }\text{ }\text{ }\text{ }\text{ }\text{ }\text{ }\text{ }\text{ }\text{ }\text{ }\text{ }\text{ }\text{ }\text{ }\text{ }\text{ }\text{ }\text{ }\text{ }\langle \tau\cos(\psi)\cos(\theta_{\text{pl}}),\tau\sin(\psi)\cos(\theta_{\text{pl}})|^{\text{ph-pl}}_\text{H} \\
&\text{ }\text{ }\text{ }\text{ }\text{ }\text{ }\text{ }\text{ }\text{ }\text{ }\text{ }\text{ }\text{ }\text{ }\text{ }\text{ }\text{ }\text{ }\text{ }\text{ }\text{ }\text{ }\text{ }\text{ }\text{ }\text{ }\text{ }\text{ }\text{ }\text{ }\text{ }\text{ }\text{ }\text{ }\text{ }\text{ }\otimes |\tau \sin(\theta_{\text{pl}})\rangle\langle \tau \sin(\theta_{\text{pl}})|^{\text{ph}}_\text{V}\\
&\equiv \int d^2\tau P(\tau,\bar{n}_\text{H},\bar{n}_\text{V}) |\tau c_{\psi}c_{\theta_{\text{pl}}},\tau s_{\psi}c_{\theta_{\text{pl}}}\rangle\langle\tau c_{\psi}c_{\theta_{\text{pl}}},\tau s_{\psi}c_{\theta_{\text{pl}}}|^{\text{ph-pl}}_\text{H} \otimes |\tau s_{\theta_{\text{pl}}}\rangle\langle\tau s_{\theta_{\text{pl}}}|^{\text{ph}}_\text{V}
\end{aligned}
\end{equation}
where $\bar{n}_{\text{H}/\text{V}}$ denotes the mean photon number of the H/V polarization such that $\bar{n} = \bar{n}_{\text{H}}+\bar{n}_{\text{V}}$, and $\theta_{\text{pl}}$ is the polarization angle such that $\cos(\theta_{\text{pl}}) = \sqrt{\bar{n}_{\text{H}}/(\bar{n}_{\text{H}} + \bar{n}_{\text{V}})}$. In the last line, we are using the notations $c_x \equiv \cos(x)$, $s_x \equiv \sin(x)$, and $P(\tau,\bar{n}_\text{H},\bar{n}_\text{V}) \equiv \frac{1}{(\bar{n}_\text{H}+\bar{n}_\text{V})\pi}e^{-\frac{|\tau|^2}{\bar{n}_\text{H} + \bar{n}_\text{V}}}$. It should be noted that this is the common P-function for a thermal state with mean-photon-number given by $\bar{n} = \bar{n}_\text{H}+\bar{n}_\text{V}$.

As a first step, we would like to study the spatial intensity of this state in the far field. We assume that our source is approximately monochromatic, and for the state immediately after the slits ($z=0$), the positive frequency part of the electric field operator $\hat{E}^{(+)}(\vec{x})$ is given (proportionally) in terms of operator-valued distributions by $\hat{E}^{(+)}(\vec{x})=\hat{a}_{\text{H}}(\vec{x})\Vec{e}_{\text{H}}+\hat{a}_{\text{V}}(\vec{x})\Vec{e}_{\text{V}}$, where we have $\left[\hat{a}_{\text{H}}(\vec{x}), \hat{a}_{\text{H}}^{\dagger}\left(\vec{x}^{\prime}\right)\right]=\left[\hat{a}_{\text{V}}(\vec{x}), \hat{a}_{\text{V}}^{\dagger}\left(\vec{x}^{\prime}\right)\right]=\delta\left(\vec{x}-\vec{x}^{\prime}\right)$, $\left[\hat{a}_{\text{H}}(\vec{x}), \hat{a}_{\text{V}}^{\dagger}\left(\vec{x}^{\prime}\right)\right]=0$, and $\Vec{e}_{\text{i}}\cdot\Vec{e}_{\text{j}} = \delta_{ij}$. Furthermore, we can define an arbitrary mode of the field by the annihilation operator $\hat{a}_{f}$ as
\begin{equation}
    \hat{a}_{f}=\int d\Vec{x}\left[f_{\text{H}}(\vec{x}) \hat{a}_{\text{H}}(\vec{x})+f_{\text{V}}(\vec{x}) \hat{a}_{\text{V}}(\vec{x})\right].
\end{equation}
Here, we require that $\left|f_{\text{H}}(\vec{x})\right|^{2}+\left|f_{\text{V}}(\vec{x})\right|^{2}=1$, which ensures that $\left[\hat{a}_{f}, \hat{a}_{f}^{\dagger}\right]=1$. Because of this commutation relation, we can construct normalized Fock states using these mode operators. This allows us to describe the modes of $\hat{\rho}_{\text{ph-pl}}$ immediately after the double-slit as follows:
\begin{equation}
    \begin{aligned}
\hat{a}_{\text{H}}^{\text{ph}} & =\frac{1}{\sqrt{A}} \int d\Vec{x} \text{ rect}\left[\frac{x - d/2}{w}\right]\text{rect}\left[\frac{y}{h}\right] \varphi(x, y) \hat{a}_{\text{H}}(\vec{x}), \\
\hat{a}_{\text{H}}^{\text{pl}} & =\frac{1}{\sqrt{A}} \int d\Vec{x} \text{ rect}\left[\frac{x + d/2}{w}\right]\text{rect}\left[\frac{y}{h}\right] \varphi(x, y) \hat{a}_{\text{H}}(\vec{x}), \\
\hat{a}_{\text{V}}^{\text{ph}} & =\frac{1}{\sqrt{A}} \int d\Vec{x} \text{ rect}\left[\frac{x - d/2}{w}\right]\text{rect}\left[\frac{y}{h}\right] \varphi(x, y) \hat{a}_{\text{V}}(\vec{x}).
\end{aligned}
\end{equation}

\noindent We have defined here the distance between the slits as $d$, the width of each slit as $w$, the height of each slit as $h$ and $A=wh$. Furthermore, rect($\cdot$) represents the  rectangular function which is commonly used in Fourier optics. Here we set the first slit centered at $x=d/2$ and second slit centered at $x=-d/2$. The function $\varphi(x, y)$ represents the random phase imposed by the rotating ground glass, which satisfies $\varphi^{*}(x, y) \varphi(x, y)=1$. For ease of notation, we will write
\begin{equation}
    \begin{aligned}
& \int d\Vec{x} \text{ rect}\left[\frac{x - d/2}{w}\right]\text{rect}\left[\frac{y}{h}\right] \equiv \int_{S_{1}} d\Vec{x}, \\
& \int d\Vec{x} \text{ rect}\left[\frac{x + d/2}{w}\right]\text{rect}\left[\frac{y}{h}\right] \equiv \int_{S_{2}} d\Vec{x}.
\end{aligned}
\end{equation}

\noindent Here, $S_{1}$ represents the integration over the region of the first slit and $S_{2}$ represents the integration over the region of the second slit. Since $d\gg w$, we can see that $\left[\hat{a}_{\text{H}}^{\text{ph}}, \hat{a}_{\text{H}}^{\text{pl}\dagger}\right]=0$. The spatial intensity distribution in the far field should agree with that obtained from Fourier optics, so we can approximate the modes in the far field as
\begin{equation}
    \begin{aligned}
\hat{a}_{\text{H}}^{\text{ph}} & =\frac{1}{\sqrt{A}} \int d\Vec{k} \int_{S_{1}} d\Vec{x} e^{i \vec{x} \cdot \vec{k}} \varphi\left(x, y\right) \hat{a}_{\text{H}}(\vec{k}), \\
\hat{a}_{\text{H}}^{\text{pl}} & =\frac{1}{\sqrt{A}} \int d\Vec{k} \int_{S_{2}} d\Vec{x} e^{i \vec{x} \cdot \vec{k}} \varphi\left(x, y\right) \hat{a}_{\text{H}}(\vec{k}), \\
\hat{a}_{\text{V}}^{\text{ph}} & =\frac{1}{\sqrt{A}} \int d\Vec{k} \int_{S_{1}} d\Vec{x} e^{i \vec{x} \cdot \vec{k}} \varphi\left(x, y\right) \hat{a}_{\text{V}}(\vec{k}).
\end{aligned}
\end{equation}

\noindent We note that $\hat{a}_{\text{H}}^{\text{ph}}$ and $\hat{a}_{\text{H}}^{\text{pl}}$ will still commute in the far field due to the disjoint nature of $S_1$ and $S_2$. Since $w \ll d, h$ and since $\varphi(x, y)$ is approximately constant within the slit regions, we can write $\varphi\left(d/2, y\right)=\varphi_{1}(y)$ and $\varphi\left(-d/2, y\right)=\varphi_{2}(y)$. Under this approximation, the modes become
\begin{equation}
    \begin{aligned}
    \hat{a}_{\text{H}}^{\text{ph}}&=\frac{w}{\pi\sqrt{A}} \int d\Vec{k} e^{i \beta k_x} \operatorname{sinc}\left(\frac{k_x}{\alpha}\right) \int_{-\frac{h}{2}}^{\frac{h}{2}} d y e^{i y k_y} \varphi_{1}\left(y\right) \hat{a}_{\text{H}}(\vec{k}),\\
\hat{a}_{\text{H}}^{\text{pl}} & =\frac{w}{\pi \sqrt{A}} \int d\Vec{k} e^{-i \beta k_x} \operatorname{sinc}\left(\frac{k_x}{\alpha}\right) \int_{-\frac{h}{2}}^{\frac{h}{2}} d y e^{i y k_y} \varphi_{2}\left(y\right) \hat{a}_{\text{H}}(\vec{k}), \\
\hat{a}_{\text{V}}^{\text{ph}} & =\frac{w}{\pi \sqrt{A}} \int d\Vec{k} e^{i \beta k_x} \operatorname{sinc}\left(\frac{k_x}{\alpha}\right) \int_{-\frac{h}{2}}^{\frac{h}{2}} d y e^{i y k_y} \varphi_{1}\left(y\right) \hat{a}_{\text{V}}(\vec{k}).
\end{aligned}
\end{equation}
Here, we have defined $\beta=\frac{\pi d}{\lambda D}, \alpha=\frac{\lambda D}{\pi w}$ where $\lambda$ is the wavelength of our light and $D$ is the distance between the slits and the far field. This form will allow us to calculate the intensity distribution in the far-field. However, we should note that to make predictions with the random phase function $\varphi(x, y)$, we will need to extend our measurement theory. The standard measurement postulate for quantum electrodynamics tells us that the expectation value for an operator $\hat{O}$ can be predicted by computing $\langle\hat{O}\rangle=\operatorname{Tr}\left[\hat{\rho}_{\text{ph-pl}} \hat{O}\right]$, but this quantity is no longer necessarily intelligible because it can contain factors of the random phase $\varphi(x, y)$. To remedy this, we recall that the random phases $\varphi(x, y)$ obey Gaussian statistics, which means that $\varphi(x, y)$ is the element of some ensemble, indexed by a variable that we will call $\varepsilon$, such that $\left\langle\varphi(\vec{x}) \varphi^{*}\left(\vec{x}^{\prime}\right)\right\rangle_{\varepsilon}=e^{-\frac{|\vec{x}-\vec{x}|^{2}}{\sigma}}$ for some constant $\sigma$ \cite{magana2016hanbury}. Here, $\langle\cdot\rangle_{\varepsilon}$ represents the ensemble average over $\varepsilon$. 

 We can then combine the standard measurement postulate of quantum electrodynamics with the statistical average over $\varepsilon$ to produce a model for measurements as follows. Given an observable $\hat{O}$, a quantum state $\hat{\rho}$, and an ensemble of random phases $\varphi_{\varepsilon}(\vec{x})$ such that $\left\langle\varphi_{\varepsilon}(\vec{x}) \varphi_{\varepsilon}^{*}\left(\vec{x}^{\prime}\right)\right\rangle_{\varepsilon}=e^{-\frac{\left|\vec{x}-\vec{x}^{\prime}\right|^{2}}{\sigma}}$, we assert that the expectation value of $\hat{O}$ is given by $\langle\hat{O}\rangle=\langle\operatorname{Tr}[\hat{\rho} \hat{O}]\rangle_{\varepsilon}.$ Therefore, we can make predictions using the spatial characteristics of our system's modes. The mean photon number operator at $k_y=0$ can be written as
\begin{equation}
    \hat{n}(\vec{k})\equiv\hat{a}_{\text{H}}^{\dagger}(k_x) \hat{a}_{\text{H}}(k_x)+\hat{a}_{\text{V}}^{\dagger}(k_x) \hat{a}_{\text{V}}(k_x).
\end{equation}
Since we have the following commutation relations (where we have used the shorthands $\int_{-\frac{h}{2}}^{\frac{h}{2}} d y=\int d y$ and $k_x = k$)
\begin{equation}
    \begin{aligned}
{\left[\hat{a}_{\text{H}}(k), \hat{a}_{\text{H}}^{\text{ph}\dagger}\right] } & =\frac{w}{\pi \sqrt{A}} e^{-i \beta k} \operatorname{sinc}\left(\frac{k}{\alpha}\right) \int d y \varphi_{1}^{*}(y), \\
{\left[\hat{a}_{\text{H}}(k), \hat{a}_{\text{H}}^{\text{pl}\dagger}\right] } & =\frac{w}{\pi \sqrt{A}} e^{i \beta k} \operatorname{sinc}\left(\frac{k}{\alpha}\right) \int d y \varphi_{2}^{*}(y), \\
{\left[\hat{a}_{\text{V}}(k), \hat{a}_{\text{V}}^{\text{ph}\dagger}\right] } & =\frac{w}{\pi \sqrt{A}} e^{-i \beta k} \operatorname{sinc}\left(\frac{k}{\alpha}\right) \int d y \varphi_{1}^{*}(y),
\end{aligned}
\end{equation}
we can compute that
\begin{equation}
    \begin{aligned}
        \label{Eq:mean}
         \langle\hat{n}(k)\rangle =&  \int d^2\tau \frac{1}{\bar{n}_\text{H}\pi}e^{-\frac{|\tau|^2}{\bar{n}_\text{H}}}|\tau|^2 \frac{w^{2}}{\pi^{2} A} \operatorname{sinc}^{2}\left(\frac{k}{\alpha}\right)\bigg[ \cos^2(\psi) \int d y \int d y^{\prime}\left\langle\varphi_{1}^{*}(y) \varphi_{1}\left(y^{\prime}\right)\right\rangle_{\varepsilon} \\
        & + \sin^2(\psi)\int d y \int d y^{\prime}\left\langle\varphi_{2}^{*}(y) \varphi_{2}\left(y^{\prime}\right)\right\rangle_{\varepsilon}\\
        & + e^{-2i\beta k}\cos(\psi)\sin(\psi) \int d y \int d y^{\prime}\left\langle\varphi_{1}^{*}(y) \varphi_{2}\left(y^{\prime}\right)\right\rangle_{\varepsilon} \\
        & + e^{2i\beta k}\cos(\psi)\sin(\psi)\int d y \int d y^{\prime}\left\langle\varphi_{2}^{*}(y) \varphi_{1}\left(y^{\prime}\right)\right\rangle_{\varepsilon}\bigg]\\
        & +\int d^2\tau \frac{1}{\bar{n}_\text{V}\pi}e^{-\frac{|\tau|^2}{\bar{n}_\text{V}}}|\tau|^2\left(\frac{w^{2}}{\pi^{2} A} \operatorname{sinc}^{2}\left(\frac{k}{\alpha}\right) \int d y \int d y^{\prime}\left\langle\varphi_{1}^{*}(y) \varphi_{1}\left(y^{\prime}\right)\right\rangle_{\varepsilon}\right) .
    \end{aligned}
\end{equation}
\noindent From Eq. (\ref{Eq:mean}), we note that the value of $\int d y \int d k^{\prime}\left\langle\varphi_{1}^{*}(y) \varphi_{1}\left(y^{\prime}\right)\right\rangle_{\varepsilon}$ will be larger than $\int d y \int d y^{\prime}\left\langle\varphi_{1}^{*}(y) \varphi_{2}\left(y^{\prime}\right)\right\rangle_{\varepsilon}$, therefore we rescale the intensity by the first value (effectively defining it as 1) and then write the second value as $\gamma$. We also note that $\int d^2\tau \frac{1}{\bar{n}_\text{i}\pi}e^{-\frac{|\tau|^2}{\bar{n}_\text{i}}}|\tau|^2 = \bar{n}_\text{i}$. The mean photon number can then be re-written as
\begin{equation}
    \begin{aligned}
\langle\hat{n}(k)\rangle = &   \frac{w^{2}}{\pi^{2} A} \operatorname{sinc}^{2}\left(\frac{k}{\alpha}\right) \Big(\bar{n}_{\text{V}} + \bar{n}_{\text{H}}\big[1 + \gamma\sin(2\psi)\cos(2\beta k)\big]\Big).
\end{aligned}
\end{equation}
 
\noindent We can see that the far-field intensity pattern exhibits the familiar $\operatorname{sinc}^{2}(k/\alpha)$ envelope. Furthermore, the interference between the $\text{H}$-polarization parts of the photonic mode and the plasmonic mode will yield the $\cos(2\beta k)$ modulation. Furthermore, the excitation strength of the surface plasmons, which is quantified by $\sin(2\psi)$, modifies the visibility of the interference pattern. These features are in agreement with the double-slit experiment \cite{shih2020introduction}.

The second-order correlation operator 
\begin{equation}
\hat{G}^{(2)}\left(k_{1}, k_{2}\right) \equiv \hat{E}^{(-)}(k_1)\hat{E}^{(-)}(k_2)\hat{E}^{(+)}(k_2)\hat{E}^{(+)}(k_1)
\end{equation}
can be expanded, yielding
\begin{equation}
\begin{aligned}
    \hat{G}^{(2)}\left(k_{1}, k_{2}\right)=&\text{ }\hat{a}_{\text{H}}^{\dagger}\left(k_{1}\right) \hat{a}_{\text{H}}^{\dagger}\left(k_{2}\right) \hat{a}_{\text{H}}\left(k_{2}\right) \hat{a}_{\text{H}}\left(k_{1}\right)+\hat{a}_{\text{V}}^{\dagger}\left(k_{1}\right) \hat{a}_{\text{V}}^{\dagger}\left(k_{2}\right) \hat{a}_{\text{V}}\left(k_{2}\right) \hat{a}_{\text{V}}\left(k_{1}\right)\\
    &+\hat{a}_{\text{H}}^{\dagger}\left(k_{1}\right) \hat{a}_{\text{V}}^{\dagger}\left(k_{2}\right) \hat{a}_{\text{V}}\left(k_{2}\right) \hat{a}_{\text{H}}\left(k_{1}\right)+\hat{a}_{\text{V}}^{\dagger}\left(k_{1}\right) \hat{a}_{\text{H}}^{\dagger}\left(k_{2}\right) \hat{a}_{\text{H}}\left(k_{2}\right) \hat{a}_{\text{V}}\left(k_{1}\right).
\end{aligned}
\end{equation}
The expectation value of this operator is very complicated. However, given an integral over some combination of $\varphi_{j}(y)$ functions that does not reach 1 in the integration region, the value of that integral will be negligible. Therefore, only two types of integral will contribute to $\langle\hat{G}^{(2)}\left(k_{1}, k_{2}\right)\rangle$, and they are 
\begin{equation}
    \begin{aligned}
    &\int d y \int d y^{\prime} \int d y^{\prime \prime} \int d y^{\prime \prime \prime}\left\langle\varphi_{j}^{*}(y) \varphi_{j}^{*}\left(y^{\prime}\right) \varphi_{j}\left(y^{\prime \prime}\right) \varphi_{j}\left(y^{\prime \prime \prime}\right)\right\rangle_{\varepsilon},\\
&\int d y \int d y^{\prime} \int d y^{\prime \prime} \int d y^{\prime \prime \prime}\left\langle\varphi_{1}^{*}(y) \varphi_{2}^{*}\left(y^{\prime}\right) \varphi_{1}\left(y^{\prime \prime}\right) \varphi_{2}\left(y^{\prime \prime \prime}\right)\right\rangle_{\varepsilon}. 
\end{aligned}
\end{equation}

\noindent Once again, we normalize the second-order correlation function by taking the first integral to be 1. To compute the second integral, we notice that the integrand attains the value 1 half as many times as the first integral. Therefore, we can approximate its value as $1/2$. The second-order correlation $\langle\hat{G}^{(2)}\left(k_{1}, k_{2}\right)\rangle$ can then be computed as
\begin{equation}
    \begin{aligned}
        \left\langle\hat{G}^{(2)}\left(k_{1}, k_{2}\right)\right\rangle =& \frac{w^{4}}{\pi^{4} A^{2}} \operatorname{sinc}^{2}\left(\frac{k_{1}}{\alpha}\right) \operatorname{sinc}^{2}\left(\frac{k_{2}}{\alpha}\right)\bigg(\bigg[1 - \frac{1}{2}\sin^2(2\psi)\sin^2(\beta[k_1-k_2])\bigg]\cdot 2\bar{n}^2_{\text{H}} \\
       &+ \bigg[1 - \frac{1}{2}\sin^2(2\psi)\bigg]\cdot4\bar{n}_{\text{H}}\bar{n}_{\text{V}} + 2\bar{n}^2_{\text{V}}\bigg).
    \end{aligned}
\end{equation}
\noindent With the second-order correlations $\langle\hat{G}^{(2)}\left(k_{1}, k_{2}\right)\rangle$, as well as the mean photon number, we can calculate the second-order coherence function $g^{(2)}\left(k_1, k_2\right)=\langle \hat{G}^{(2)}(k_1,k_2)\rangle /\langle\hat{n}(k_1)\rangle\langle\hat{n}(k_2)\rangle$. However, we note that one should use the same approximation for calculations of the mean photon number by considering $\gamma \approx 0$. This $g^{(2)}\left(k_{1}, k_{2}\right)$ is then able to predict the fringes observed in our experiment. 

Next, we will extend this theory to make predictions for post-selective measurements. This is accomplished by using an equivalent expression for the state $\hat{\rho}_{\text{pl-ph}}$ which describes the photon statistics seen by the detectors. For the single-detector case, our model predicts a mean photon number of $\langle\hat{n}(k)\rangle$ at position $k$. Furthermore, the auto-correlation function $g^{(2)}(k, k)=2$ at every point $k$. A natural density matrix for describing the detectors' photon statistics is therefore given by:
\begin{equation}
    \hat{\rho}_{\text{ph-pl}}\left(k_1, k_2\right)=\hat{\rho}_{a} \otimes \hat{\rho}_{b}+\left(1 - \zeta \sin^2(\beta(k_1-k_2))\right)\left(\hat{\rho}_{J}-\hat{\rho}_{a} \otimes \hat{\rho}_{b}\right),
\end{equation}
\noindent where $\zeta$ is a parameter which depends on $\bar{n}_{\text{V}}$, the efficiency with which photons are converted to plasmons, and the QGS model. Also,  
\begin{equation}
    \hat{\rho}_{J}=\int d^2\tau \frac{1}{\bar{n}_{\text{H}}\pi}e^{-\frac{|\tau|^2}{\bar{n}_{\text{H}}}}|\tau\cos(\theta),\tau\sin(\theta)\rangle\langle \tau\cos(\theta),\tau\sin(\theta)|
\end{equation}
\noindent is a joint photon distribution between the two detectors whose P-function possess the same algebraic form as that of Eq. (S2), but where $\bar{n}$ and $\theta$ are in this case defined by $\bar{n}\cos^2(\theta) = \langle\hat{n}(k_1)\rangle$ and $\bar{n}\sin^2(\theta) = \langle\hat{n}(k_2)\rangle$. Also, $\hat{\rho}_{a}=\operatorname{Tr}_{b}\left[\hat{\rho}_{J}\right], \hat{\rho}_{b}=\operatorname{Tr}_{a}\left[\hat{\rho}_{J}\right]$ ($a, b$ are the modes of the detectors at $k_{1}, k_{2}$ respectively). Essentially, our model causes the photon distributions at points in phase with each other to be a joint distribution, while points that are out of phase with each other are weighted more towards a disjoint distribution. We can then calculate the post-selected $\tilde{g}^{(2)}\left(x_{1}, x_{2}\right)$ using the operators $\hat{n}_{an_1}\equiv |n_1\rangle\langle n_1|_a, \hat{n}_{bn_2}\equiv |n_2\rangle\langle n_2|_b$ to be:
\begin{equation}
\begin{aligned}
\tilde{g}^{(2)}\left(k_{1},k_{2}\right)&=\operatorname{Tr}\left[\hat{\rho}_{\text{ph-pl}}\left(k_{1}, k_{2}\right) \hat{n}_{a n_1} \hat{n}_{b n_2}\right]/\left(\operatorname{Tr}\left[\hat{\rho}_{\text{ph-pl}}\left(k_{1}, k_{2}\right) \hat{n}_{a n_1}\right]\operatorname{Tr}\left[\hat{\rho}_{\text{ph-pl}}\left(k_{1}, k_{2}\right) \hat{n}_{b n_2}\right]\right) \\
& =1+\left(1 - \zeta \sin^2(\beta(k_1-k_2))\right)\times \\
&\text{ }\text{ }\text{ }\text{ }\text{ }\text{ }\text{ }\text{ }\left(\left(\begin{array}{c}
n_1+n_2 \\
n_1
\end{array}\right)\left[\frac{\left(1+\cos ^{2}(\theta) \bar{n}\right)^{n_1+1}\left(1+\sin ^{2}(\theta) \bar{n}\right)^{n_2+1}}{(1+\bar{n})^{n_1+n_2+1}}\right]-1\right)\\
&=  1 + \left(1 - \zeta \sin^2(\beta(k_1-k_2))\right)\Big[\tilde{g}_{\text{th}}^{(2)}(n_1,n_2) - 1\Big],
\end{aligned}
\end{equation}
where here $\tilde{g}_{\text{th}}^{(2)}(n_1,n_2)$ represents the multiparticle coherence function for the joint thermal distribution $\hat{\rho}_J$ (see details in subsequent section). Intuitively, one can think of $\tilde{g}_{\text{th}}^{(2)}(n_1,n_2) - 1$ as the coherence term containing information about the photon number statistics, and $1 - \zeta \sin^2(\beta(k_1-k_2))$ as the coherence term containing information about the double slit structure. This expression, as well as this density matrix's predictions for $\operatorname{Tr}\left[\hat{\rho}_{\text{ph-pl}}\left(k_{1}, k_{2}\right) \hat{n}_{a N}\right]$, are in fairly good agreement with the observations made in our experiment. However, observations of $\tilde{g}^{(2)}\left(k_{1},k_{2}\right)$ do not simply show the predicted $\sin^2\left(\beta\left(k_{1}-k_{2}\right)\right)$ modulation. There is also a $\sinc^2\left(\frac{\left(k_{1}-k_{2} + k'\right)}{\sigma}\right)$ modulation present for some coefficients $k'$ and $\sigma$ which depend on $n_1,n_2$. Our quantum theory is not able to explain this modulation, but this is a common issue which has arisen in other experiments \cite{mehringer2018photon,magana2016hanbury,scarcelli2004two,liu2009n}. The commonplace solution to this conundrum is to impose the observed modulation so that the remaining qualities of the data can be fitted to the predicted $\tilde{g}^{(2)}\left(k_{1},k_{2}\right)$, so the actual expression which we will use for fitting is $\sinc^2\left(\frac{\left(k_{1}-k_{2} + k'\right)}{\sigma}\right)\tilde{g}^{(2)}\left(k_{1},k_{2}\right)$. One interesting thing to note is that, when this modulation is imposed classically, it is typically done with $k' = 0$. However, the $n_1,n_2$ dependence of $k'$ in the post-selected case is an interesting and uniquely quantum behavior. In the next section, we will provide a numerically-validated derivation for this envelope in the case of the classical $g^{(2)}(k_1,k_2)$. In doing so, we are able to provide a strong motivation for the imposed envelope in our expression for $\tilde{g}^{(2)}(k_1,k_2)$ derived from the above theory.

\section{Derivation of the Diffraction Envelope}

In the previous section, we were able to use a quantum mechanical description for our source after the plasmonic interaction to deduce an approximation for its multiphoton wavepacket statistics in the far-field. However, this approach was unable to explain the observed envelope in either the intensity correlations or the multiparticle coherence. In this section, we will present a derivation of the $g^{(2)}(k_1,k_2)$ function which does exhibit the observed envelope. Consequently, we will gain some insight into why the envelope is lost upon quantization, and this will justify our choice to use the this modulation in our quantum approach.

We can describe thermal light $E^{(+)}_{\text{th}}(\Vec{x})$ as a light source whose intensity correlations $\braket{E^{(-)}_{\text{th}}(\Vec{x}_1)E^{(+)}_{\text{th}}(\Vec{x}_2)}$ follow a Gaussian profile. Specifically,
\begin{equation}
    \braket{E^{(-)}_{\text{th}}(\Vec{x}_1)E^{(+)}_{\text{th}}(\Vec{x}_2)} = e^{-\frac{|\Vec{x_1}-\Vec{x}_2|^2}{s}}
\end{equation}
for some $s > 0$. Furthermore, we assume that the complex quantities $E^{(+)}_{\text{th}}(\Vec{x})$ obey the complex-Gaussian moment theorem. Importantly, this theorem implies that
\begin{equation}
\begin{aligned}
    \braket{E^{(-)}_{\text{th}}(\Vec{x}_1)E^{(-)}_{\text{th}}(\Vec{x}_2)E^{(+)}_{\text{th}}(\Vec{x}_2)E^{(+)}_{\text{th}}(\Vec{x}_1)} =& \braket{E^{(-)}_{\text{th}}(\Vec{x}_1)E^{(+)}_{\text{th}}(\Vec{x}_1)}\braket{E^{(-)}_{\text{th}}(\Vec{x}_2)E^{(+)}_{\text{th}}(\Vec{x}_2)}\\& + \braket{E^{(-)}_{\text{th}}(\Vec{x}_1)E^{(+)}_{\text{th}}(\Vec{x}_2)}\braket{E^{(-)}_{\text{th}}(\Vec{x}_2)E^{(+)}_{\text{th}}(\Vec{x}_1)}.
\end{aligned}    
\end{equation}
The electric field describing the thermal input to the system with polarization angle $\theta_{\text{pl}}$ is then given by
\begin{equation}
    E^{(+)}_0(\Vec{x}) = E^{(+)}_{\text{th}}(\Vec{x})\Vec{e}_{\text{H}}\cos(\theta_{\text{pl}}) + E^{(+)}_{\text{th}}(\Vec{x})\Vec{e}_{\text{V}}\sin(\theta_{\text{pl}}),
\end{equation}
where $\Vec{e}_{\text{H/V}}$ are the H/V polarization vectors. The metallic slit will impose a rectangular mask onto this field of the form $\text{rect}\left[\frac{x - d/2}{w}\right]\text{rect}\left[\frac{y}{h}\right]$ where $d$ is the distance between the two plasmonic slits, $w$ is the width of each slit, and $h$ is the height of each slit. Furthermore, the plasmonic scattering will cause part of the H-polarized component of $E^{(+)}_0(\Vec{x})$ to relocate to the second slit. As such, the total field after the plasmonic interaction is given by
\begin{equation}
    E^{(+)}(\Vec{x}) = E^{(+)}_{\text{ph}}(\Vec{x}) + E^{(+)}_{\text{pl}}(\Vec{x})
\end{equation}
where
\begin{equation}
    \begin{aligned}
        E^{(+)}_{\text{ph}}(\Vec{x}) &= \text{rect}\left[\frac{x - d/2}{w}\right]\text{rect}\left[\frac{y}{h}\right]\Big(E^{(+)}_{\text{th}}(\Vec{x})\Vec{e}_{\text{H}}\cos(\theta_{\text{pl}})\cos(\psi) + E^{(+)}_{\text{th}}(\Vec{x})\Vec{e}_{\text{V}}\sin(\theta_{\text{pl}})\Big),\\
        E^{(+)}_{\text{pl}}(\Vec{x}) &= \text{rect}\left[\frac{x + d/2}{w}\right]\text{rect}\left[\frac{y}{h}\right]E^{(+)}_{\text{th}}(\Vec{x})\Vec{e}_{\text{H}}\cos(\theta_{\text{pl}})\sin(\psi).
    \end{aligned}
\end{equation}
Here, $\psi$ is the plasmonic splitting angle. In the far-field, we can use Fourier optics to obtain that
\begin{equation}
    E^{(+)}(\Vec{k}) = \int d\Vec{x} e^{i\Vec{x}\cdot\Vec{k}}E^{(+)}(\Vec{x}).
\end{equation}
Then, using the same shorthand notations in Eq. (S5), we can compute the far-field correlation term as
\begin{equation}
    \begin{aligned}
        \braket{E^{(-)}(\Vec{k}_1)E^{(+)}(\Vec{k}_2)} =& \int_{S_1}d\Vec{x}\int_{S_1}d\Vec{x}' e^{i\Vec{k_1}\cdot(\Vec{x}-\Vec{x}')}\braket{E_{\text{th}}^{(-)}(\Vec{x})E_{\text{th}}^{(+)}(\Vec{x}')}(\cos^2(\theta_{\text{pl}})\cos^2(\psi) + \sin^2(\theta_{\text{pl}}))\\
        &+ \int_{S_1}d\Vec{x}\int_{S_2}d\Vec{x}' e^{i\Vec{k_1}\cdot(\Vec{x}-\Vec{x}')}\braket{E_{\text{th}}^{(-)}(\Vec{x})E_{\text{th}}^{(+)}(\Vec{x}')}(\cos^2(\theta_{\text{pl}})\cos(\psi)\sin(\psi))\\
        &+ \int_{S_2}d\Vec{x}\int_{S_1}d\Vec{x}' e^{i\Vec{k_1}\cdot(\Vec{x}-\Vec{x}')}\braket{E_{\text{th}}^{(-)}(\Vec{x})E_{\text{th}}^{(+)}(\Vec{x}')}(\cos^2(\theta_{\text{pl}})\cos(\psi)\sin(\psi))\\
        &+ \int_{S_2}d\Vec{x}\int_{S_2}d\Vec{x}' e^{i\Vec{k_1}\cdot(\Vec{x}-\Vec{x}')}\braket{E_{\text{th}}^{(-)}(\Vec{x})E_{\text{th}}^{(+)}(\Vec{x}')}(\cos^2(\theta_{\text{pl}})\sin^2(\psi)).
    \end{aligned}
\end{equation}
The coherence length $s$ is assumed to be small enough that the second and third integrals will be approximately $0$, leaving only the first and the last. Furthermore, we only wish to study the case where $k_{1y} = k_{2y} = 0$, and so we will use shorthands $k_{1x}\equiv k_1, k_{2x}\equiv k_2$. Our goal is then to evaluate the quantity
\begin{equation}
    q_j(k_1,k_2) = \int_{S_j}d\Vec{x}\int_{S_j}d\Vec{x}'e^{ik\cdot(x-x')}\braket{E_{\text{th}}^{(-)}(x)E_{\text{th}}^{(+)}(x)} = \int_{S_j}d\Vec{x}\int_{S_j}d\Vec{x}'e^{ik\cdot(x-x')} e^{-\frac{|x-x'|^2}{\sigma}},
\end{equation}
such that the correlation term becomes
\begin{equation}
    \braket{E^{(-)}(k_1)E^{(+)}(k_2)}\propto q_1(k_1,k_2)(\cos^2(\theta_{\text{pl}})\cos^2(\psi) + \sin^2(\theta_{\text{pl}})) + q_2(k_1,k_2)(\cos^2(\theta_{\text{pl}})\sin^2(\psi)).
\end{equation}
Here, we numerically evaluate the $q_j(k_1,k_2)$ terms, and we are able to show that the resulting intensity $\braket{E^{(-)}(k)E^{(+)}(k)}$ is in agreement with our quantum prediction for $\braket{\hat{n}(k)}$ derived in the previous section. Furthermore, we can use the Gaussian moment theorem to obtain the second-order correlation function via
\begin{equation}
    g^{(2)}(k_1,k_2) = 1 + \frac{\braket{E^{(-)}(k_1)E^{(+)}(k_2)}\braket{E^{(-)}(k_2)E^{(+)}(k_1)}}{\braket{E^{(-)}(k_1)E^{(+)}(k_1)}\braket{E^{(-)}(k_2)E^{(+)}(k_2)}}.
\end{equation}
Numerical evaluation of this quantity indeed produces the same sinusoidal modulation which was predicted with the quantum model, but notably it is also able to predict the presence of the envelope function, which can be approximated by a function of the form $\sinc^2((k_1-k_2)/\sigma)$ without loss of generality. 

The key difference between this model and our approach in Chapter 4 is the coupling of the random phase with the random intensity. Here, they are grouped together into the same variable $E_{\text{th}}^{(+)}(\Vec{x})$ which both satisfies the Gaussian moment theorem and possesses a Gaussian correlation profile. The intensity fluctuations are then responsible for the $g^{(2)}$ attaining its maximum value of $2$, and the phase fluctuations are responsible for the sinusoidal modulation. It is thus the coupling between these two random quantities which allows for the envelope. However, quantizing the system requires that the random intensity fluctuations occur within the confines of a Hilbert space, while the random phase fluctuations remain a part of some classical ensemble, since they are only an aspect of the source's mode structure. Therefore, such a quantum description of the model necessitates a decoupling of the random intensity from the random phase, and this is why the envelope function does not directly shows up in the quantum theory. As such, we use an envelope in the form of a $\sinc^2((k_1-k_2)/\sigma)$ function in order to properly map this theoretical approach to our experiment. In the case of multiparticle coherence, there is sometimes an observed shifting in the envelope. We expect that this interesting behavior, which has no classical counterpart, likely arises from an inherent coupling between the random intensity and the random phase together with the plasmonic scattering. Nevertheless, this motivates the use of the $\sinc^2((k_1-k_2+k')/\sigma)$ envelope to model the observed multiparticle coherence in Chapter 4.

\section{Conditional Detection of Vacuum Events}
To calculate the probability of detecting vacuum events, we model our system using five beam splitters. Among them, only BS$_1$ truly exists in the experimental setup, which separates the pre-selection arm; while BS$_2$ is used to simulate the plasmonic interaction. Finally, BS$_3$, BS$_4$ and BS$_5$ are fictitious beam splitters, which are used to model the losses in one pre-selection arm and two post-selection arms. More specifically, for these fictitious beam splitters, the photons transmit through each of the beam splitters correspond to those received by the actual detectors in the experiment; whereas the photons reflected from the beam splitters are considered as losses. This approach establishes six distinct outcomes for all input photons, allowing us to calculate the probability of obtaining $n_1, n_2, n_3, n_4, n_5$ and $n_6$ across these six modes as
\begin{equation}
\begin{split}
\label{Eq:condition_prob}
P(n_1,n_2,n_3,n_4,n_5,n_6)=&\sum_{k}^{n}\binom{k}{n}\Gamma(n+\frac{1}{2}-k)\Gamma(\frac{1}{2}+k)\times\frac{1}{\pi}\frac{\Bar{n}^n}{(\Bar{n}+1)^{n+1}}\frac{1}{n_1!n_2!n_3!n_4!n_5!n_6!} \\
&(\sin\theta_1\cos\theta_4)^{2n_1}(\cos\theta_1\sin\theta_2\cos\theta_5)^{2n_2}(\cos\theta_1\cos\theta_2\cos\theta_3)^{2n_3} \\
&(\sin\theta_1\sin\theta_4)^{2n_4}(\cos\theta_1\sin\theta_2\sin\theta_5)^{2n_5}(\cos\theta_1\cos\theta_2\sin\theta_3)^{2n_6}.
\end{split}
\end{equation}

Here, $n_1, n_2$ and $n_3$ represent the number of photons obtained by the actual detectors; while $n_4, n_5$ and $n_6$ represent the losses. For simplicity, we define $n=n_1+n_2+n_3+n_4+n_5+n_6$. Additionally, $\theta_1, \theta_2, \theta_3, \theta_4$ and $\theta_5$ represent coefficients that elucidate the relationship between transmitted and reflected photons across the five beam splitters. Furthermore, $\bar{n}$ is the mean photon number characterizing the overall beam intensity; and $\Gamma(x)$ represents the Euler gamma function. In the experiment, we obtain the probabilities of getting $n_1, n_2$ and $n_3$. Therefore, by fitting our experimental data to Eq. (\ref{Eq:condition_prob}), we can determine the values of the parameters $\theta_1, \theta_2, \theta_3, \theta_4, \theta_5$ and $\Bar{n}$. We can thus estimate the probabilities of detecting vacuum events from an arbitrary input. For example, $P(0,0,0,0,0,0)$ denotes the probability of getting vacuum events when the input is vacuum; while $P(0,0,0,1,0,0)+P(0,0,0,0,1,0)+P(0,0,0,0,0,1)$ signifies the probability of encountering vacuum events when the input is a single photon. In this case, the single photon does not arrive at any of the three actual detectors. On the basis of this approach and our experimental data, we can calculate the probabilities shown in Table. \ref{tab}. Notably, the unconditional probabilities (without any pre- or post-selections) correspond to the data shown in Fig. 1b; whereas the conditional probabilities are detected with one pre-selection arm and two post-selection arms. From Table. \ref{tab} we demonstrate that pre- and post-selection enhances the probability of detecting vacuum events. 

In the first column of Table. \ref{tab}, we show the photon-number distribution of detecting vacuum events. If we list all the possible inputs and sum the probabilities in the first column all together, we will obtain the highest pillar in Fig. 1b, which is around 10\%. In the second column, we implement the pre- and post-selection to get the conditional probabilities by dividing the photon-number distribution obtained from different input with the probability of getting vacuum in all the three detectors. 

\begin{table}[!htbp]
\centering
\caption{\label{tab}Estimated probabilities of vacuum events using multiparticle input.}
{\def\arraystretch{1}\tabcolsep=8pt
\begin{tabular}{cccc}
\hline
Input  & Unconditional Probability  & Conditional Probability   \\ \hline
Vacuum  & 7.775\%  & 82.83\%  \\
1 particle  & 1.553\%  & 14.22\%  \\
2 particles  & 0.3103\%  & 2.442\%  \\
3 particles  & 0.06198\%  & 0.4193\%  \\ \hline
\end{tabular}
}
\end{table}

\section{Properties of the degree of second-order coherence \texorpdfstring{$\tilde{g}^{(2)}$}{g(2)}}

Post-selection is the process by which the data collected through a PNR detection is filtered in favor of a particular number event. For example, if one were to post-select on $7$-photon events, then all data points without $7$ photons would be given a value of $0$ and all data points with $7$ photons would be given a value of $1$. In this way, post-selective measurements can be used to study the dynamics of particular multiphoton wavepackets. The mathematical operator which describes a post-selective measurement on $N$ photons is simply the Fock-state projection operator $|N \rangle\langle N|$, since it has the desired property that its eigenvalues are $1$ for the $N$-photon Fock state and $0$ for all other Fock states. Post-selective measurements are more interesting when there is more than one mode present, so let's suppose that our state has two modes $a$ and $b$. A helpful shorthand for the post-selection operators is given by:
\begin{equation}
\begin{aligned}
    \hat{n}_{aN} =&\text{ } |N \rangle_a\langle N|_a\otimes \mathbb{I}_b,\\
    \hat{n}_{bM} =&\text{ } \mathbb{I}_a \otimes |M \rangle_b\langle M|_b.
\end{aligned}
\end{equation}
The advantage to considering a state with at least two modes is that we can now study the correlations between multiphoton wavepackets. This is accomplished by means of the wavepacket correlation function, given by
\begin{equation}
    \begin{aligned}
        \tilde{g}^{(2)}(N,M) = \frac{\text{Tr}\left[\hat{\rho}\hat{n}_{aN}\hat{n}_{bM}\right]}{\text{Tr}\left[\hat{\rho}\hat{n}_{aN}\right]\text{Tr}\left[\hat{\rho}\hat{n}_{bM}\right]}
    \end{aligned}
\end{equation}
where $\hat{\rho}$ is the two-mode quantum state in question. For a concrete example of why this quantity is interesting (which is also of substantial relevance to this particular chapter), let us consider the case of a thermal state which has passed through a beam-splitter of angle $\theta$ (which takes the annihilation operator $\hat{a}$ to the annihilation operator $\hat{a}\cos(\theta) + \hat{b}\sin(\theta)$). The initial state is given by
\begin{equation}
    \hat{\rho}_0 = \sum_{n=0}^\infty \frac{\bar{n}^n}{(1+\bar{n})^{n+1}}|n,0\rangle\langle n,0|.
\end{equation}
After the beam-splitter, it will become
\begin{equation}
    \hat{\rho} = \sum_{n=0}^\infty \frac{\bar{n}^n}{(1+\bar{n})^{n+1}}\sum_{k,l=0}^n \sqrt{\binom{n}{k}\binom{n}{l}}\cos^{k+l}(\theta)\sin^{2n-k-l}(\theta)|k,n-k\rangle\langle l,n-l|.
\end{equation}

Recall that the intensities in the arms of the beam-splitter's output are $\bar{n}\cos^2(\theta)$ and $\bar{n}\sin^2(\theta)$, and that the correlation function is given by $g^{(2)} = 2$. Due to the fact that $g^{(2)} > 1$, a thermal state is commonly thought of as a classical state of light. However, as we shall soon see, post-selection paints a different story about the classicality of a thermal state's multiphoton subsystems. One can show that
\begin{equation}
    \begin{aligned}
        \text{Tr}\left[\hat{\rho}\hat{n}_{aN}\right] &= \frac{\bar{n}^N}{(1+\bar{n}\cos^2(\theta))^{N+1}}\cos^{2N}(\theta),\\
        \text{Tr}\left[\hat{\rho}\hat{n}_{bM}\right] &= \frac{\bar{n}^M}{(1+\bar{n}\sin^2(\theta))^{M+1}}\sin^{2M}(\theta),\\
        \text{Tr}\left[\hat{\rho}\hat{n}_{aN}\hat{n}_{bM}\right] &= \frac{\bar{n}^{N+M}}{(1+\bar{n})^{N+M+1}}\binom{N+M}{N}\cos^{2N}(\theta)\sin^{2M}(\theta),\\
        \tilde{g}_{\text{th}}^{(2)}(N,M) &= \binom{N+M}{N}\left[\frac{(1+\bar{n}\cos^2(\theta))^{N+1}(1+\bar{n}\sin^2(\theta))^{M+1}}{(1+\bar{n})^{N+M+1}}\right].
    \end{aligned}
\end{equation}
This wavepacket correlation function has the interesting property that, while $\tilde{g}_{\text{th}}^{(2)}(N,M)$ is larger than $1$ for $N = M$, it is smaller than $1$ when $N$ and $M$ are very different from one-another. Therefore, these post-selective measurements have given us access to the underlying multiphoton scattering which govern the photon-number statistics of thermal light. This particular expression is relevant to our study of plasmons, as the input beam follows thermal photon-number statistics and the density matrix we use to make post-selective predictions is built on this model of a thermal state which has undergone a beam-splitter transformation.

\section{Spectral-spatial response of the plasmonic sample}

The spatial interference fringes generated by our plasmonic sample across various wavelengths are depicted in Fig. \ref{Fig5} \textbf{b} to \textbf{c}. We explored this response using full-wave electromagnetic simulations. These simulations were executed utilizing a Maxwell's equation solver based on the finite difference time domain method (Lumerical FDTD). To accurately capture the behavior, we accounted for the dispersion of the materials constituting the structure by incorporating their frequency-dependent permittivities. The spectral-spatial response of our plasmonic sample were indeed captured by Eq. (S11) of this appendix or Eq. (2) in Chapter 4. The permittivity values crucial to our analysis were sourced from specific references: the permittivity of the gold film was extracted from reference \cite{Johnson1972}. For the glass substrate (BK7), we relied on the manufacturer's specifications, ensuring precision in our computations. Likewise, the permittivity of the index matching fluid (Cargille oil Type B 16484) was determined through meticulous extrapolation based on the manufacturer's specification.

\chapter{Supplementary information for Chapter 5 - Conditional multiparticle quantum plasmonic sensing}

In this appendix, we present the details of FDTD simulation, as well as the derivation
of the degree of second-order coherence, particle statistics, and conditioned signal-to-noise ratio, as the supplementary information for Chapter 5.

\section{FDTD simulation}
The design of the plasmonic structure given in Fig. \ref{Fig9}\textbf{a} is simulated with a 2-D FDTD simulations by a \SI{130}{\micro\meter} domain in $x$ direction and  \SI{8}{\micro\meter} along the $y$ direction. The boundary condition is satisfied via the perfect matching layers to efficiently absorb the light scattered by the structure. Besides, the simulations time was long enough so that all energy in the simulation domain was completely decayed. The upper clad is made of CYTOP, a polymer with refractive index that closely matches the refractive index of 1.33. The mesh size was as small as $0.03 $ nm along $x$ direction and where we have highly confined field propagation. To create the propagating plasmonic modes, we use a pair of mode sources in both sides of the central slit. The generated SP modes propagate toward the central slit where they interfere. The near-fields along a linear line underneath the nanostrucutre were extracted and used for the far-field analysis. The coupled light to the mode $\hat{e}$, i.e. $T_{\text{ph}}$, was  calculated by the power flow through to the same linear line beneath the slit normalized to the input power.   To have a realistic estimation of the subtracted light, the mode $\hat{d}$  was first propagated for a distance of  10$\lambda$ (8.1 \SI{}{\micro\meter}) along the gold-glass interface and then a grating coupling efficiency of 36\% was considered to out couple the plasmonic mode to the free space  \cite{biosensordeleon}. The out-coupling was done far from the slit to avoid interactions of slit near-fields with fields of the assumed grating.

\begin{figure}[ht!]
	\centering \includegraphics[width=1\textwidth]{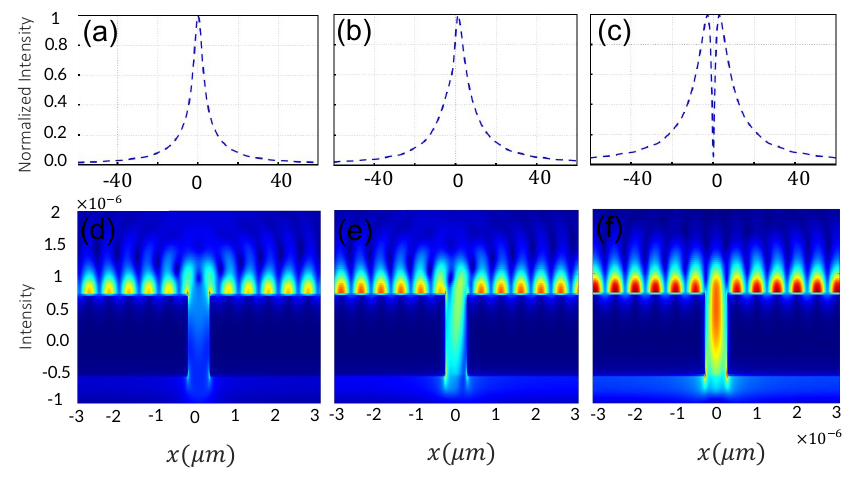}
	\caption[Normalized near-field intensity distribution scattered by a plasmonic nanoslit]{Normalized near-field intensity distribution scattered by a plasmonic nanoslit. The blue dashed line indicates the interference pattern produced by the field transmitted through the 320-nm-wide slit, this corresponds to mode $\hat{e}$. The panels from (a) to (c) are obtained for $\varphi=0$, $\varphi = \pi/2$ and $\varphi = \pi$ respectively. The dashed line in all plots represents the intensity distribution of the fields transmitted through the slit indicative of dipolar and quadrupolar near-field symmetry for $\varphi=0$ and $\varphi = \pi$.  Panels (d),(e) and (f) depict the near-field intensity distribution $ | E|  $ corresponding to panels (a)-(c) respectively. } 
	\label{Fig12}
\end{figure}

\section{Derivation of the degree of second-order coherence, particle statistics, and conditioned SNR}

First, we calculate the second-order correlation function $g^{(2)}_L (0)$ associated to the L-plasmon-subtracted light field. We assume a thermal light field with Bose-Einstein statistics described by$\rho_{\text{th}}= \sum_{n=0}^{\infty} \text{p}_{\text{pl}}(n) {\ket n} {\bra n}$, where $\text{p}_\text{pl}(n)=\bar n^{n}/(1+\bar n)^{1+n}$. The subtraction of L-plasmon(s) from a single-mode thermal field gives
\begin{equation}
\rho_{L}=\frac{(\hat{a})^{L} \rho (\hat{a}^{\dagger})^{L} }{Tr((\hat{a})^{L} \rho (\hat{a}^{\dagger})^{L})} = \sum_{n=0}^{\infty} \frac{(n+L)! }{n! L!} \frac{\bar{n}^{n} }{(1+\bar{n})^{L+n+1}} {\ket n} {\bra n}=\text{p}_{\text{pl}}(n) {\ket n} {\bra n}
\label{eqgsup1}
\end{equation}
The second-order correlation function of a single-mode field is given by
\begin{equation}
g^{(2)}(0)=\frac{\langle\hat{a}^{\dagger}\hat{a}^{\dagger}\hat{a}\hat{a}\rangle}{\langle\hat{a}^{\dagger}\hat{a}\rangle^{2}}= \frac{\langle\hat{n}(\hat{n}-1)\rangle}{\langle\hat{n}\rangle^{2}} = \frac{\langle \hat{n}\rangle^{2} - \langle \hat{n}\rangle }{\langle \hat{n}\rangle^{2}}
\label{eqgsup2}
\end{equation}

We can now calculate each element in Eq. \ref{eqgsup2}. We have
\begin{equation}
\langle\hat{n}^2\rangle=\sum_{n=0}^{\infty}n^2\text{p}_{\text{pl}}(n)=(L+1)\bar{n}[(L+2)\bar{n}+1].
\label{eqgsup3}
\end{equation}

Similarly,
\begin{equation}
\langle\hat{n}\rangle=\sum_{n=0}^{\infty}n\text{p}_{\text{pl}}(n)=(L+1)\bar{n}.
\label{eqgsup4}
\end{equation}

Combining Eq. \ref{eqgsup2}, Eq. \ref{eqgsup3} and Eq. \ref{eqgsup4}, we obtain
\begin{equation}
g^{(2)}_L (0)=\frac{L+2}{L+1},
\label{eqgsup5}
\end{equation}
which is independent of the mean occupation number $\bar{n}$ of the input thermal field.

Now we derive Eq. \ref{eqg5}. First, we note that in our calculation, we assume that mode $\hat{a}$ and mode $\hat{c}$ come from the same input source. Following similar approaches to those presented in \cite{OmarOptica17},  for the lossless case, the mean occupation number of mode $\hat{e}$ is given by $\bar{n}_e=\bar{n}\xi\cos^2(\varphi/2)$. Here, $\bar{n}$ is the mean occupation number in the input modes $\hat{a}$ and $\hat{c}$, and $\xi$ represents the normalized transmission of the plasmonic tritter \cite{safari2019measurement}. However, we need to consider that the plasmonic structure induces loss, and we have non-unity detection efficiency. As discussed in Chapter 5, conditional measurements will change the mean occupation number of the mode $\hat{e}$. We first consider the situation in which no plasmons are subtracted (no conditional measurement is implemented). In this case, the average occupation number of mode $\hat{e}$ is simply modulated by the loss $\gamma$ of the plasmonic tritter, and the quantum efficiency $\eta_{\text{ph}}$ of the detector,
\begin{equation}
\bar{n}_e=\bar{n}\gamma\xi\eta_{\text{ph}}\cos^2(\frac{\varphi}{2}).
\label{eqgsup6}
\end{equation}

In this case, since no conditional measurement is made, the particle statistics are preserved. Therefore, the standard deviation is the same to that of a thermal field,
\begin{equation}
\Delta n_e=\sqrt{\bar{n}_e+\bar{n}_e^2}.
\label{eqgsup7}
\end{equation}

Therefore, the signal-to-noise ratio (SNR) is given by
\begin{equation}
\text{SNR}=\frac{\bar{n}_e}{\Delta n_e}=\frac{\bar{n}_e}{\sqrt{\bar{n}_e+\bar{n}_e^2}}=\frac{\sqrt{\bar{n}\gamma\xi\eta_{\text{ph}}\cos^2(\frac{\varphi}{2})}}{\sqrt{1+\bar{n}\gamma\xi\eta_{\text{ph}}\cos^2(\frac{\varphi}{2})}}.
\label{eqgsup8}
\end{equation}

Now we consider the conditional subtraction of plasmons. The $L$-plasmon subtracted state $\rho_e(L)$ of mode $\hat{e}$ is conditioned on detection of $L$ plasmon(s) in mode $\hat{d}$ \cite{Allevi2010},
\begin{equation}
\rho_e(L)=\frac{1}{\text{p}_d(L)}\text{Tr}_d[\rho\mathrm{II}\otimes\Pi_L(\eta_{\text{pl}})].
\label{eqgsup9}
\end{equation}

Specifically, $\text{p}_d(L)$ is the probability of measuring $L$ plasmon(s) in mode $\hat{d}$. Since the transformation of the plasmonic tritter preserves the particle statistics, mode $\hat{d}$ still possesses thermal statistics,
\begin{equation}
\text{p}_d(L)=\frac{(\bar{n}_d)^n}{(1+\bar{n}_d)^{n+1}},
\label{eqgsup10}
\end{equation}
where $\bar{n}_d=\bar{n}\gamma\xi\eta_{\text{pl}}\sin^2(\varphi/2)$. Additionally, without loss of generality, we describe the initial state $\rho$ before conditional measurements as
\begin{equation}
\rho=\sum_{n=0}^{\infty}\text{p}_\text{pl}(n)\sum_{k,l=0}^{n}A_k^n(\xi)A_l^n(\xi)\ket{n-k}\bra{n-l}\otimes\ket{k}\bra{l},
\label{eqgsup11}
\end{equation}
which describes the two-mode state after the reduced plasmonic tritter transformation. We note that this reduced plasmonic tritter transformation is similar to the beam splitter transformation, therefore $A_k^n(\xi)=\sqrt{\binom{n}{k}\xi^{n-k}(1-\xi)^k}$.

Finally, the positive-operator-valued measure (POVM) of a realistic photon-counting device with quantum efficiency $\eta$ is given by \cite{Allevi2010}:
\begin{equation}
\Pi_L(\eta)=\sum_{m=0}^{\infty}B_{m,L}(\eta)\ket{m}\bra{m},
\label{eqgsup12}
\end{equation}
in which $B_{m,L}(\eta)=\binom{m}{l}\eta^L(1-\eta)^{m-L}$. Combining the above equations, we have
\begin{equation}
\rho_e(L)=\frac{1}{\text{p}_d(L)}\sum_{m=0}^{\infty}\sum_{n=0}^{\infty}B_{m,L}(\eta_{\text{pl}})\text{p}_\text{pl}(m+n)[A_m^{m+n}(\xi)]^2\ket{n}\bra{n}.
\label{eqgsup13}
\end{equation}

Then we can calculate the conditional mean occupation number using Eq. \ref{eqgsup13},
\begin{equation}
\bar{n}_e= \bar{n} \gamma \xi \eta_{\text{ph}}\cos^{2}{(\frac{\varphi}{2})}[\frac{(L+1)}{1+\bar{n}\gamma(1-\xi) \eta_{\text{pl}}\cos^{2}{(\frac{\varphi}{2})}}].
\label{eqgsup14}
\end{equation}

Similarly, one can calculate the standard deviation of the number of detection events of mode $\hat{e}$, when conditioned on the detection of $L$ plasmons,
\begin{equation}
\Delta n_e=\frac{\bar{n}_e}{\sqrt{\frac{(1+L)\bar{n}\gamma\xi\eta_{\text{ph}}\cos^{2}{(\frac{\varphi}{2})}}{1+\bar{n}\gamma(\xi\eta_{\text{ph}}+(1-\xi)\eta_{\text{pl}})\cos^{2}{(\frac{\varphi}{2})}}}}.
\label{eqgsup15}
\end{equation}

Finally, the $L$-plasmon subtracted signal-to-noise ratio (SNR) is given by
\begin{equation}
\text{SNR}=\frac{\bar{n}_e}{\Delta n_e}=\sqrt[]{\frac{(1+L) \bar n \gamma\eta_{\text{ph}} \xi \cos^{2}(\frac{\varphi}{2})}{1+ \bar n  \gamma(\xi\eta_{\text{ph}}+(1-\xi)\eta_{\text{pl}}) \cos^{2}(\frac{\varphi}{2})}}.
\label{eqgsup16}
\end{equation}

\chapter{Supplementary information for Chapter 6 - Multiphoton quantum imaging using natural light}

In this appendix, we present the probability of observing multiphoton events, and a detailed derivation of the equations presented in Chapter 6. 

\section{Probability of observing multiphoton events}

In this section, we provide additional experimental results. In Table S1, we report the values associated with the probability of measuring specific multiphoton events from a thermal light beam. Specifically, Table S1 presents the probability of observing a particular number of photons under the post-selection scheme of Fig. 2. Similarly, Table S2 presents the joint probabilities of measuring a particular number of photons in two separate arms under the post-selection scheme of Fig. 3. Lastly, Table S3 presents the probability of observing a particular number of photons in the second arm under the photon-subtraction scheme of Fig. 4.

\begin{table}[!htbp]
\centering
\caption{The measured probability of post selection.}
\label{tab:sitable1}
\begin{tabular}{|p{1.25cm}<{\centering}|p{1.35cm}<{\centering}|p{1.35cm}<{\centering}|p{1.35cm}<{\centering}|p{1.35cm}<{\centering}|p{1.35cm}<{\centering}|p{1.35cm}<{\centering}|p{1.35cm}<{\centering}|p{1.35cm}<{\centering}|}
\hline
$\bar{n}$&$|0\rangle\langle0|$& $ |1\rangle\langle1|$ &$ |2\rangle\langle2| $ &$ |3\rangle\langle3|$& $|4\rangle\langle4|$& $ |5\rangle\langle5|$& $ |6\rangle\langle6|$& $ |7\rangle\langle7|$\\
\hline
$0.8 $&$55.17\%$ &$26.11 \%$ &$10.76\%$ &$4.51\%$& $ 1.95\%$& $ 0.87\%$ & $ 0.40\%$&$ 0.18\%$\\
\hline
\end{tabular}
\end{table}

\begin{table}[!htbp]
\centering
\caption{The measured probability of correlation between the two arm in the source.}
\label{tab:sitable2}
\begin{tabular}{|p{2.5cm}<{\centering}|p{1.5cm}<{\centering}<{\centering}|p{1.2cm}<{\centering}|p{1.2cm}<{\centering}|p{1.2cm}<{\centering}|p{1.2cm}<{\centering}|p{1.2cm}<{\centering}|p{1.2cm}<{\centering}|p{1.2cm}<{\centering}|p{1.2cm}<{\centering}|}
\hline
\diagbox{Arm1}{Arm2} & $|0\rangle\langle0|$& $ |1\rangle\langle1|$ &$ |2\rangle\langle2| $ &$ |3\rangle\langle3|$& $|4\rangle\langle4|$& $ |5\rangle\langle5|$& $ |6\rangle\langle6|$& $ |7\rangle\langle7|$\\
\hline
$|0\rangle\langle0|$&$15.99\%$ &$9.40 \%$ &$4.26\%$ &$1.78\%$& $ 0.73\%$& $ 0.28\%$ & $ 0.11\%$&$ 0.04\%$\\
\hline
$ |1\rangle\langle1|$ &$7.53\%$ &$7.05 \%$ &$4.64\%$ &$2.65\%$& $ 1.41\%$& $ 0.68\%$ & $ 0.33\%$&$ 0.14\%$\\
\hline
$ |2\rangle\langle2| $ &$2.70\%$ &$3.68 \%$ &$3.25\%$ &$2.39\%$& $ 1.54\%$& $ 0.92\%$ & $ 0.50\%$&$ 0.28\%$\\
\hline
$ |3\rangle\langle3|$&$0.89\%$ &$1.64 \%$ &$1.86\%$ &$1.70\%$& $ 1.33\%$& $ 0.94\%$ & $ 0.61\%$&$ 0.37\%$\\
\hline
$|4\rangle\langle4|$&$0.27\%$ &$0.65 \%$ &$0.93\%$ &$1.02\%$& $ 0.93\%$& 
$ 0.77\%$ & $ 0.57\%$&$ 0.40\%$\\
\hline
$ |5\rangle\langle5|$&$0.08\%$ &$0.24 \%$ &$0.42\%$ &$0.54\%$& $ 0.58\%$& 
$ 0.54\%$ & $ 0.46\%$&$ 0.36\%$\\
\hline
$ |6\rangle\langle6|$&$0.02\%$ &$0.08 \%$ &$0.18\%$ &$0.26\%$& $ 0.32\%$& 
$ 0.35\%$ & $ 0.34\%$&$ 0.29\%$\\
\hline
$ |7\rangle\langle7|$&$0.007\%$ &$0.02 \%$ &$0.06\%$ &$0.12\%$& $ 0.16\%$& 
$ 0.19\%$ & $ 0.21\%$&$ 0.20\%$\\
\hline
\end{tabular}
\end{table}

\begin{table}[!htbp]
\centering
\caption{The measured probability of photon subtraction.}
\label{tab:table2}
\begin{tabular}
{|p{1.5cm}<{\centering}|p{1.5cm}<{\centering}|p{1.5cm}<{\centering}|p{1.5cm}<{\centering}|p{1.5cm}<{\centering}|}
\hline
$\bar{n}$&$N=0$&$N=1$& $ N=2$ &$ N=3 $\\
\hline
$0.08 $&$97\%$ &$2.1 \%$ &$0.3\%$& $0.05\%$\\
\hline
\end{tabular}
\end{table}

\section{Detailed derivation of equations}

Here we provide a detailed derivation of the equations presented in Chapter 6. The initial quantum state of our signal is a weak, single-mode thermal state. Written explicitly in the Fock basis, our initial thermal state of light is represented by 
\begin{equation}
    \hat{\rho}_{0} = \sum_{n=0}^\infty \frac{\bar{n}^n}{(1+\bar{n})^{n+1}}\left|n\rangle\langle n\right|,
\end{equation}
where $\bar{n}$ is the mean number of photons of the state. In our experiment, we uniformly illuminate an object with this state, producing a signal with a new state that has a different mode structure. The mode information is contained within the annihilation operator $\hat{a}$ which obeys $\hat{a}|n\rangle = \sqrt{n}|n-1\rangle$, defined in terms of the operator-valued distribution $\hat{a}(\Vec{\boldsymbol{x}})$ by
\begin{equation}
    \hat{a} = \int d^2 x f(\Vec{\boldsymbol{x}})\hat{a}(\Vec{\boldsymbol{x}}),
\end{equation}
where $\left[\hat{a}(\Vec{\boldsymbol{x}}),\hat{a}^\dagger(\Vec{\boldsymbol{x}}')\right] = (2\pi)^2\delta(\Vec{\boldsymbol{x}}-\Vec{\boldsymbol{x}}')$ is the canonical commutation relation and $f(\Vec{\boldsymbol{x}})$ is the transverse profile of the beam. This expression assumes that the light is strongly peaked around a particular frequency, and in this case a transverse positional description can be used.

In our experiment, we uniformly illuminate an object using the thermal state, which forms an image that we would like to measure. We do this by discretizing the transverse spatial profile of the mode into $X$ squares which we will call pixels. This is equivalent to the transformation taking $\hat{a}$ to $\sum_{i=1}^X \lambda_i \hat{A}_i$ where $\hat{A}_i$ is the annihilation operator for the mode at the $i^{\text{th}}$ pixel and $\lambda_i$ is its weight. Since the object was illuminated uniformly, $\lambda_i$ will be either $0$, representing a pixel with no light, or some constant value, representing a pixel with light, such that $\sum_{i=1}^X|\lambda_i|^2 = 1$. It is important to note, however, that this theory will also apply for non-uniform illuminations. This allows us to define the ideal image vector $\Vec{\boldsymbol{s}}_0\in \mathbb{R}^X$ where each component $s_{0,i}$ is equal to $|\lambda_i|^2\bar{n}$.

We now collect random combinations of these pixels onto a single-pixel camera that employs photon-number-resolving detection. We will see later how this allows for image reconstructions which use fewer measurements than traditional methods require. We will perform $M$ such measurements, and each random selection of pixels will be represented by the covector $\Vec{\boldsymbol{Q}}_t\in\mathbb{R}^{X*}$ which consists of zeros and ones. It follows that, after the signal has been filtered by this covector, the resulting mode operator of the signal will be given by $\hat{a}_t=\sum_{i=0}^X Q_{t,i}\lambda_i' \hat{A}_i$ where $\lambda_i' = \lambda_i/\sqrt{\sum_{i=0}^\infty Q_{t,i}|\lambda_i|^2}$ is the re-normalized weight of each pixel. The quantum state of the signal after this filtering process is therefore thermal, with a mean-photon-number given by $\bar{n}_t = \Vec{\boldsymbol{Q}}_t\cdot \Vec{\boldsymbol{s}}_0$, and can be written as
\begin{equation}
    \hat{\rho}_{\boldsymbol{Q},t} = \sum_{n=0}^\infty \frac{\bar{n_t}^n}{(1+\bar{n}_t)^{n+1}}\left|n\rangle\langle n\right|.
\end{equation}
Here we are using the label $\boldsymbol{Q}$, which represents the matrix of pixel filtrations and is defined by $\boldsymbol{Q} = \bigoplus_{t=1}^M \Vec{\boldsymbol{Q}}_t$. We can simultaneously write all such density matrices as
\begin{equation}
    \boldsymbol{\hat{\rho}}_{\boldsymbol{Q}} = \bigoplus_{t=1}^M \hat{\rho}_{\boldsymbol{Q},t},
\end{equation}
which can be thought of as a vector of density matrices. 

We now use the Gluaber-Sudarshan $P$ function representation of the quantum state, written in terms of coherent states $|\alpha\rangle$, and given by
\begin{equation}
    \boldsymbol{\hat{\rho}}_{\boldsymbol{Q}} = \bigoplus_{t=1}^M \int d^2\alpha \frac{1}{\pi \bar{n}_t}e^{-\frac{|\alpha|^2}{\bar{n}_t}}\left|\alpha\rangle\langle\alpha\right|.
\end{equation}
Before measuring the state with our photon-number-resolving detector, we will send it through a fiber-coupler in order to produce a second mode that can be used for the photon-subtraction technique which we will discuss later. After this transformation, represented by taking annihilation operator $\hat{a}_t$ to annihilation operators $\hat{a}_t\cos(\theta) + i \hat{b}_t\sin(\theta)$ where $\theta$ is the beam-splitter angle, the state is given by
\begin{equation}
    \boldsymbol{\hat{\rho}}_{\boldsymbol{Q}} = \bigoplus_{t=1}^M \int d^2\alpha \frac{1}{\pi \bar{n}_t}e^{-\frac{|\alpha|^2}{\bar{n}_t}}\left|\alpha\cos(\theta),i\alpha\sin(\theta)\rangle\langle\alpha\cos(\theta),i\alpha\sin(\theta)\right|_{a,b}.
\end{equation}
We will use the labels $a,b$ to represent the two output modes of the fiber-coupler.

From here, we make use of two photon-number-resolving detectors, one in each arm, to perform measurements. The primary difficulty of this measurement scheme is that the signal's strength is comparable to the noise of our two detectors, and the measurement techniques which we will employ are meant to alleviate the effects of that noise. The impacts of noise and detector efficiencies can be modeled with the photocounting technique, by which for a given state $\hat{\rho} = \sum_{n=0}^\infty p(n,m)\left|n\rangle \langle m\right|$, its diagonal matrix elements $p_{\text{noise}}(n,n)$ with dark counts $\nu$ and detector efficiency $\eta$ accounted for can be computed by
\begin{equation}
    p_{\text{loss}}(n,n) = \left\langle:\frac{(\eta\hat{n}+\nu)^n}{n!}e^{-(\eta\hat{n} + \nu)}:\right\rangle,
\end{equation}
where $:\cdot:$ is the normal-ordering prescription. In our case, these diagonal elements can be computed for dark counts $\nu_{a/b}$ and detector efficiencies $\eta_{a/b}$ as
\begin{equation}
    \begin{aligned}
               \Vec{\boldsymbol{p}}_{\boldsymbol{Q}}(n, m)=&\bigoplus_{t=1}^M\left\langle: \frac{\left(\eta_a \hat{n}_a+\nu_a\right)^n}{n !} e^{-\left(\eta_a \hat{n}_a+\nu_a\right)} \otimes \frac{\left(\eta_b \hat{n}_b+\nu_b\right)^m}{m !} e^{-\left(\eta_b \hat{n}_b+\nu_b\right)}:\right\rangle\\
               =&\bigoplus_{t=1}^M \int d^2\alpha \frac{1}{\pi \bar{n}_t}e^{-\frac{|\alpha|^2}{\bar{n}_t}} \frac{\left(\eta_a |\alpha|^2\cos^2(\theta)+\nu_a\right)^n}{n !} e^{-\left(\eta_a |\alpha|^2\cos^2(\theta)+\nu_a\right)}\\
               &\times\frac{\left(\eta_b |\alpha|^2\sin^2(\theta)+\nu_b\right)^m}{m !} e^{-\left(\eta_b |\alpha|^2\sin^2(\theta)+\nu_b\right)} \\
               =& \bigoplus_{t=1}^M\frac{e^{-\nu_a-\nu_b}}{n! m!}\sum_{i=0}^n\sum_{j=0}^m \binom{n}{i}\binom{m}{j}\eta_a^i\eta_b^j\nu_a^{n-i}\nu_b^{n-j}\cos^{2i}(\theta)\sin^{2j}(\theta)\\
               &\times\int d^2\alpha \frac{|\alpha|^{2i+2j}}{\pi \bar{n}_t}e^{-\frac{|\alpha|^2}{\bar{n}_t} - \eta_a|\alpha|^2\cos^2(\theta) - \eta_b|\alpha|^2\sin^2(\theta)} \\
               =&\bigoplus_{t=1}^M\frac{e^{-\nu_a-\nu_b}}{\bar{n}_t n! m!}\sum_{i=0}^n\sum_{j=0}^m \binom{n}{i}\binom{m}{j}(i+j)!\\
               &\times\frac{\eta_a^i\eta_b^j\nu_a^{n-i}\nu_b^{m-j}}{\left(\frac{1}{\bar{n}_t}+\eta_a\cos^2(\theta)+\eta_b\sin^2(\theta)\right)^{1+i+j}}\cos^{2i}(\theta)\sin^{2j}(\theta).
   \end{aligned}
\end{equation}
Unfortunately, the finite sum in the last line of this expression does not have a nice analytical form. However, since it is a finite sum, these diagonal matrix elements can be easily calculated numerically. 
When the signal $\bar{n}_t$ is absent, we will detect only the noise. In this case, the joint probability of noise event is given by $p_n(k,l) = p_{n,a}(k)p_{n,b}(l)$, where $p_{n,i}(k) = e^{-\nu_i}\frac{\nu_i^k}{k!}$. We note that our ability to reliably reconstruct the signal's mode profile from our measurements hinges on each of the $M$ measurements in arm $a$ being clearly distinguishable from its background noise. In other words, the signal-to-noise ratio for each measurement should be as high as possible. Let us now discuss two methods for accomplishing this.

The first method is that of post-selection (Fock-projection) in arm $a$. This method does not utilize arm $b$, so that arm will always be traced out here. The signal-to-noise ratio in the case where we post-select on $N$ photons in arm $a$ can be represented by a vector, and is written as
\begin{equation}
    \overrightarrow{\textbf{SNR}}_{\text{post}}(N) = \frac{\sum_{m=0}^\infty \Vec{\boldsymbol{p}}_{\boldsymbol{Q}}(N,m)}{p_{n,a}(N)}=\frac{\Vec{\boldsymbol{p}}_{\boldsymbol{Q}}(N)}{p_{n,a}(N)}.
\end{equation}
Numerical evaluations of this quantity show that each component of the signal-to-noise ratio vector is increasing in an approximately exponential fashion with respect to $N$. This shows that we can greatly reduce the impact of noise on our data by post-selecting on high photon numbers.

The other method showcased in this chapter is that of photon-subtraction, by which we first make a post-selective measurement in arm $b$ on $N$ photons and then measure the photon events in arm $a$. The conditional intensity in arm $a$ can then be written as $\langle\hat{\boldsymbol{n}}_a\rangle_N = \bigoplus_{t=0}^M\left(\sum_{k=0}^\infty k p_{\boldsymbol{Q},t}(k,N)\right)/\left(\sum_{k=0}^\infty p_{\boldsymbol{Q},t}(k,N)\right)$, where the factor in the denominator is due to the renormalization of the state after the measurement in arm $b$. Similarly, the noise measurement can be written as $\langle\hat{n}_a\rangle_{N,0} = \bigoplus_{t=0}^M\left(\sum_{k=0}^\infty k p_n(k,N)\right)/\left(\sum_{k=0}^\infty p_n(k,N)\right)$. By taking this approach, the resulting signal-to-noise ratio seen in arm $a$ can be represented by
\begin{equation}
    \overrightarrow{\textbf{SNR}}_{\text{sub}}(N) = \frac{\langle\hat{\boldsymbol{n}}_a\rangle_N}{\langle\hat{n}_a\rangle_{N,0}}.
\end{equation}
In contrast to the post-selection case, each component in this vector increases in an approximately linear fashion with respect to $N$. While this may be less desirable when compared to the exponential trend of post-selection in arm $a$, it is useful when precise post-selective measurements in arm $a$ cannot be made. For instance, if $\bar{n}_t$ is very large, then we can choose $\theta$ to be very small so that photon-number-resolution can be made accurate in arm $b$. This would allow us to increase the signal-to-noise ratio in arm $a$ through photon subtraction while making the more-precise measurement of intensity in that arm.

Finally, our measurements in arm $a$ will be used to form a reconstruction of the signal vector, $\Vec{\boldsymbol{s}}_0$. This is accomplished using the compressive sensing (CS) technique, by which the reconstructed image, represented by $\Vec{\boldsymbol{s}}\in \mathbb{R}^X$, is found by minimizing the following quantity with respect to the dummy-vector $\Vec{\boldsymbol{s}}'\in \mathbb{R}^X$:
\begin{equation}
     \sum_{i=0}^X \lVert\nabla s_i'\rVert_{l_1} + \frac{\mu}{2}\lVert \boldsymbol{Q}\Vec{\boldsymbol{s}}' - \langle\hat{\boldsymbol{n}}\rangle\rVert_{l_2}.
\end{equation}
Here, $\langle\hat{\boldsymbol{n}}\rangle$ could be replaced with either of the previously-described quantities,  $\Vec{\boldsymbol{p}}_{\boldsymbol{Q}}(N)$ or $\langle\hat{\boldsymbol{n}}_a\rangle_N$. Moreover, the $1$- and $2$-norm are denoted by $\lVert\cdot\rVert_{l_1}$ and $\lVert\cdot\rVert_{l_2}$, respectively. The discrete gradient operator is described by $\nabla$, and the penalty factor by $\mu$. The value of $\Vec{\boldsymbol{s}}'$ which minimizes this quantity is then the value which we ascribe to $\Vec{\boldsymbol{s}}$. Accurate reconstruction of this image vector, such that $\Vec{\boldsymbol{s}}$ agrees with $\Vec{\boldsymbol{s}}_0$, is sensitive to background noise, and so by reducing the impact of that noise as much as possible via either of the two methods described above, we can attain a more reliable image of the signal.





\backmatter


\bibliographystyle{nature_edited}
\bibliography{main}





\chapter{Vita}

Mingyuan Hong was born in Harbin, Heilongjiang Province, China, 1997. He received a Bachelor of Science degree in Atomic and Molecular Physics from the University of Science and Technology of China in June 2019, where he conducted undergraduate research on quantum dot single-photon sources and surface plasmons under the supervision of Prof. Chao-Yang Lu.

In August 2019, he began the Ph.D. program in the Department of Physics \& Astronomy at Louisiana State University, under the supervision of Prof. Omar S. Maga\~{n}a-Loaiza. His research focused on experimental quantum optics, including quantum plasmonics, quantum imaging, multiparticle quantum coherence, and quantum metrology. During his doctoral studies, he contributed to multiple peer-reviewed publications and presented his work at national and international conferences.

\end{document}